\documentclass[a4paper]{article}
\pdfoutput=1 

\usepackage{jcappub} 

\usepackage[T1]{fontenc} 
\usepackage[utf8]{inputenc}

\usepackage{amsmath}
\usepackage{amssymb}
\usepackage{bm}
\usepackage{bbold}
\usepackage[compat=1.0.0]{tikz-feynman}
\usepackage{multirow}

\usepackage{graphicx}
\usepackage{hyperref}
\hypersetup{colorlinks=true, linkcolor=teal, urlcolor=blue, citecolor=red}
\usepackage[american]{babel}
\usepackage{enumerate}
\usepackage{mathtools}
\usepackage{subcaption}
\usepackage{soul}
\setstcolor{red}
\usepackage{makebox}
\usepackage{comment}

\newcommand{\U}[1]{\mathrm{U}(1)_{\mathrm{#1}}}			

\renewcommand{\(}{\left(}
\renewcommand{\)}{\right)}

\newcommand{\del}{\partial}
\newcommand{\abs}[1]{\left| #1 \right| }
\newcommand{\mean}[1]{\left \langle #1 \right \rangle }

\usepackage{pifont}

\newcommand\varpm{\mathbin{\vcenter{\hbox{%
  \oalign{\hfil$\scriptstyle\hspace{-0.2ex}+\hspace{-0.2ex}$\hfil\cr
          \noalign{\kern-.5ex}
          $\scriptscriptstyle({-})$\cr}%
}}}}

\newcommand\scalemath[2]{\scalebox{#1}{\mbox{\ensuremath{\displaystyle #2}}}}

\newcommand\varmp{\mathbin{\vcenter{\hbox{%
  \oalign{\hfil$\scriptstyle\hspace{-0.2ex}-\hspace{-0.2ex}$\hfil\cr
          \noalign{\kern-.5ex}
          $\scriptscriptstyle({+})$\cr}%
}}}}

\newcommand{\red}[0]{\color{red}}

\usepackage[capitalise]{cleveref}

\crefname{section}{Sec.}{Secs.}
\crefname{table}{Tab.}{Tabs.}
\crefname{figure}{Fig.}{Figs.}
\crefname{equation}{Eq.}{Eqs.}
\crefname{appendix}{Appendix\ }{Appendix\ }

\graphicspath{{figures/}}

\allowdisplaybreaks

\title{Gravitational echoes of lepton number symmetry breaking with light and ultralight Majorons}

\author[a,b]{Andrea Addazi}
\author[c,b]{Antonino Marcian\`o}
\author[d,e]{Ant\'onio~P.~Morais}
\author[f]{Roman Pasechnik}
\author[g]{Jo\~ao Viana}
\author[c,h]{Hao Yang}

\affiliation[a]{Center for Theoretical Physics, College of Physics Science and Technology, Sichuan University, 610065 Chengdu, China.}
\affiliation[b]{Laboratori Nazionali di Frascati INFN, Frascati (Rome), Italy.}
\affiliation[c]{Center for Field Theory and Particle Physics \& Department of Physics, Fudan University, 200433 Shanghai, China.}
\affiliation[d]{Theoretical Physics Department, CERN, 1211 Geneva 23, Switzerland.}
\affiliation[e]{Departamento de F\'{i}sica da Universidade de Aveiro and Centre  for  Research  and  Development  in  Mathematics  and  Applications  (CIDMA),
	Campus de Santiago, 3810-183 Aveiro, Portugal.}
\affiliation[f]{Department of Physics, Lund University,
	221 00 Lund, Sweden.}
\affiliation[g]{Centro de F\'{i}sica Te\'{o}rica e Computacional, Faculdade de Ci\^{e}ncias,Universidade de Lisboa,
	 Campo Grande, Edif\'{i}cio C8 1749-016 Lisboa, Portugal.}
\affiliation[h]{Department of Astronomy, School of Physics and Astronomy, Shanghai Jiao Tong University, 200240 Shanghai, China.}

\emailAdd{addazi@scu.edu.cn}
\emailAdd{marciano@fudan.edu.cn}
\emailAdd{a.morais@cern.ch}
\emailAdd{roman.pasechnik@hep.lu.se}
\emailAdd{jfvvchico@hotmail.com}
\emailAdd{hyang19@fudan.edu.cn}

\abstract{We formulate a version of the low-scale Majoron model equipped with an inverse seesaw mechanism featuring lepton-number preserving dimension-6 operators in the scalar potential. Contrary to its dimension-4 counterpart, we find that the model can simultaneously provide light and ultralight Majorons, neutrino masses and their mixing, while featuring strong first-order cosmological phase transitions associated to the spontaneous breaking of the lepton number and the electroweak symmetries in the early Universe. We show by a detailed numerical analysis that under certain conditions on the parameter space accounted for in collider physics, the model can be probed via the primordial gravitational wave spectrum potentially observable at LISA and other planned facilities.}

\begin{document}

\begin{flushright}
CERN-TH-2023-054\\
\vskip1cm
\end{flushright}
	
	\maketitle
	\flushbottom

\section{Introduction}
\label{Sect:intro}

With the discovery of Gravitational Waves (GWs) by the LIGO detectors in 2015 \cite{LIGOScientific:2016aoc} a new era of multi-messenger astronomy has started. The most popular example to date was the observation of gravitational ripples by the LIGO and Virgo collaborations resulting from a neutron star binary merger \cite{LIGOScientific:2017vwq}, accompanied by a faint electromagnetic counterpart detected seconds later by the gamma-ray telescopes Fermi-GRB and INTEGRAL \cite{LIGOScientific:2017zic}. However, the possibilities for multi-messenger signals are by no means exhausted with the gravitational and electromagnetic (EM) interactions. Furthermore, the observation of a neutrino flux together with GWs and an EM signal from a supernova explosion would offer a weak interaction component to the observed event, enabling us to collect more information and broadening the scope of our understanding about such phenomena.

Another type of gravitational footprints that current and future experiments will be looking for traces back to the early moments of our Universe and is expected to be reflected in the form of a stochastic background. This can be a manifestation of e.g.~inflationary dynamics \cite{Mukhanov:1990me}, cosmic strings \cite{Vilenkin:1984ib} or strong first-order phase transitions (SFOPTs)\cite{Kosowsky:1992rz}, which we study in detail in this article. The latter can be responsible for a stochastic GW background generated by expanding and colliding vacuum bubbles of an energetically favoured vacuum configuration, which is expanding in a Universe filled up with a false vacuum phase. If the bubble wall velocity does not runaway, which can happen for very strong FOPTs, where the released latent heat during the transition, or equivalently the difference in the trace anomaly $\alpha$, is very large \cite{Ellis:2019oqb} and the wall velocity $v_\mathrm{w} \to 1$, the dominant contribution to the primordial GW spectrum results from the sound waves (SW) component. Furthermore, if $v_\mathrm{w}$ is larger than the Chapman-Jouget limit, i.e.~$v_\mathrm{w} > v_\mathrm{J}$, the peak frequency and amplitude of the primordial GW background is largely dominated by supersonic detonations \cite{Hindmarsh:2017gnf}, as we will consider in our numerical analysis.

The details of the phase-transitions are dependent on the underlying Particle Physics model. For example, in the Standard Model (SM), with the Higgs boson mass observed by the ATLAS and CMS experiments \cite{ATLAS:2015yey}, an electroweak (EW) scale SFOPT is not possible. However, the inclusion of a SM singlet in the scalar potential is sufficient to enhance the electroweak phase-transition to a strong one. Typically, the stronger the transition, the larger the GW peak amplitude and for scales not too far from the EW/TeV scale the frequency range can be in the reach of the \textit{Laser Interferometer Space Antenna} (LISA) mission \cite{LISA:2017pwj,Barausse:2020rsu}. While collider experiments are undoubtedly the preferred source for directly measuring new particles and couplings, GW interferometers such as LISA can potentially offer a gravitational portal to probe New Physics (NP) scenarios complementary to collider experiments. In some cases, GW detectors can even go beyond the reach of current and future collider experiments, both in the very-high or ultra-low energy limits \cite{Pati:1974yy,Croon:2018kqn,Bustillo:2020syj}.

In this article we study a class of Majoron models equipped with a spontaneously broken global lepton-number symmetry $\U{L}$ and a neutrino inverse seesaw mechanism. We investigate under which circumstances the Majoron is long-lived or even stable, thus becoming a potential Dark Matter (DM) candidate. We attribute the emergence of a seesaw scale $\Lambda$ to some unknown physics, which is currently beyond the reach of collider experiments, parameterizing the effects of such an unknown physics via $\U{L}$-preserving dimension-6 operators in the scalar potential. For a possible explicit breaking of a global $\U{L}$ via higher-dimensional operators induced by gravitational effects, see e.g.~Ref.~\cite{Akhmedov:1992hi}.

In this work, however, we solely consider a constrained set of the $\U{L}$-preserving dimension-6 operators such that the only explicit $\U{L}$ breaking effects come in the form of a soft Majoron mass term. The model yields a generic low-scale inverse seesaw mechanism, with a possibility to explain the smallness of the light active neutrino mass scale. As previously discussed by some of the authors, the standard dimension-4 Majoron model \cite{Addazi:2019dqt} cannot simultaneously predict visible GW in the reach of LISA while offering a possible good candidate for DM. This follows from the fact that the size of the scalar quartic portal coupling needed to enhance the potential barrier between false and true vacua is too large to comply with invisible Higgs decays at the LHC \cite{CMS:2022qva}, if the Majoron mass is lighter than a half of the Higgs boson mass. On the contrary, due to a richer structure of the dimension-6 operator extension of the model, a light Majoron state is found to be in full consistency with experimental bounds on invisible Higgs boson decays. Finally, the model features SFOPTs, which can produce primordial stochastic GW signals potentially observable at future experimental facilities such as LISA, or a next generation of interferometers such as BBO or DECIGO.

The article is organised as follows. In \cref{Sec:neut}, we review three canonical Majoron models, highlight their features and discuss whether low-scale lepton number symmetry breaking is to be expected. In \cref{Sec:model} we discuss the extended inverse seesaw model with dimension-six operators in the scalar sector, focusing on the observables that will be studied in the numerical analysis. In \cref{sec:Eff+GW} we review the basic details of the thermal effective potential and the spectrum of Gravitational Waves, in \cref{sec:Res} we present and discuss our results before concluding in \cref{sec:con}. 

\section{Setting the stage: Which Majoron model?}
\label{Sec:neut}

Majoron models are well motivated frameworks engineered to offer a mechanism for neutrino mass generation. These are typically equipped with a global $\mathrm{U}(1)_\mathrm{L}$ lepton number symmetry, broken by the VEV of a complex SM singlet scalar, $\sigma$, whose CP-odd component is the Majoron. Among the possible realizations it is worth briefly discussing the basic properties of the type-I (T1S), inverse (IS) and extended inverse seesaw (EIS) scenarios whose quantum numbers and particle content are shown in \cref{tab:models}.
\begin{table}[htb!]
    \begin{center}
        \begin{tabular}{c|ccccc|c} 
             & $L^i$ & $\nu_\mathrm{R}^i$ & $S^i$ & $\sigma$ & $H$ & Model \\
            \hline
            \multirow{3}{4em}{$\mathrm{U}(1)_\mathrm{L}$} & $1$ & $1$ & $\times$ & $-2$ & $0$ & T1S \\ 
            & $1$ & $1$ & $0$ & $-1$ & $0$ & IS\\ 
            & $1$ & $1$ & $-1$ & $2$ & $0$ & EIS \\ 
        \end{tabular}
        \caption{ \footnotesize Quantum numbers of the scalar and neutrino sectors of three distinct realizations of Majoron models. In the first line neutrinos masses are generated via a standard type-I seesaw mechanism whereas the second and third depict the inverse and extended inverse seesaw cases respectively. The $\times$ in the first line indicates the absence of the $S$ field while $i=1,2,3$ is a family index. As usual, $L$ and $H$ are electroweak lepton and Higgs doublets, $\nu_\mathrm{R}$ and $S$ denote SM-singlet neutrinos, whereas $\sigma$ is a SM complex scalar singlet.}
        \label{tab:models}  
    \end{center}
\end{table}

For each of the three Majoron models \cite{Boulebnane:2017fxw}, the corresponding neutrino-sector Lagrangian can be written as
\begin{equation}
\begin{aligned}
    \mathcal{L}_\nu^\mathrm{T1S} =& y_\nu^{ij} \overline{L}_i \Tilde{H} \nu_{\mathrm{R}j} + y_\sigma^{ij} \Bar{\nu}_{\mathrm{R}i}^c \nu_{\mathrm{R}j} \sigma
    + \mathrm{h.c.}\,,
    \\
    \mathcal{L}_\nu^\mathrm{IS} =& y_\nu^{ij} \overline{L}_i \Tilde{H} \nu_{\mathrm{R}j} + y_\sigma^{ij} \Bar{S}_i^c \nu_{\mathrm{R}j} \sigma +\Lambda^{ij} \Bar{S}_{i}^c S_j
    + \mathrm{h.c.}\,,
    \\
     \mathcal{L}_\nu^\mathrm{EIS} =& y_\nu^{ij} \overline{L}_i \Tilde{H} \nu_{\mathrm{R}j} + y_\sigma^{ij} \Bar{S}_i^c S_j \sigma + 
    y_\sigma^{\prime ij} \Bar{\nu}_{\mathrm{R}i}^c \nu_{\mathrm{R}j} \sigma^\ast +\Lambda^{ij} \Bar{\nu}_{\mathrm{R}i}^c S_j
    + \mathrm{h.c.}\,,
\end{aligned}
    \label{eq:Lnu}
\end{equation}
where
\begin{equation}
    L_i =
    \begin{pmatrix}
\nu_{\mathrm{L}i} \\
e_{\mathrm{L}i} 
\end{pmatrix}
\qquad
\textrm{and}
\qquad
\Tilde{H} \equiv i \tau_2 H^\dagger\,.
\end{equation}
The mass matrices, written in the basis $\left\{\Bar{\nu}_{\mathrm{L}i},\Bar{\nu}_{\mathrm{R}i}^c,\Bar{S}^c_i\right\} \otimes \left\{\nu_{\mathrm{L}j},\nu_{\mathrm{R}j},S_j\right\}$ are given, in a block compact form, as
\begin{equation}
\bm{M_\nu}^\mathrm{T1S} = 
\left( \begin{array}{cc} 
0        &  \tfrac{v_h}{\sqrt{2}} \bm{y_\nu} \\  
\tfrac{v_h}{\sqrt{2}} \bm{y_\nu}  & \tfrac{v_\sigma}{\sqrt{2}}  \bm{y_\sigma}
\end{array} \right) \,,
\quad
\bm{M_\nu}^\mathrm{IS} = 
\left( \begin{array}{ccc} 
0        &  \tfrac{v_h}{\sqrt{2}} \bm{y_\nu} & 0 \\  
\tfrac{v_h}{\sqrt{2}} \bm{y_\nu}  & 0 &  \tfrac{v_\sigma}{\sqrt{2}}  \bm{y_\sigma} \\
0 &  \tfrac{v_\sigma}{\sqrt{2}}  \bm{y_\sigma}  &  \bm{\Lambda}
\end{array} \right) \,,
\quad
    \bm{M_\nu}^\mathrm{EIS} = 
\left( \begin{array}{ccc} 
0        &  \tfrac{v_h}{\sqrt{2}} \bm{y_\nu} & 0 \\  
\tfrac{v_h}{\sqrt{2}} \bm{y_\nu}  & \tfrac{v_\sigma}{\sqrt{2}}  \bm{y^\prime_\sigma} & \bm{\Lambda} \\
0 & \bm{\Lambda}  & \tfrac{v_\sigma}{\sqrt{2}}  \bm{y_\sigma}
\end{array} \right) \,,
\label{eq:Mnu}
\end{equation}
such that the light neutrino masses scale as
\begin{equation}
\bm{m}^\mathrm{T1S}_\nu \approx \frac{1}{\sqrt{2}} \frac{\bm{y_\nu}^2}{\bm{y_\sigma}} \frac{v_h^2 }{v_\sigma}\,,
\qquad
\bm{m}^\mathrm{IS}_\nu \approx \frac{\bm{y_\nu}^2}{\bm{y_\sigma}^2} \frac{\bm{\Lambda} v_h^2 }{v_\sigma^2}\,,
\qquad
\bm{m}^\mathrm{EIS}_\nu \approx \frac{\bm{y_\nu}^2 \bm{y_\sigma}}{2 \sqrt{2}}\frac{v_h^2 v_\sigma}{\bm{\Lambda}^2}\,,
\label{eq:nu-light}
\end{equation}
where matrix product is implicit. If one assumes that the Yukawa couplings are all of a comparable size, in particular that the $\bm{y_\nu}$ are not tuned to be extremely small, the scale of active neutrino masses will be essentially driven by $v_\sigma$ and $\Lambda$ (the latter for the IS and EIS). This implies that the following relations are in order:
\begin{itemize}
    \item $v_\sigma \gg v_h$ for the T1S;
    \item $v_\sigma \gg v_h$ and/or $\Lambda \ll v_h$ for the IS;
    \item $v_\sigma \sim v_h$ and $\Lambda \gg v_h$ for the EIS.
\end{itemize}
The T1S with Majoron requires a $\mathrm{U}(1)_\mathrm{L}$ breaking scale well above the electroweak one which is out of the sensitivity reach of LISA frequencies and possibly even future experiments such as the Einstein Telescope or the Cosmic Explorer. For the considered IS model, while a small $\Lambda$ can relax the size of $v_\sigma$, the fact that the $\Lambda^{ij} \Bar{S}_{i}^c S_j$ operator is $\mathrm{U}(1)_\mathrm{L}$ preserving one does not expect it to be tiny. Therefore, it is natural and rather more elegant to consider $v_\sigma$ in the multi-TeV regime, such that GWs from FOPTs in the IS model are likely above the sensitivity reach of LISA. Last but not least, the EIS model is the best candidate scenario to naturally accommodate $\mathrm{U}(1)_\mathrm{L}$ lepton number symmetry breaking at the EW-TeV scale, thus at the reach of the LISA frequency range, provided that either a large $\Lambda$ scale or a small $\bm{y_\nu}^2 \bm{y_\sigma}$ product, or a combination of both, offer the needed suppression to generate the neutrino mass scale. Furthermore, with $\Lambda > v_h$ the inclusion of dimension-6 operators is well motivated in the EIS model as opposed to the IS one where $\Lambda$ is preferred to be lighter than $v_h$.

In the remainder of this article we will then focus on the EIS model as a well motivated scenario to study the interplay between gravitational echoes at LISA and their implications for collider physics and the properties of the neutrino sector. 

In addition to three light neutrinos, the EIS model features six heavy ones $N^\pm_{1,2,3}$, whose masses are
\begin{equation}
    \bm{m}_{N^\pm} \approx \bm{\Lambda} \pm \frac{v_\sigma}{2\sqrt{2}} \left(\bm{y_\sigma} + \bm{y^\prime_\sigma} \right)\,,
    \label{eq:nu-heavy}
\end{equation}
when expanded to second order in $v_h \ll \Lambda$ and first order in $v_\sigma \ll \Lambda$. Without loss of generality for this article's goals and for cleaner numerical calculations, one chooses $y_{\sigma\,i} \gg y^\prime_{\sigma\,i} \to 0$ as well as a flavour diagonal basis.

While the heavy neutrinos masses are essentially dominated by the $\Lambda$-scale, the lightness of active neutrino results from a combination of such a new-physics scale and the sizes of the Yukawa couplings $y_\nu$ and $y_\sigma$. In our analysis we have chosen to assign the normal hierarchy among light neutrinos as a phenomenological input\footnote{We have also checked the case of the inverted hierarchy and found no difference in the results. Therefore, all our conclusions in this article hold for the inverted scenario.}. Last but not least, it is convenient to invert $m_\nu^\mathrm{EIS}$ in \cref{eq:nu-light} recasting it as
\begin{equation}
    y_\sigma^i = 2 \sqrt{2} \frac{m_{\nu_i} \Lambda^2}{v_h^2 v_\sigma y_{\nu_i}^2}\,,
    \label{eq:y-sig}
\end{equation}
where the the parameters on the right-hand-side are used as input in our numerical analysis. Note that for ease of notation one has dropped the EIS label in $m_{\nu_i}$ in \cref{eq:y-sig} and anywhere else in the remainder of this article.


\section{The 6D Majoron model}
\label{Sec:model}

The standard dimension-4 Majoron model, equipped with an extended inverse seesaw mechanism, offers a great description for the generation of neutrino masses and mixing. One of the allowed lepton number symmetry conserving terms is the fermion bilinear $\Lambda\nu_\mathrm{R}^c S$ according to the transformation properties in \cref{tab:models}. It is therefore legitimate to ask what is the origin and size of the $\Lambda$ scale, whose relevance for active neutrino mass generation, in particular their smallness, is crucial. While in our previous work we have solely assumed such a parameter to lie in the sub-TeV regime, in the current analysis we go one step forward allowing it to be in a range from 10 TeV up to 1 PeV. The unknown nature of the UV physics beyond such a scale is encoded in effective $\U{L}$-preserving dimension-6 operators in the scalar sector.

One of the main goals of this work is to understand whether a light or ultralight Majoron can coexist with an observable spectrum of primordial GWs, triggered by the breaking of the $\U{L}$ symmetry at finite temperature. Our aim is to use show how future GW information can be used to extract the preferred sizes for the $\Lambda$ scale and the Yukawa couplings. In other words, the potential observation (or lack of it) of primordial GWs at LISA in the coming decade can provide us concrete hints about the scale of new physics and in particular of neutrino mass generation. The same philosophy can be applied to other observables, such as the trilinear Higgs coupling, the mass of new scalars and the $\U{L}$ breaking scale as we discuss in \cref{sec:Res}.

\subsection{The scalar potential}
\label{Sec:scalar_pot}

Our model contains two scalar fields, an electroweak doublet $H$ and a complex singlet $\sigma$ whose quantum numbers under the $\U{L}$ symmetry can be found in \cref{tab:models} (third line). The tree-level scalar potential can then be written as
\begin{equation}
    V_{_0} (H,\sigma) = V_{_\mathrm{SM}}(H) + V_{_{4\mathrm{D}}}(H,\sigma) + V_{_{6\mathrm{D}}}(H,\sigma) + V_{_\mathrm{soft}} (\sigma)\,,
    \label{eq:V0}
\end{equation}
with
\begin{equation}
    \begin{aligned}
    V_{_\mathrm{SM}}(H) &= \mu_h^2 H^{\dagger}H + \lambda_h(H^{\dagger}H)^2\,,
    \\
    V_{_{4\mathrm{D}}}(H,\sigma) &= \mu_{\sigma}^2 \sigma^{\dagger}\sigma + \lambda_{\sigma}(\sigma^{\dagger}\sigma)^2 + \lambda_{\sigma h}H^{\dagger}H \sigma^{\dagger}\sigma\,,
    \\
    V_{_{6\mathrm{D}}}(H,\sigma) &= \frac{\delta_0}{\Lambda^2}(H^{\dagger}H)^3 + \frac{\delta_2}{\Lambda^2}(H^{\dagger}H)^2\sigma^{\dagger}\sigma + \frac{\delta_4}{\Lambda^2}H^{\dagger}H(\sigma^{\dagger}\sigma)^2
    +
    \frac{\delta_6}{\Lambda^2}(\sigma^{\dagger}\sigma)^3\,,
    \\
    V_{_\mathrm{soft}} (\sigma) &= \frac{1}{2}\mu_b^2 \left(\sigma^2 + \sigma^{\ast 2} \right)\,.
    \label{eq:V-parts}
    \end{aligned}
\end{equation}
The $H$ and $\sigma$ fields can be expanded in terms of their real valued components as
\begin{equation}
\begin{aligned}
H = \dfrac{1}{\sqrt{2}} 
\begin{pmatrix}
\omega_1 + i \omega_2  \\
\phi_h + h + i \eta
\end{pmatrix}\,,	
\qquad
\sigma = \dfrac{1}{\sqrt{2}} \left( \phi_\sigma + h^\prime + i J \right)\,,	
\end{aligned}
\end{equation}	
with $h$ and $h^\prime$ denoting radial quantum fluctuations around the classical field configurations $\phi_h$ and $\phi_\sigma$, whereas $\omega_{1,2}$, $\eta$ and $J$ represent Goldstone modes. While $\omega_{1,2}$ and $\eta$ are eaten by longitudinal degrees of freedom of the $W$ and $Z$ bosons upon electroweak symmetry breaking, the $\U{L}$ generators are global, implying that $J$, the Majoron, is a physical real scalar field present in the particle spectrum. As we will see below, the Majoron explicitly acquires its mass from the last term in \cref{eq:V-parts}, $V_{_\mathrm{soft}} (\sigma)$, which preserves a remnant $\mathbb{Z}_2 \subset \U{L}$ symmetry in the scalar sector such that the potential becomes invariant under the transformation $J \to -J$.

The EFT description presented in this article is valid at energy scales below $\Lambda$, such that vacuum stability conditions are considered only in the range of applicability of the EFT, \textit{i.e.}~for field values below $\Lambda$. In our numerical analysis, we use the public software tool \texttt{CosmoTransitions}~\cite{Wainwright:2011kj}, tailored to find global and local minima via a phase tracing algorithm. In the case of parameter space points with unbounded from below directions, the action of the tunneling path does not converge and the point is rejected. In order to maximize the number of viable points, we use as first guess the usual tree-level boundedness from below conditions $ \lambda > 0$, $\lambda_\sigma > 0 $ and $\lambda_{\sigma h} > -2 \sqrt{\lambda \lambda_\sigma}$.

In our previous work \cite{Addazi:2019dqt} we concluded that observable GWs at the reach of forthcoming space-based interferometers is typically favoured by a not too small quartic portal coupling $\lambda_{\sigma h} \gtrsim 0.1$. However, LHC constraints from invisible Higgs decays \cite{CMS:2022qva} disfavour $\lambda_{\sigma h}$ values larger than order $\mathcal{O}(0.01)$, posing strong constraints on Majoron dark matter production via the freeze-out mechanism \cite{Cline:2013gha}. In this article we investigate whether effects coming from new physics above the electroweak (EW) scale and associated to the neutrino sector modify our previous conclusions allowing for visible GWs at LISA or future planned experiments, such as BBO or DECIGO. In our analysis one considers that the scale $\mu_\sigma$ is not too far from the EW scale and below $1~\mathrm{TeV}$, while $\Lambda$ is set to lie between $10~\mathrm{TeV}$ and $1000~\mathrm{TeV}$. In the scalar sector such effects are parametrized by $V_{_{6\mathrm{D}}}(H,\sigma)$ while in the fermion sector, the scale $\Lambda$ is linked to the generation of neutrino masses as we discuss in \cref{Sec:neut}.

The classical field configurations $\phi_h$ and $\phi_\sigma$ acquire their vacuum expectation values (VEVs) when the scalar potential $V_{_0} (\phi_h,\phi_\sigma)$ is extremized 
\begin{equation}
   \mean{\frac{\partial V_{_0}}{\partial \phi_{\alpha}}}_{\rm vac}=0 \,, \qquad
\mean{\phi_h}_{\rm vac}\equiv v_h\simeq 246\, {\rm GeV}\,, \qquad 
\mean{\phi_\sigma}_{\rm vac}\equiv v_\sigma\,,
\end{equation}
from where we obtain the minimization conditions that read as
\begin{equation}
    \begin{aligned}
    \mu_h^2 &= -v_h^2 \lambda_h  - \frac12 v_\sigma^2 \lambda_{\sigma h} - \frac34 \frac{v_h^4 \delta_0}{\Lambda^2} - \frac12 \frac{v_h^2 v_\sigma^2 \delta_2}{\Lambda^2} - \frac14 \frac{v_\sigma^4 \delta_4}{\Lambda^2}\,,
    \\
    \mu_\sigma^2 &= - v_\sigma^2 \lambda_\sigma - \mu_b^2 -\frac12 v_h^2 \lambda_{\sigma h} - \frac14 \frac{v_h^4 \delta_2}{\Lambda^2} - \frac12 \frac{v_h^2 v_\sigma^2 \delta_4}{\Lambda^2} - \frac34 \frac{v_\sigma^4 \delta_6}{\Lambda^2}\,.
    \end{aligned}
    \label{eq:tad}
\end{equation}
Taking the Hessian matrix and using the tadpole expressions in \cref{eq:tad}, we can cast the mass matrix of the CP-even states as
\begin{equation}
    \bm{M}^2 = 
\left( \begin{array}{cc} 
M_{hh}^2         &  M_{\sigma h}^2 \\  
M_{\sigma h}^2  & M_{\sigma \sigma}^2
\end{array} \right) \,,
\label{eq:hess}
\end{equation}
with
\begin{equation}
\begin{split}
    M_{hh}^2 = 2 v_h^2 \lambda_h + &\frac{3 v_h^4 \delta_0}{\Lambda^2} + \frac{v_h^2 v_\sigma^2 \delta_2}{\Lambda^2}\,, \qquad
    M_{\sigma \sigma}^2 = 2 v_\sigma^2 \lambda_\sigma + \frac{v_h^2 v_\sigma^2 \delta_4}{\Lambda^2} + \frac{3 v_\sigma^4 \delta_6}{\Lambda^2}\,,
    \\
    M_{\sigma h}^2 &= v_h v_\sigma \lambda_{\sigma h} + \frac{v_h^3 v_\sigma \delta_2}{\Lambda^2} + \frac{v_h v_\sigma^3 \delta_4}{\Lambda^2}\,.
    \label{eq:hess_elem}
\end{split}
\end{equation}
We can now rotate $\bm{M}$ to the mass eigenbasis as follows 
\begin{equation}
\bm{m}^2 = {O^\dagger}_{i}{}^{m} M_{mn}^2 O^{n}{}_{j} = 
\begin{pmatrix}
m_{h_1}^2 & 0 \\ 
0   & m_{h_2}^2 
\end{pmatrix}\,,
\qquad
\text{with}
\qquad
\bm{O} = 
\begin{pmatrix}
\cos \alpha_h & \sin \alpha_h \\ 
-\sin \alpha_h   & \cos \alpha_h 
\end{pmatrix}\,,
\end{equation}
such that the physical basis vectors $h_1$ and $h_2$ are obtained in terms of the gauge eigenbasis ones, $h$ and $h^\prime$, as follows:
\begin{equation}
\begin{pmatrix}
h_1 \\
h_2 
\end{pmatrix}
=
\bm{O}
\begin{pmatrix}
h \\
h^\prime 
\end{pmatrix}\,.
\label{eq:trans}
\end{equation}
In what fallows we identify the $h_1$ with a SM-like Higgs boson with mass $125.09~\mathrm{GeV}$, while $h_2$ is a new visible scalar that can either be heavier or lighter than the Higgs. Upon rotation to the mass eigenbasis one obtains
\begin{equation}
    m_{h_{1,2}}^2 = \frac12 \left[ M_{hh}^2 + M_{\sigma \sigma}^2 \pm \left( M_{hh}^2 - M_{\sigma \sigma}^2 \right) \sec (2 \alpha_h) \right]
    \quad
    \text{with}
    \quad
    \cot (2 \alpha_h)= \frac12  \frac{M_{hh}^2 - M_{\sigma \sigma}^2}{M_{\sigma h}^2} \,.
    \label{eq:mass}
\end{equation}
As in our numerical analysis we will use both the scalar mixing angle $\cos \alpha_h$ and the physical masses $m_{h_{1,2}}$ as input parameters, it is convenient to invert \cref{eq:mass} and recast it as
\begin{equation}
    M_{hh,\sigma \sigma}^2 = \frac12 \left[ m_{h_1}^2 + m_{h_2}^2 \pm \left( m_{h_1}^2 - m_{h_2}^2 \right) \cos (2 \alpha_h) \right]
    \quad \text{and} \quad
    M_{\sigma h}^2 = \frac12 \left( m_{h_1}^2 - m_{h_2}^2 \right) \sin (2 \alpha_h)\,,
    \label{eq:inputs1}
\end{equation}
which will be used to determine the elements of the Hessian matrix in terms of the physical masses and mixing angle. In the CP-odd sector, the mass of the pseudo-Goldstone boson, the Majoron, is simply given by
\begin{equation}
    m_J^2 = -2 \mu_b^2\,,
    \label{eq:mub}
\end{equation}
implying that $\mu_b^2 < 0$.

\subsection{Inverted equations for scalar couplings}
\label{sec:dark}

The main objective of this article is to show that it is possible to reconcile light and ultralight Majorons with observable GWs at LISA when effects from dimension-6 operators are sizeable. It is therefore necessary to reevaluate how such operators modify the invisible Higgs decay rate and confront it with experimental data.

The invisible Higgs decay rate to a pair of Majorons reads as \citep{Hall:2009bx}
\begin{equation}
    \Gamma\left( h_1 \to J J \right) = \frac{1}{32 \pi} \frac{\left(\lambda_{J J h_1}^{(0)}\right)^2}{m_{h_1} } \sqrt{1 - 4\dfrac{m_J^2}{m_{h_1}^2}}\,,
    \label{eq:inv}
\end{equation}
with $\lambda_{J J h_1}^{(0)}$ the effective triple Higgs-Majoron coupling expressed in the mass eigenbasis as
\begin{equation}
    \lambda_{J J h_1}^{(0)} = \frac{v_h}{\Lambda^2} \left[ (v_h^2 \delta_2 + v_\sigma^2 \delta_4 + \Lambda^2 \lambda_{\sigma h}) \cos \alpha_h + v_\sigma (v_h^2 \delta_4 + 3 v_\sigma^2 \delta_6 +2 \Lambda^2 \lambda_\sigma) \sin \alpha_h \right]\,,
    \label{eq:Lhjj}
\end{equation}
and where the superscript $(0)$ denotes tree-level accuracy. The latest observed (expected) upper bound on the Higgs invisible decay branching ratio as reported by the CMS experiment \cite{CMS:2022qva} is $0.18$ $(0.10)$ at the $95\%$ confidence level, and can be written in terms of the decay width as
\begin{equation}
    \mathrm{Br}(h_1 \to J J) = \frac{\Gamma \left( h_1 \to J J \right)}{\Gamma \left(h_1 \to J J \right) + \Gamma \left(h_1 \to \mathrm{SM} \right)}\,,
    \label{eq:BRinv}
\end{equation}
where $\Gamma \left( h_1 \to \mathrm{SM}  \right) = 4.07~\mathrm{MeV}$. In the absence of dimension-6 operators, that is $\delta_{0,2,4,6} = 0$, \cref{eq:Lhjj} would simply reduce to $\lambda_{J J h_1}^{(0)} = \tfrac12 v_h \lambda_{\sigma h} \cos \alpha_h$, requiring small values of the portal coupling $\lambda_{\sigma h} \lesssim \mathcal{O}(0.01)$ in order to comply with experimental data. On the contrary, for the considered model, it is the combination $\tfrac{v_h^2}{\Lambda^2} \delta_2 + \tfrac{v_\sigma^2}{\Lambda^2} \delta_4 +  \lambda_{\sigma h}$ that has the dominant role instead of the individual couplings. As it shown in the numerical analysis, this feature is crucial for the generation of observable primordial GW signals compatibly with light/ultralight Majorons. Such an analysis also indicates the preferred scale $\Lambda$, identified with the heavy neutrino masses. In other words, a potential observation (or lack of it) of primordial GWs at LISA can be seen as a gravitational probe for the underlying mechanism of neutrino mass generation and its scale.

In total, one considers five physical observables as input parameters, $m_{h_1}$, $m_{h_2}$, $\alpha_h$, $\mathrm{Br}\left( h_1 \to J J \right)$ and $m_J$, which means that it is possible to find five Lagrangian parameters that can be determined in terms of the physical ones. One of such parameters is $\mu_b^2$, which is related to the physical Majoron mass $m_J$ via \cref{eq:mub}. One can also express $\lambda_{\sigma h}$, $\lambda_\sigma$, $\lambda_h$, and $\delta_6$ in terms of the above physical observables. To do this we first use the expression for $\lambda_{h_1 J J}^{(0)}$ in \cref{eq:Lhjj}, replacing it in $\Gamma\left( h_1 \to J J \right)$ as given in \cref{eq:inv}. Substituting now the decay width in \cref{eq:BRinv}, it is convenient to recast $\mathrm{Br}(h_1 \to J J)$ as follows
\begin{equation}
    A(\mathrm{Br}) = \left[\delta_2 + \frac{v_h \left(v_\sigma^2 \delta_4 + \Lambda^2 \lambda_{\sigma h} \right) + v_\sigma \left( v_h^2 \delta_4 +3 v_\sigma^3 \delta_6 + 2 \Lambda^2 \lambda_\sigma \right) \tan \alpha_h}{v_h^3}\right] \cos \alpha_h\,,
    \label{eq:ABr}
\end{equation}
where, on the left-hand-side $A(\mathrm{Br})$ is defined as
\begin{equation}
    A(\mathrm{Br}) \equiv \pm 4 \sqrt{2 \pi} \left( 1 - 4\tfrac{m_J^2}{m_h^2} \right) m_h^{3/2} \frac{\Lambda^2}{v_h^3} \sqrt{\frac{\mathrm{Br}(h \to J J) \Gamma(h \to \mathrm{SM}) }{\left[1 - \mathrm{Br}(h \to J J)\right] (m_h^2 - 4 m_J^2)}}\,.
    \label{eq:ABr-deff}
\end{equation}
If we now combine \cref{eq:ABr} with \cref{eq:hess_elem,eq:mass}, we obtain the following closed set of formulas for the theory couplings:
\begin{equation}
    \begin{aligned}
    \lambda_{\sigma h} =& \frac{\tan \left(2 \alpha _h\right) \left(M_{hh}^2-M_{\sigma
   \sigma }^2\right)}{2 v_h v_{\sigma }}-\frac{\delta _2 v_h^2+\delta _4 v_{\sigma }^2}{\Lambda ^2}\,,\\ 
    \lambda_{\sigma} =& -\frac{2 A(\mathrm{Br}) v_h^3 v_{\sigma } \csc \left(\alpha _h\right)+\Lambda ^2 \sec \left(2
   \alpha _h\right) \left(M_{\sigma \sigma }^2-M_{hh}^2\right)+\Lambda ^2
   \left(-M_{hh}^2+M_{\sigma \sigma }^2-2 M_{\sigma \sigma }^2 v_{\sigma }\right)}{4 \Lambda ^2 \left(v_{\sigma }-1\right) v_{\sigma }^2} \\
   &+ \frac{\delta _4 v_h^2}{2 \Lambda^2}\,,
    \\
    \lambda_h =& \frac{1}{2} \left(\frac{M_{hh}^2}{v_h^2}-\frac{3 \delta _0 v_h^2+\delta _2
   v_{\sigma }^2}{\Lambda ^2}\right)\,,
    \\
    \delta_6 =& \frac{2 A(\mathrm{Br}) v_h^3 v_{\sigma } \csc \left(\alpha _h\right)-\Lambda ^2 \left(\sec \left(2 \alpha _h\right)
   \left(M_{hh}^2-M_{\sigma \sigma }^2\right)+M_{hh}^2+M_{\sigma \sigma }^2\right)}{6(v_{\sigma } -1) v_{\sigma }^4 }\,,
    \end{aligned}
    \label{eq:coup_inputs}
\end{equation}
with the elements of the Hessian matrix $M_{hh,\sigma \sigma}^2$ expressed in terms of the physical scalar masses $m_{h_{1,2}}$ and the mixing angle $\alpha_h$, as in \cref{eq:inputs1}. The remaining parameters of the scalar potential, including the singlet VEV $v_\sigma$, are kept unconstrained.

The inverted equations obtained above are derived from tree-level relations. They are used as a first approximation in order to keep the invisible Higgs decay branching fraction under control in the sense that it allows one use $\mathrm{Br}\left( h_1 \to J J \right)$ as an input parameter. However, one-loop corrections to $\lambda_{h_1 J J}^{(0)}$ coupling can be important \cite{Camargo-Molina:2016moz} and modify $\mathrm{Br}(h_1 \to J J)$. These are calculated in the effective potential approach using the formalist developed in \cite{Camargo-Molina:2016moz}. This is justified in the zero external momentum limit, $p^2 \to 0$, \cite{Ellis:1990nz} when the kinematical suppression of the $p^2 \neq 0$ corrections from heavy states is $m_h^2 / (4 m_\mathrm{heavy}^2) < 0.1$. The one-loop corrected coupling reads as
\begin{equation}
    \lambda_{J J h_1} = \lambda_{J J h_1}^{(0)} + \frac{1}{16 \pi^2} \left(\lambda_{J J h_1}^N + \lambda_{J J h_1}^{h_2}\right)\,,
    \label{eq:hJJ}
\end{equation}
with the superscript $N$ and $h_2$ denoting contributions from heavy neutrinos and the CP-even heavy Higgs boson when $m_{h_1}^2 / (4 m_{h_2}^2) < 0.1$ \cite{Ellis:1990nz}. Expressions for $\lambda_{J J h_1}^N$ and $ \lambda_{J J h_1}^{h_2}$ are given in \cref{app:tri}.

\subsection{Majoron decays}
\label{Sec:Jgg}

It is beyond the scope of the article to study the dark properties of the Majoron, however, it is instructive to identify whether it can be stable, unstable or long-lived by verifying its decay rate to photons \cite{Garcia-Cely:2017oco,Heeck:2019guh} and neutrinos \cite{McDonald:1993ex}.

At two-loop level, the decay rate of the Majoron to a pair of photons can be expressed as \cite{Heeck:2019guh}
\begin{equation}
    \Gamma(J\to\gamma\gamma) = \frac{|g_{J\gamma\gamma}|^2 m_J^3}{64\pi}\,,
    \qquad
    \text{with}
    \qquad
    g_{J\gamma\gamma} = g_{J\gamma\gamma}^{(1)} + g_{J\gamma\gamma}^{(2)}\,.
\end{equation}
There are two contributions to the coupling, which take the form
\begin{equation}
    \begin{aligned}
    g_{J\gamma\gamma}^{(1)} =& \frac{\alpha}{8 \pi^3 v_h^2 v_\sigma}\sum\limits_l (\bm{M_D} \bm{M_D}^\dagger)_{ll} F\left(\frac{m_J^2}{4 m_l^2}\right)\,,
    \\
    g_{J\gamma\gamma}^{(2)} =& \frac{\alpha}{8 \pi^3 v_h^2 v_\sigma} \mathrm{tr}(\bm{M_D} \bm{M_D}^\dagger)\sum\limits_f N_c^f Q_f^2 T_3^f F\left(\frac{m_J^2}{4 m_f^2}\right)\,,
    \end{aligned}
\end{equation}
where $N_c^f$, $T_3^f$, $Q_f$ denote color, isospin and electric charge of the of fermion $f$ respectively. Note that the index $l$ goes over lepton flavors while $f$ goes over charged fermion flavors. The Dirac mass matrix is $\bm{M_D} = \frac{v_h}{\sqrt{2}} \bm{y_\nu}$, and the loop function reads as
\begin{equation}
    F(x) \equiv - \frac{1}{4x}\left\{\log\left[1 - 2 x + 2 \sqrt{x(x-1)}\right]\right\}^2 - 1.
\end{equation}

In a flavour diagonal basis, the interaction between Majorons and neutrinos can be parameterized as \cite{Boulebnane:2017fxw,Escudero:2019gvw}
\begin{equation}
    \mathcal{L} = \frac{i}{2} \lambda_{\nu_j} J \overline{\nu}_j \gamma_5 \nu_j\,,
    \label{eq:Jnn}
\end{equation}
with $j = 1,2,3$ denoting active neutrinos from the lightest, $m_1$, to the heaviest, $m_3$, and $\lambda_{\nu_j} \equiv m_j/v_\sigma$.
One of the predictions of Majoron models is that it can also decay to neutrinos if kinematically allowed. Such a decay takes place at tree-level and, for the case of the extended inverse seesaw model, its total decay width to neutrinos is given by
\begin{equation}
    \Gamma\left( J \to \nu \nu \right) = \frac{m_J}{16 \pi v^2_\sigma} \sum_i \left(m^2_{\nu_i} \sqrt{1 - \frac{4 m^2_{\nu_i}}{m_J^2}}\right)\,.
    \label{eq:Jnunu}
\end{equation}

\subsection{Higgs trilinear coupling}
\label{Sec:hhh}

The presence of SFOPTs can result from sizeable scalar triliniear interactions \cite{Noble:2007kk} responsible for inducing a potential barrier between the false and true vacua. Particularly relevant is the Higgs boson triple coupling, $\lambda_{h_1 h_1 h_1}$, which can simultaneously be probed at colliders and have an impact in the spectrum of primordial GWs \cite{Biekotter:2022kgf}. For instance, at the LHC, the dominant process to measure $\lambda_{h_1 h_1 h_1}$ is the gluon fusion into a pair of Higgs bosons \cite{Chala:2018ari,Baglio:2012np,Huang:2015tdv,DiMicco:2019ngk,Abouabid:2021yvw}.

Similarly to $\lambda_{h_1 J J}$, the calculation of the Higgs trilinear coupling is performed at one-loop in the effective potential approach. One can express it according to
\begin{equation}
    \lambda_{h_1 h_1 h_1} = \lambda_{h_1 h_1 h_1}^{(0)} + \frac{1}{16 \pi^2} \left(\lambda_{h_1 h_1 h_1}^t + \lambda_{h_1 h_1 h_1}^N + \lambda_{h_1 h_1 h_1}^{h_2}\right)
    \label{eq:tri1}
\end{equation}
with the superscript $t$, $N$ and $h_2$ denoting one-loop contributions from the top quark, heavy neutrinos and the second CP-even Higgs boson when $m_{h_1}^2 / (4 m_{h_2}^2) < 0.1$. Note that we also include the top quark it in the calculation as the kinematical suppression from $p^2 \neq 0$ corrections is $m_h^2 / (4 m_t^2) \approx 0.1$. We show in \cref{app:tri} expressions for the r.h.s of \cref{eq:tri1}.

\section{Thermal effective potential and Gravitational Waves}
\label{sec:Eff+GW}

As the universe expands and cools down, the configuration of its temperature dependent vacuum state changes, typically giving rise to transitions between symmetric and broken phases. In this article we specialize to the case of the electroweak and lepton number symmetries, and study the impact on the stochastic primordial gravitational wave background.

\subsection{The one-loop $T$-dependent effective potential}

The shape of the effective potential depends on the symmetries and on the field content of the underlying theory. Therefore, for the purpose of exploring the features of phase-transitions in the model under consideration, we construct the one-loop temperature dependent effective potential in the following form \cite{Quiros:1999jp,Curtin:2016urg}:
\begin{equation}
V_{\rm eff}(T) = V_0 + V^{(1)}_{\rm CW} + \Delta V(T) + V_{\rm ct}\,,
\label{eq:eff-pot}
\end{equation}
where $V_0$ denotes the classical (tree-level) potential as given in \cref{eq:V0}, $V^{(1)}_{\rm CW}$ is the zero-temperature, one-loop, Coleman-Weinberg (CW) potential, $\Delta V(T)$ describes the leading order thermal corrections and $V_{\rm ct}$ is the counter-term potential.

The CW potential, expressed in the Landau gauge, reads as
\begin{equation}
V^{(1)}_{\rm CW}  = \sum_i (-1)^{F_i} n_i \frac{m_i^4(\phi_\alpha)}{64 \pi^2} \left( \log\left[ \frac{m_i^2(\phi_\alpha)}{Q^2}\right] - c_i \right) \,,
\end{equation}
where $m_i^2(\phi_\alpha)$ is the $\phi_\alpha$-field 
dependent mass of the particle $i$, $n_i$ is the number of degrees of freedom  (d.o.f.'s)
for a given particle $i$, $F=0(1)$ for bosons (fermions), $Q$ is the renormalization scale and,  in the $\overline{\rm MS}$-scheme, the constant $c_i$ is equal to $3/2$ for each d.o.f.~of scalars, fermions and longitudinally polarised gauge bosons, and to $1/2$ for transversely polarised gauge boson d.o.f.'s.

In what follows we fix the renormalization scale to the EW VEV $Q \equiv v_h$. In our analysis the scalar boson masses, the $\U{L}$ breaking VEV $v_\sigma$ and nucleation temperatures are not far from $v_h$. Therefore, a renormalization improved treatment as in scenarios with large separation of scales, as e.g.~in scenarios with classical conformal symmetry \cite{Kierkla:2022odc,Chataignier:2018kay} is unnecessary in the EIS model.

For a simpler analysis we require that the tree-level minimum conditions and masses are identical to their one-loop values.  The countertem potential is then introduced as
\begin{equation}
    V_{\rm ct} = \frac{\del V_0}{\del p_i} \delta_{p_i}\,,
\end{equation}
with $p_i$ denoting the parameters in $V_0$ and $\delta_{p_i}$ the corresponding parameter counterterms. The renormalization conditions necessary to fulfil the above requirements read as
\begin{equation}
    \left \langle 
     \frac{\del V_\mathrm{ct}}{\del \phi_i} \right \rangle = \left \langle  - \frac{\del V^{(1)}_{\rm CW}}{\del \phi_i} \right \rangle \,,
     \qquad
    \left \langle 
     \frac{\del^2 V_\mathrm{ct}}{\del \phi_i \del \phi_j} \right \rangle = \left \langle  - \frac{\del^2 V^{(1)}_{\rm CW}}{\del \phi_i \del \phi_j} \right \rangle \,,
     \qquad
     \text{with}
     \qquad
     \phi_i = \left\{\phi_h , \phi_\sigma \right\}\,,
\end{equation}
with the following solutions
\begin{equation}
    \begin{aligned}
       \delta_{\mu_h^2} &= \frac12 \frac{\del^2 V^{(1)}_{\rm CW}}{\del v_h^2} 
       +
       \frac12 \frac{v_\sigma}{v_h} \frac{\del^2 V^{(1)}_{\rm CW}}{\del v_h \del v_\sigma}
       -
       \frac32 \frac{1}{v_h} \frac{\del V^{(1)}_{\rm CW}}{\del v_h}
       +
       a \frac34 \frac{v_h^4}{\Lambda^2}
       +
       b \frac12 \frac{v_h^2 v_\sigma^2}{\Lambda^2}
       + 
       c \frac14 \frac{v_\sigma^4}{\Lambda^2}\,,
       \\
       \delta_{\mu_\sigma^2} &= \frac12 \frac{\del^2 V^{(1)}_{\rm CW}}{\del v_\sigma^2}
       +
       \frac12 \frac{v_h}{v_\sigma} \frac{\del^2 V^{(1)}_{\rm CW}}{\del v_h \del v_\sigma}
       -
       \frac32 \frac{1}{v_\sigma} \frac{\del V^{(1)}_{\rm CW}}{\del v_\sigma}
       +
       b \frac14 \frac{v_h^4}{\Lambda^2}
       +
       c \frac12 \frac{v_h^2 v_\sigma^2}{\Lambda^2}
       +
       d \frac34 \frac{v_\sigma^4}{\Lambda^2}
       -
       f\,,
       \\
       \delta_{\lambda_h} &= -\frac12 \frac{1}{v_h^2} \frac{\del^2 V^{(1)}_{\rm CW}}{\del v_h^2}
       +
       \frac12 \frac{1}{v_h^3} \frac{\del V^{(1)}_{\rm CW}}{\del v_h}
       -
       a \frac32 \frac{v_h^2}{\Lambda^2}
       -
       b \frac12 \frac{v_\sigma^2}{\Lambda^2}\,,
       \\
       \delta_{\lambda_\sigma} &= -\frac12 \frac{1}{v_\sigma^2} \frac{\del^2 V^{(1)}_{\rm CW}}{\del v_\sigma^2}
       +
       \frac12 \frac{1}{v_\sigma^3} \frac{\del V^{(1)}_{\rm CW}}{\del v_\sigma}
       -
       c \frac12 \frac{v_h^2}{\Lambda^2}
       -
       d \frac32 \frac{v_\sigma^2}{\Lambda^2}\,,
       \\
       \delta_{\lambda_{\sigma h}} &= - \frac{1}{v_h v_\sigma} \frac{\del^2 V^{(1)}_{\rm CW}}{\del v_h \del v_\sigma}
       -
       b \frac{v_h^2}{\Lambda^2}
       -
       c \frac{v_\sigma^2}{\Lambda^2}\,,
       \\
       \delta_{\delta_0} &= a\,,
       \quad
       \delta_{\delta_2} = b\,,
       \quad
       \delta_{\delta_4} = c\,,
       \quad
       \delta_{\delta_6} = d\,,
       \quad
       \delta_{\mu_b^2} = f\,,
    \end{aligned}
\end{equation}
with $a$, $b$, $c$, $d$ and $f$ being arbitrary dimensionless constants. In our numerical analysis we fix $a=b=c=d=f=0$.

One-loop thermal corrections are given by \cite{Quiros:1999jp}
\begin{equation}
\Delta V(T) = \frac{T^4}{2 \pi^2} \left\{ \sum_{b} n_b J_B\left[\frac{m_b^2(\phi_\alpha)}{T^2}\right] - \sum_{f} n_f J_F\left[\frac{m_f^2(\phi_\alpha)}{T^2}\right] \right\}\,,
\label{finite_T_correction}
\end{equation} 
where $J_B$ and $J_F$ are the thermal integrals for bosons and fermions, respectively, provided by
\begin{align} \label{eq:JBJF}
J_{B/F}(y^2) = \int_0^\infty d x \, x^2 \log\left( 1 \mp \exp [ - \sqrt{x^2 + y^2}] \right)\,.
\end{align}
Corrections for the first non-trivial order of the thermal expansion $\sim(m/T)^2$ have the following form
\begin{align}
\Delta V^{(1)}(T)|_{\rm L.O.} = \frac{T^2}{24} \left\{ {\rm Tr}\left[ M_{\alpha\beta}^2(\phi_\alpha) \right] + 
\sum_{i=W,Z,\gamma} n_i m_i^2(\phi_\alpha) + 
\sum_{i = \mathrm{f}_i} \frac{n_i}{2} m_i^2(\phi_\alpha) \right\} \,,
\label{eq:DV_LO}
\end{align}
where in the last sum all SM fermions plus six heavy neutrinos, $N^+_{1,2,3},N^-_{1,2,3}$, are implicit. The first term in \cref{eq:DV_LO} denotes the trace of the field-dependent scalar Hessian matrix $M_{\alpha\beta}^2(\phi_\alpha)$, which is a basis invariant quantity. In practical calculations, it is convenient to use the diagonal elements of the gauge eigenbasis mass form that, for the considered model, are given by $M_{hh}^2$ and $M_{\sigma \sigma}^2$ in \cref{eq:hess_elem} with the replacement of the VEVs by the classical field configurations $v_h \to \phi_h$ and $v_\sigma \to \phi_\sigma$. The $n_i$ coefficients in \cref{eq:DV_LO} represent the number of d.o.f for a given particle, as indicated by the sums. In particular, for the SM gauge bosons ($W, Z$ and transversely polarised 
photon $\gamma$) we have
\begin{equation}
n_W = 6, \qquad n_Z = 3, \qquad n_\gamma = 2 \,,
\end{equation}
whereas for scalars and the longitudinally polarized photon $(A_L)$ we have
\begin{equation}
n_s = 6, \qquad n_{A_L} = 1\,,
\end{equation}
while for fermions
\begin{equation}
n_{u,d,c,s,t,b} = 12, \qquad n_{e,\mu,\tau} = 4\,, \qquad n_{\nu_{1,2,3}} = n_{N^\pm_{1,2,3}} = 2 \,.
\end{equation}
The appearance of $T^2$ terms in the thermal expansion signals the restoration of symmetries broken at zero temperature. Furthermore, such a restored symmetry by $T^2$ contributions to the effective potential can lead to the breakdown of perturbation theory in a close vicinity of the critical temperature. To address this, an all-order resummation process via the so called daisy or ring diagrams is required \cite{Dolan:1973qd,Parwani:1991gq,Arnold:1992rz,Espinosa:1995se}. In practice, such a procedure can be achieved taking into account finite temperature corrections to the field-dependent masses entering in the effective potential \cref{eq:eff-pot} as follows
\begin{eqnarray}
 \mu_\alpha^2(T) = \mu_\alpha^2 + c_\alpha T^2 \,.
 \label{eq:mu-T}
\end{eqnarray}
The $c_\alpha$ coefficients can be calculated from \cref{eq:DV_LO} as
\begin{equation}
    c_\alpha = \dfrac{\del {\Delta V^{(1)}(T,\phi_h, \phi_\sigma)}^2}{\del^2 \phi_\alpha}\,,
\end{equation}
where for the 6D EIS model one has
\begin{equation}
    \begin{aligned}
       c_h &= \frac{3}{16} g^2 + \frac{1}{16} {g'}^2 + 
\frac12 \lambda_h + \frac{1}{12} \lambda_{\sigma h}+ 
\frac14 \sum_q y_q^2+ \frac{1}{12} \sum_\ell y_{\ell}^2 + \frac{1}{24} K_\nu + K_\Lambda^h\,, \\
c_\sigma &= \frac13\lambda_\sigma + \frac16 \lambda_{\sigma h} + \frac{1}{24} K_\sigma + K_\Lambda^\sigma \,,
    \end{aligned}
    \label{eq:coeff_thermalmasses} 
\end{equation}
with $q$ and $\ell$ denoting SM quarks and charged leptons respectively. Notice that, in practice, only the third generation Yukawa couplings play a sizeable role. In \cref{eq:coeff_thermalmasses} one also defines
\begin{equation}
    \begin{aligned}
    K_\nu &= \sum_{i=1}^3 y_{\nu_i}^\mathrm{eff} \quad \textrm{with} \quad y_{\nu_i}^\mathrm{eff} = \frac{\phi_h \phi_\sigma}{2} \frac{y_{\nu_i}^2 y_{\sigma_i}}{\Lambda^2} \quad \textrm{and} \quad m_{\nu_i}(\phi_h) = \frac{\phi_h}{\sqrt{2}} y_{\nu_i}^\mathrm{eff} \qquad  K_\sigma = \sum_{i = 1}^3 y_{\sigma_i}^2 \\
    K_\Lambda^h &= \frac{3 \phi_h^2}{\Lambda^2} \delta_0 + \frac{\phi_h^2 + \phi_\sigma^2}{4 \Lambda^2}\delta_2 + \frac{\phi_\sigma^2}{6 \Lambda^2} \delta_4 \qquad K_\Lambda^\sigma = \frac{\phi_h^2}{4 \Lambda^2} \delta_2 + \frac{\phi_h^2}{6 \Lambda^2} \delta_4 + \frac{\phi_\sigma^2}{2 \Lambda^2} \delta_4 + \frac{9 \phi_\sigma^2}{4 \Lambda^2} \delta_6\,.
    \end{aligned}
\end{equation}

Similarly, the temperature dependence of the vector boson masses at leading order is introduced by adding $T^2$ corrections to the diagonal terms of the gauge boson Hessian matrix. Note that only the longitudinal polarizations, including that of the photon, $\{W^+_L,W^-_L,Z_L,A_L\}$, receive finite corrections in a thermal medium such that the mass spectrum is obtained upon diagonalization of the $T^2$-corrected mass form
\begin{eqnarray}
M_{\rm gauge}^{2}(\phi_{h};T) = M_{\rm gauge}^{2}(\phi_{h}) + \frac{11}{6}T^2 
\left( 
\begin{array}{cccc} 
g^2 & 0 & 0 & 0 \ \\  
0 & g^2 & 0 & 0 \ \\
0 & 0 & g^2 & 0 \ \\
0 & 0 & 0 & {g'}^2 
\end{array} 
\right) \,,
\end{eqnarray}
whose eigenvalues of the zero-temperature mass matrix $M_{\rm gauge}^{2}(\phi_{h})$ read as
\begin{equation}
\begin{aligned}
&m_W^2(\phi_{h}) = \frac{\phi_h^2}{4} g^2 \,,
&m_Z^2(\phi_{h}) = \frac{\phi_h^2}{4} (g^2+{g'}^2) \,.
\end{aligned}
\label{eq:mZW}
\end{equation}
Rotating to the physical basis one obtains the following mass spectrum
\begin{eqnarray}
&& m_{W_L}^2(\phi_{h};T) = m_W^2(\phi_{h}) + \frac{11}{6}g^2T^2\,, \\
&& m_{Z_L,A_L}^2(\phi_{h};T) = \frac{1}{2}m_Z^2(\phi_{h}) + 
\frac{11}{12}(g^2+{g'}^2)T^2 \pm {\cal D} \,,
\end{eqnarray}
with the field-dependent $W,Z$ boson masses given in \cref{eq:mZW}, and
\begin{equation}
    {\cal D}^2 = \Big(\frac{1}{2}m_Z^2(\phi_{h}) + \frac{11}{12}(g^2+{g'}^2)T^2 \Big)^2 - 
\frac{11}{12} g^2{g'}^2 T^2 \Big( \phi_h^2 + \frac{11}{3}T^2 \Big) \,.
\end{equation}
New techniques accounting for a great improvement in the calculation of sizeable higher order thermal effects, in particular considering the resummed effective field theory constructed from dimensional reduction \cite{Ekstedt:2022bff,Croon:2020cgk,Schicho:2022wty,Niemi:2021qvp,Schicho:2021gca}, were recently developed. We leave for future work the inclusion of such methods in our studies.

\subsection{Dynamics of the Phase Transition}

Phase transitions can be described as dynamical processes occurring via non-perturbative solutions of the equations of motion. For the low-$T$ regime they are essentially realized through quantum tunneling, or instantons~\cite{Linde:1981zj,Dine:1992wr}, whereas at high temperature these processes are dominated by thermal jumps. The formalism behind both cases is identical and can be described by a classical motion in Euclidean space. In particular, the classical action reads as \citep{Coleman:1977py}
\begin{equation}
\hat{S}_3(\hat{\phi},T) = 4 \pi \int_0^\infty \mathrm{d}r \, r^2 \left\{ \frac{1}{2} 
\left( \frac{\mathrm{d}\hat{\phi}}{\mathrm{d}r} \right)^2 + V_{\rm eff}(\hat{\phi},T) \right\} \,,
\end{equation}
with the full one-loop effective potential specified in \cref{eq:eff-pot}, with $\hat{\phi}$ a particular solution found by the path that minimizes the action \citep{Coleman:1977py,Wainwright:2011kj}.

The transition rate from the false to the true vacuum is, to a good approximation, given by \cite{Ellis:2020nnr,Ellis:2022lft}
\begin{equation}
    \Gamma(T) \approx T^4 \left( \frac{\hat{S}_3}{2 \pi T} \right)^{3/2} e^{-\hat{S}_3/T}\,.
    \label{Eq:Gamma}
\end{equation}

There are three relevant temperatures characterizing the phase transition. When the Universe reaches a temperature for which multiple minima are degenerate, one defines the critical temperature $T_c$. As the Universe cools down below $T_c$, thermal fluctuations can become large enough to nucleate one true vacuum bubble per cosmological horizon. The nucleation temperature, $T_n$, is then defined as the solution of
\begin{equation}
    \int_{T_n}^{T_c} \frac{d T}{T} \frac{\Gamma(T)}{H(T)^4} = 1\,,
    \label{eq:Tn_1}
\end{equation}
with $H(T)$ the Hubble rate at temperature $T$. In an alternative definition \cite{Wang:2020jrd}, $T_n$ is determined as the temperature at which the transition rate matches the Hubble rate, i.e.
\begin{equation}
    \frac{\Gamma(T_n)}{H(T_n)^4} = 1\,,
    \label{eq:Tn}
\end{equation}
with the total Hubble rate accounting for the radiation and vacuum energy density contributions given by \cite{Brdar:2018num}
\begin{equation}
    H(T)^2 = \frac{g_\ast(T) \pi^2 T^4 }{90 M_\mathrm{Pl}^2} + \frac{\Delta V}{3 M_\mathrm{Pl}^2}\,,
    \label{eq:HT}
\end{equation}
where $g_\ast(T)$ represents the number of relativistic degrees of freedom at a temperature $T$.
In the presence of strong supercooling, where the strength of the phase transition $\alpha$ can be several orders of magnitude above $1$, the vacuum energy contribution dominates \cite{Kierkla:2022odc}. Conversely, for scenarios with slight or no supercooling, where according to the definition in \cite{Wang:2020jrd} $\alpha \lesssim 0.1$, the radiation component will dominate. \cref{eq:Tn} can then be approximated to the well known relation
\begin{equation}
    \frac{\hat{S}_3(T_n)}{T_n} \approx 140\,,
\end{equation}
valid for temperatures of the order of the EW scale and small $\alpha$. We implement \cref{eq:Tn} in our calculations neglecting the vacuum contribution when $\alpha \lesssim \mathcal{O}(0.1)$.

Finally, the percolation temperature $T_\ast$ is defined when $34\%$ of the false vacuum has transited to the true one. This condition results in the presence of a large structure, denoted as percolation cluster, where the true vacuum spans over the whole Universe such that it cannot collapse back into the false vacuum.
The probability of finding a point in the false vacuum reads as~\cite{Ellis:2020nnr}
\begin{align}
P(T) = e^{-I(T)}, & & I(T) = \frac{4\pi v_b^3}{3} \int_T^{T_c} \frac{\Gamma(T')dT'}{T'^4 H(T')}\left(\int_T^{T'}\frac{d\tilde{T}}{H(\tilde{T})}\right)^3\,,
\label{eq:Tp}
\end{align}
and the percolation temperature is obtained by solving $I(T_*) = 0.34$ or, equivalently, $P(T_*) = 0.7$.

One of the key quantities relevant for the study of primordial GWs is the so called order parameter. It can be defined as
\begin{equation}
    \frac{\Delta v_\alpha}{T_\ast} = \eta_\alpha\,, \qquad \textrm{with} \qquad \Delta v_\alpha = \abs{v_\alpha(T_\ast+\delta T) - v_\alpha(T_\ast-\delta T)}\,, \qquad \alpha = \{h,\sigma \}\,,
    \label{eq:order}
\end{equation}
such that $\Delta v_\alpha$ is the absolute value of the difference between $v_\alpha(T)$ computed before and after a phase transition, with $\delta T$ taken to be sufficiently small, i.e. $\delta T \ll T_n$. Note that $\eta_\alpha$ can be regarded as a measure of the strength of the phase transition. In particular, in the context of EW baryogenesis, a phase transition is said to be strong whenever the criterion $\eta_h \gtrsim 1$ is obeyed. However, for the case of GWs the most commonly used parameter to define the strength of the phase transition is the difference in the trace anomaly \cite{Hindmarsh:2015qta,Hindmarsh:2017gnf}
\begin{equation}
\alpha = \frac{1}{\rho_\gamma} \left[ \Delta V - \dfrac{T}{4} \left( \frac{\partial \Delta V}{\partial T} \right) \right] \,,
\label{eq:alpha}
\end{equation}
where $\Delta V = V_i - V_f$ with $V_i\equiv V_{\rm eff}(\phi^i_{h,\sigma};T_*)$ and $V_f\equiv V_{\rm eff}(\phi^f_{h,\sigma};T_*)$ the values of the effective potential in the initial (metastable) and final (stable) phases respectively, and
\begin{equation}
\rho_\gamma = g_* \frac{\pi^2}{30} T_n^4\,,  \qquad g_* \simeq 108.75 \,,
\end{equation}
is the energy density of the radiation medium at the bubble nucleation epoch found in terms of the number of relativistic d.o.f. In the determination of $g_\ast$, besides the SM particles we have only considered the Majoron $J$ and the new CP-even Higgs boson $h_2$ provided that heavy neutrinos are non-relativistic at $T_n$, i.e.~$m_{N_i} \gg T_n$.

Another relevant parameter characterizing a phase transition is the inverse time scale, which, in units of the Hubble parameter $H$ reads as
\begin{equation}
\frac{\beta}{H} = T_\ast  \left. \frac{\partial}{\partial T} \left( \frac{\hat{S}_3}{T}\right) \right|_{T_\ast}\,.
\label{eq:betaH}
\end{equation}
The $\beta/H$ parameter offers a description of the duration of the phase transition which means that to smaller values of $\beta/H$ typically corresponds a larger energy density amplitude of the stochastic spectrum of primordial GWs. And $\alpha$

\subsection{Primordial Gravitational Waves: a semi-analytical approximation}

Strong FOPTs are violent processes occurring in the early Universe and are expected to leave a signature in the form of a stochastic background of primordial GWs. In the first approximation, the primordial stochastic GW background is statistically isotropic, stationary and Gaussian. Furthermore, both the $+$ and the $\times$ polarizations are assumed to have the same spectrum and are mutually uncorrelated. The GW power spectrum is given in terms of the energy-density of the gravitational radiation per logarithmic frequency as
\begin{equation}
h^2 \Omega_{\rm GW}(f) \equiv \frac{h^2}{\rho_c} \frac{\partial \rho_{\rm GW}}{\partial \log f}\,, 
\label{eq:hGW}
\end{equation}
with $\rho_c$ the critical energy density today. This observable is independent of uncertainties in the measurement of the Hubble parameter. For details on the derivation of \cref{eq:hGW} see for example \citep{Caprini:2001nb,Figueroa:2012kw,Hindmarsh:2016lnk} and references therein.

In our analysis we are interested in a regime where vacuum bubbles undergo supersonic expansion. In turn, the leading contribution for the GW power spectrum results from sound waves (SW) \citep{Hindmarsh:2013xza} generated by the so called supersonic detonations. The most up to date understanding for the SW contribution to the SGWB production is discussed in \cite{Caprini:2019egz}. Other sources of GW production are collisions between bubble walls \cite{Kosowsky:1991ua} and magnetohydrodynamics (MHD) tubulences in the plasma \cite{Caprini:2009yp}. However, following the discussion in \cite{Hindmarsh:2017gnf,Ellis:2019oqb}, the collision component is typically very inefficient unless for the case of very strong FOPTs with runaway bubbles where the wall velocity undergoes unbounded acceleration, i.e.~$v_w \to 1$ \cite{Bodeker:2009qy,Prokopec:2018tnq,Kierkla:2022odc}. In such scenarios $\alpha$ may become orders of magnitude larger than unity and the efficiency of bubble collisions in producing gravitational radiation surpasses the sound waves one. The impact of MHD turbulence is also neglected in the current work and left for future studies when a better understand of the importance of such a component becomes available. Nonetheless, using the formulas in \cite{Caprini:2015zlo}, the MHD component is found to have no impact to the peak amplitude and frequency of the GW power spectrum. Therefore we will solely consider the sound waves contribution in the remainder of our analysis.

While it is quite challenging to provide a precise estimate for the bubble wall velocity ~\cite{Dorsch:2018pat,Moore:1995ua}, in our numerical studies we fix the it to $v_w = 0.95$. The regime of supersonic detonations is realized by further requiring $v_w > v_\mathrm{J}$, with $v_\mathrm{J}$ the Chapman-Jouguet velocity defined as
\begin{equation}
	v_\mathrm{J} = \dfrac{1}{1+\alpha} \left(c_s + \sqrt{\alpha^2 + \tfrac{2}{3} \alpha}\right)\,.
	\label{eq:vJ}
\end{equation}

The SGWB spectrum expressed in terms of the peak amplitude $h^2 \Omega_\mathrm{GW}^\mathrm{peak}$ and the spectral function reads as
\begin{equation}
	h^2 \Omega_\mathrm{GW} = h^2 \Omega_\mathrm{GW}^\mathrm{peak} \left(\dfrac{4}{7}\right)^{-\tfrac{7}{2}} \left(\dfrac{f}{f_\mathrm{peak}}\right)^3 \left[1 + \dfrac{3}{4} \left(\dfrac{f}{f_\mathrm{peak}}\right) \right]^{-\tfrac{7}{2}}\,,
	\label{eq:spectrum}
\end{equation}
where $f_\mathrm{peak}$ is the peak-frequency. Semi-analytic expressions 
for peak-amplitude and peak-frequency in terms of $\beta/H$ and $\alpha$ can be found in Ref.~\cite{Caprini:2019egz} and can be summarised as follows
\begin{equation}
\begin{aligned}
	& f_\mathrm{peak} = 26 \times 10^{-6} \left( \dfrac{1}{H R} \right) \left( \dfrac{T_\ast}{100} \right) \left( \dfrac{g_\ast}{100~\mathrm{GeV}} \right)^{\tfrac{1}{6}} \mathrm{Hz} \,, \\
    & h^2 \Omega_\mathrm{GW}^\mathrm{peak} = 1.159 \times 10^{-7} \left(\dfrac{100}{g_\ast}\right)  \left(\dfrac{HR}{\sqrt{c_s}}\right)^2 K^{\tfrac{3}{2}} \qquad \rm{for} \qquad H \tau_\mathrm{sh} = \dfrac{2}{\sqrt{3}} \dfrac{HR}{K^{1/2}} < 1 \,,
    \\
    & h^2 \Omega_\mathrm{GW}^\mathrm{peak} = 1.159 \times 10^{-7} \left(\dfrac{100}{g_\ast}\right)  \left(\dfrac{HR}{c_s}\right)^2 K^{2} \qquad \rm{for} \qquad H \tau_\mathrm{sh} = \dfrac{2}{\sqrt{3}} \dfrac{HR}{K^{1/2}} \simeq 1 \,,
    \end{aligned}
\label{eq:Opeak1}
\end{equation}
where $\tau_\mathrm{sh}$ is the fluid turnover time or the shock formation time, which quantifies the time the GW source was active. In these expressions, $c_s = 1/\sqrt{3}$ is the speed of sound, $R$ is the mean bubble separation,
\begin{equation}
	K = \dfrac{\kappa \alpha}{1 + \alpha}
	\label{eq:K}
\end{equation}
is the fraction of the kinetic energy in the fluid to the total bubble energy, and
\begin{equation}
	H R = \dfrac{H}{\beta} \left( 8 \pi \right)^{\tfrac{1}{3}} \max\left(v_b, c_s\right) \,.
	\label{eq:HR}
\end{equation}
The efficiency factor $\kappa$ is taken from the numerical fits in the appendix of~\cite{Espinosa:2010hh}. One can also express the peak amplitude in terms of the peak frequency, i.e.~$h^2 \Omega_\mathrm{GW}^\mathrm{peak} (f_\mathrm{peak})$, by solving \cref{eq:Opeak1} with respect to $HR$. The recasted form of the peak energy density amplitude reads as
\begin{equation}
\begin{aligned}
    & h^2 \Omega_\mathrm{GW}^\mathrm{peak} (f_\mathrm{peak}) = 7.835 \times 10^{-17} f_\mathrm{peak}^{-2} \left(\dfrac{100}{g_\ast}\right)^{2/3} \left( \dfrac{T_\ast}{100} \right)^2 \dfrac{K^{\tfrac{3}{2}}}{c_s} \quad \rm{for} \quad H \tau_\mathrm{sh} = \dfrac{2}{\sqrt{3}} \dfrac{HR}{K^{1/2}} < 1 \,,
    \\
    & h^2 \Omega_\mathrm{GW}^\mathrm{peak} (f_\mathrm{peak}) = 7.835 \times 10^{-17} f_\mathrm{peak}^{-2} \left(\dfrac{100}{g_\ast}\right)^{2/3} \left( \dfrac{T_\ast}{100} \right)^2 \dfrac{K^{2}}{c_s^2} \quad \rm{for} \quad H \tau_\mathrm{sh} = \dfrac{2}{\sqrt{3}} \dfrac{HR}{K^{1/2}} \simeq 1 \,,
    \end{aligned}
\label{eq:Opeak2}
\end{equation}
which is more conveniently written for our numerical analysis.


\section{Results and discussion}
\label{sec:Res}

We perform our numerical calculations with \texttt{CosmoTransitions}~\cite{Wainwright:2011kj}, using the method discussed in the appendix of~\cite{Freitas:2021yng} in order to obtain a smooth effective action $\hat{S}_3/T$. In particular, for every single point in parameter space, we interpolate the action around the nucleation temperature with a polynomial fit in order to mitigate numerical instabilities in the determination of the $\beta/H$, $T_n$ and $T_\ast$ parameters.

\subsection{Revisiting the 4D limit}
\label{sec:4D}

The analysis developed in \cite{Addazi:2019dqt} revealed that for Majorons with mass $m_J < m_{h_1}/2$, constraints from invisible Higgs decays imply a small portal coupling $\lambda_{\sigma h} \lesssim \mathcal{O}(10^{-2})$, excluding SGWB signatures in the observable region. In this subsection we improve the previous work focusing on the 4D limit of the EIS model taking $\delta_{0,2,4,6} \to 0$, and using an inversion procedure to fix $\mathrm{Br}(h_1 \to JJ) < 0.18$ as input for all generated points. One can then define:
\begin{equation}
    A^\prime(\mathrm{Br}) =  \mathrm{Sign}\left[ \left(M_{hh}^2 - M_{\sigma \sigma}^2\right) \right] 4 \sqrt{2 \pi} \frac{m_h}{v_h} \sqrt{\frac{\mathrm{Br}(h_1 \to J J) \Gamma(h \to \mathrm{SM}) }{\left[1 - \mathrm{Br}(h \to J J)\right] (m_h^2 - 4 m_J^2)^{1/2}}} \,.
    \label{eq:ABr-deff-4D}
\end{equation}
in order to obtain the 4D version of the inverted equations
\begin{equation}
    \begin{aligned}  
    \lambda_{\sigma h} =& A^\prime(\mathrm{Br}) \sec \alpha_h\,,
    \\
    \lambda_\sigma =& \tfrac12 {A^\prime}(\mathrm{Br})^2 \frac{M_{\sigma \sigma}^2 v_h^2 \cos(2 \alpha_h)^2 \csc (\alpha_h)^2 \sec (\alpha_h)^4 }{\( M_{hh}^2 - M_{\sigma \sigma}^2 \)^2}\,,
    \\
    \lambda_h =& \tfrac12 \frac{M_{hh}^2}{v_h^2}\,,
    \\
    v_\sigma =& \frac{\(M_{hh}^2 - M_{\sigma \sigma}^2\) \cos(\alpha_h)^2 \sec(2 \alpha_h) \sin(\alpha_h)}{A^\prime(\mathrm{Br}) v_h}\,.
    \end{aligned}
    \label{eq:coup_inputs-4D}
\end{equation}
In \cref{tab1} we show the ranges of the input parameters used in the scan of the revisited 4D scenario.
\begin{table*}[h!]
\centering
\begin{tabular}{@{}rcccr@{}}
& \multicolumn{3}{c}{} \\
\hline
& Parameter & Range & Distribution &\\ \hline
&  & &  &\\
& $m_{h_2}$ & $[60,\,1000]\,\text{GeV}$ & linear\\
&  & &  &\\
& $m_{J}$ & $[10^{-10}~\mathrm{eV},\,100~\mathrm{keV}]\,$ & exponential &\\
&  & &  &\\
& $\mathrm{Br}(h_1 \to J J)$ & $[10^{-15},\,0.18]$ & exponential &\\
&  & &  &\\
& $\sin{(\alpha_h)}$ & $\pm[0,\,0.24]$ & linear &\\
&  & &  &\\
\hline
\end{tabular}
\caption{\footnotesize Ranges of the input parameters in the scans for the 4D limit of the EIS model. We fix $m_{h_1} = 125.01~\mathrm{GeV}$ and $v_h = 246.22~\mathrm{GeV}$.}
\label{tab1}
\end{table*}
\begin{figure}[htb!]
\centering
\includegraphics[width=0.49\textwidth]{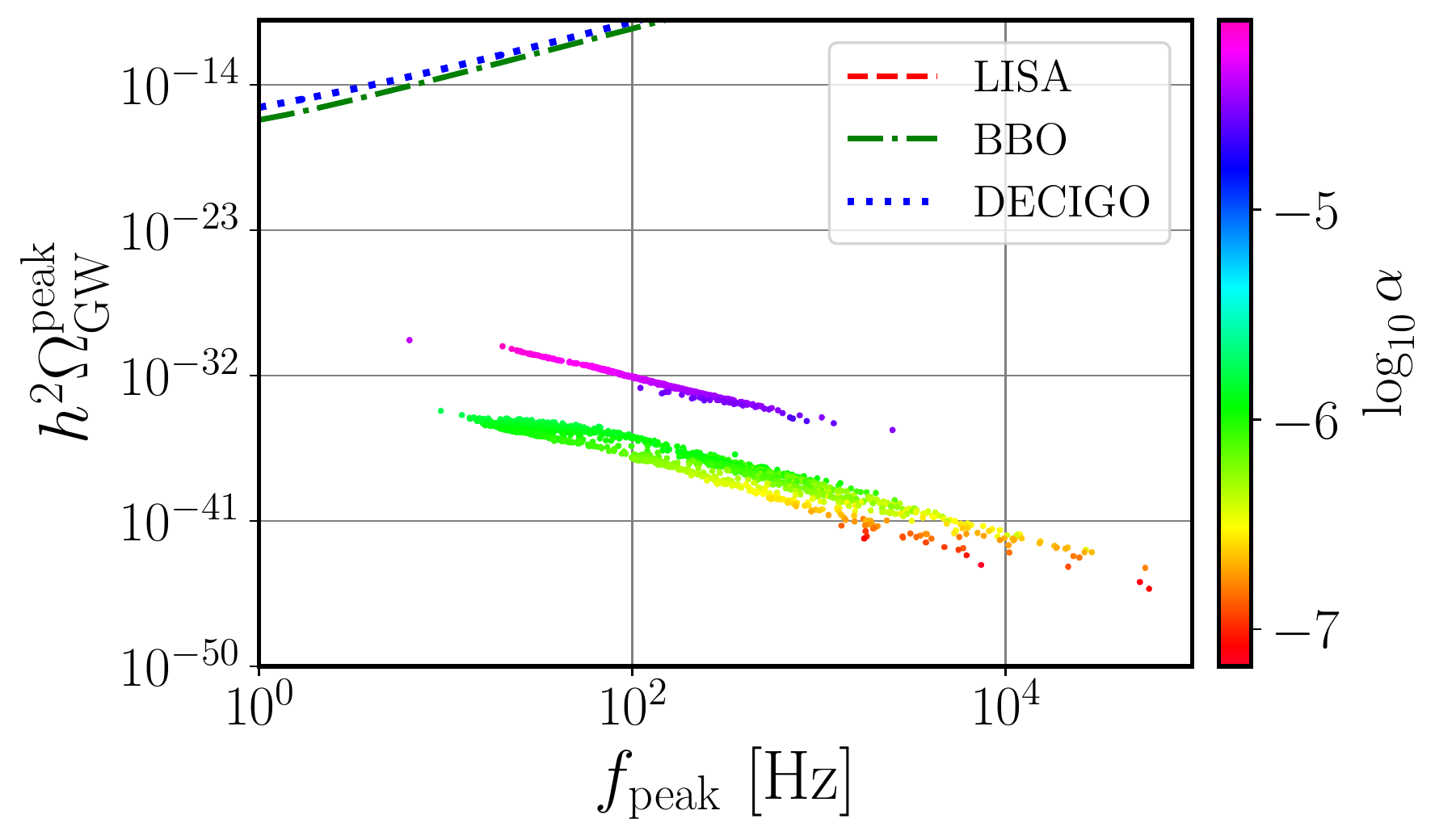}
\includegraphics[width=0.48\textwidth]{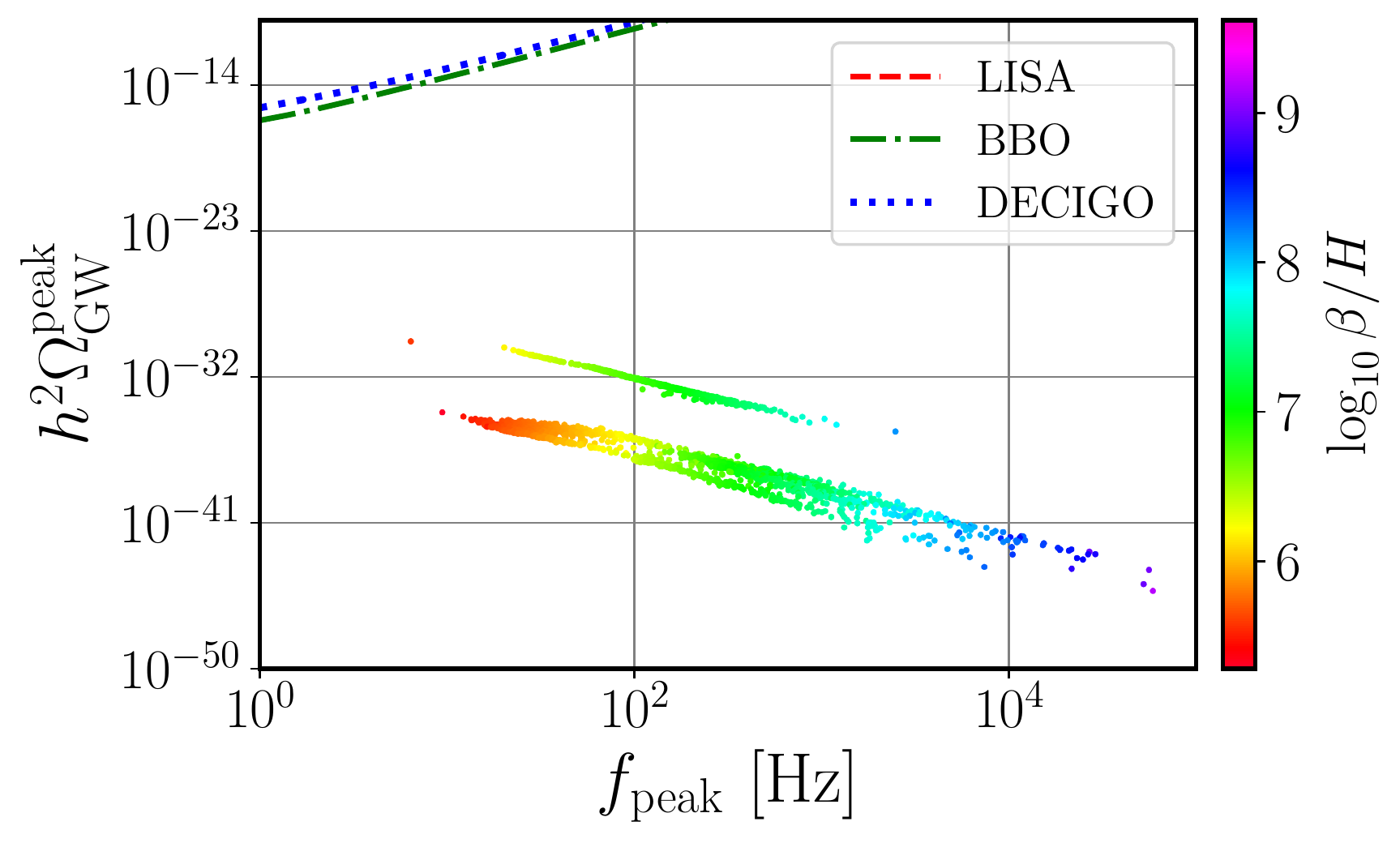}      
\caption{\footnotesize Scatter plots showing the strength (left) and inverse duration (right) of the phase transition in terms of the peak frequency and peak energy density amplitude of the SGWB for the 4D limit of the EIS model.}
\label{fig:4D}
\end{figure}
We show in \cref{fig:4D} the results obtained for the peak energy density amplitude $h^2 \Omega^\mathrm{peak}_\mathrm{GW}$ of the SGWB in terms of the peak frequency. The colour scale on the left panel represents the strength of the phase transition $\alpha$, while on the right panel it describes its inverse duration $\beta/H$. With the current improved analysis we confirm our previous results where all FOPTs are rather weak, $\alpha < 10^{-4}$, and very short lasting, $\beta/H > 10^5$. Indeed, none of the generated points is within the sensitivity reach of possible future GW detectors such as BBO or DECIGO. The maximum absolute value of the portal coupling $\lambda_{\sigma h}$ was indeed $0.01$, once again consistent with the conclusions in \cite{Addazi:2019dqt}.

\subsection{SGWB in the 6D EIS model}
\label{sec:6D}

In this section we study how the effect of UV physics, encoded in the form of dimension-6 operators in the scalar potential, can influence the properties of the phase transitions described in \cref{sec:4D}. Using the inversion equations in \cref{eq:coup_inputs} we have performed a parameter space scan with the ranges shown in \cref{tab2}.
\begin{table*}[h!]
\centering
\begin{tabular}{@{}rcccr@{}}
& \multicolumn{3}{c}{} \\
\hline
& Parameter & Range & Distribution &\\ \hline
&  & &  &\\
& $m_{h_2}$ & $[60,\,1000]\,\text{GeV}$ & linear\\
&  & &  &\\
& $m_{J}$ & $[10^{-10}~\mathrm{eV},\,100~\mathrm{keV}]\,$ & exponential &\\
&  & &  &\\
& $m_{\nu_1}$ & $[10^{-6},\,10^{-1}]\,\text{eV}$ & exponential\\
&  & &  &\\
& $\mathrm{Br}(h_1 \to J J)$ & $[10^{-15},\,0.18]$ & exponential &\\
&  & &  &\\
& $\sin{(\alpha_h)}$ & $\pm[0,\,0.24]$ & linear &\\
&  & &  &\\
& $v_\sigma$ & $[100,\,1000]~\mathrm{GeV}$ & linear &\\
&  & &  &\\
& $\Lambda$ & $[10,\,1000]~\mathrm{TeV}$ & exponential &\\
&  & &  &\\
& $\frac{{\red \delta_0}v_h^2}{2 \Lambda^2}$ & $\pm[10^{-10},\,4 \pi]$ & exponential &\\
&  & &  &\\
& $\frac{{\red \delta_2}\max(v_h^2,v_\sigma^2)}{2\Lambda^2}$ & $\pm[10^{-10},\,4 \pi]$ & exponential &\\
&  & &  &\\
& $\frac{{\red \delta_4}v_\sigma^2}{2\Lambda^2}$ & $\pm[10^{-10},\,4 \pi]$ & exponential &\\
&  & &  &\\
\hline
\end{tabular}
\caption{ \footnotesize Randomly sampled ranges of the input parameters in the scans for the 6D EIS model. We fix $m_{h_1} = 125.01~\mathrm{GeV}$ and $v_h = 246.22~\mathrm{GeV}$. In the last three lines we show the expressions implemented in our code used to calculate the parameters highlighted in {\red red} in terms of the $\Lambda$ scale and the VEVs.}
\label{tab2}
\end{table*}
In the last three lines of \cref{tab2} we sample the values of $\delta_{0,2,4}$ requiring that the effective quartic couplings $\tfrac{\delta_0 v_h^2}{2\Lambda^2}$, $\tfrac{\delta_2 \max(v_h^2,v_\sigma^2)}{2\Lambda^2}$ and $\tfrac{\delta_4 v_\sigma^2}{2\Lambda^2}$ are perturbative. Notice that the negative signs in the sampling range of such dimension-six couplings is not problematic provided that all FOPT scenarios found with \texttt{CosmoTransitions} feature a stable vacuum, at least for energy scales below $\Lambda$. The neutrino masses are sampled according to a normal ordering, \textit{i.e.}
\begin{equation}
\begin{aligned}
        m_{\nu_2}^2 &= m_{\nu_1}^2 + \Delta m_{21}^2\qquad \text{with} \qquad \Delta m_{21}^2 = 8\times 10^{-5}~\mathrm{eV}^2\,, \\
        m_{\nu_3}^2 &= m_{\nu_2}^2 + \Delta m_{32}^2\qquad \text{with} \qquad \Delta m_{32}^2 = 3\times 10^{-3}~\mathrm{eV}^2 \,.
\end{aligned}
\end{equation}

\begin{figure}[htb!]
\centering
\includegraphics[width=0.49\textwidth]{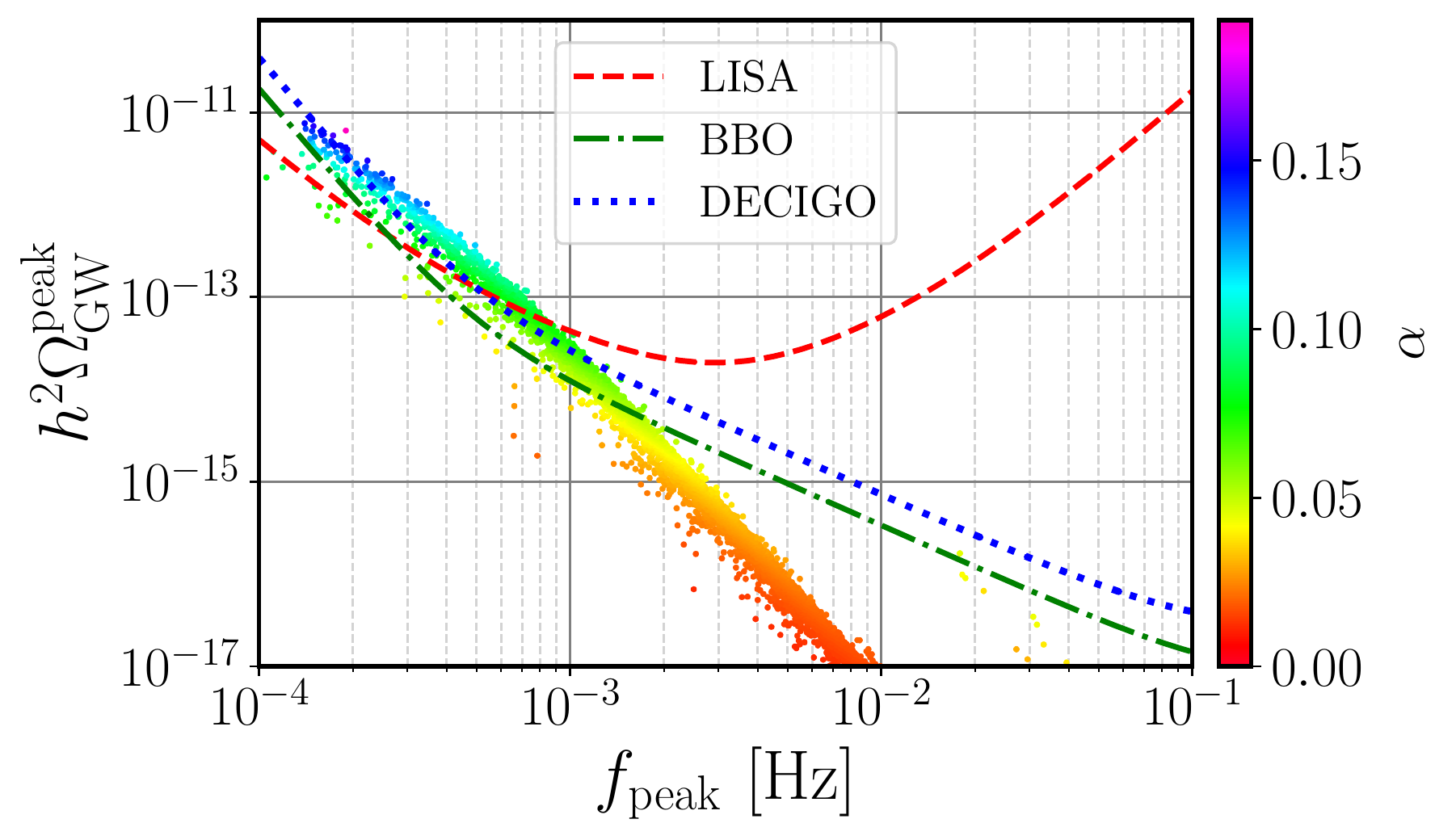}
\includegraphics[width=0.48\textwidth]{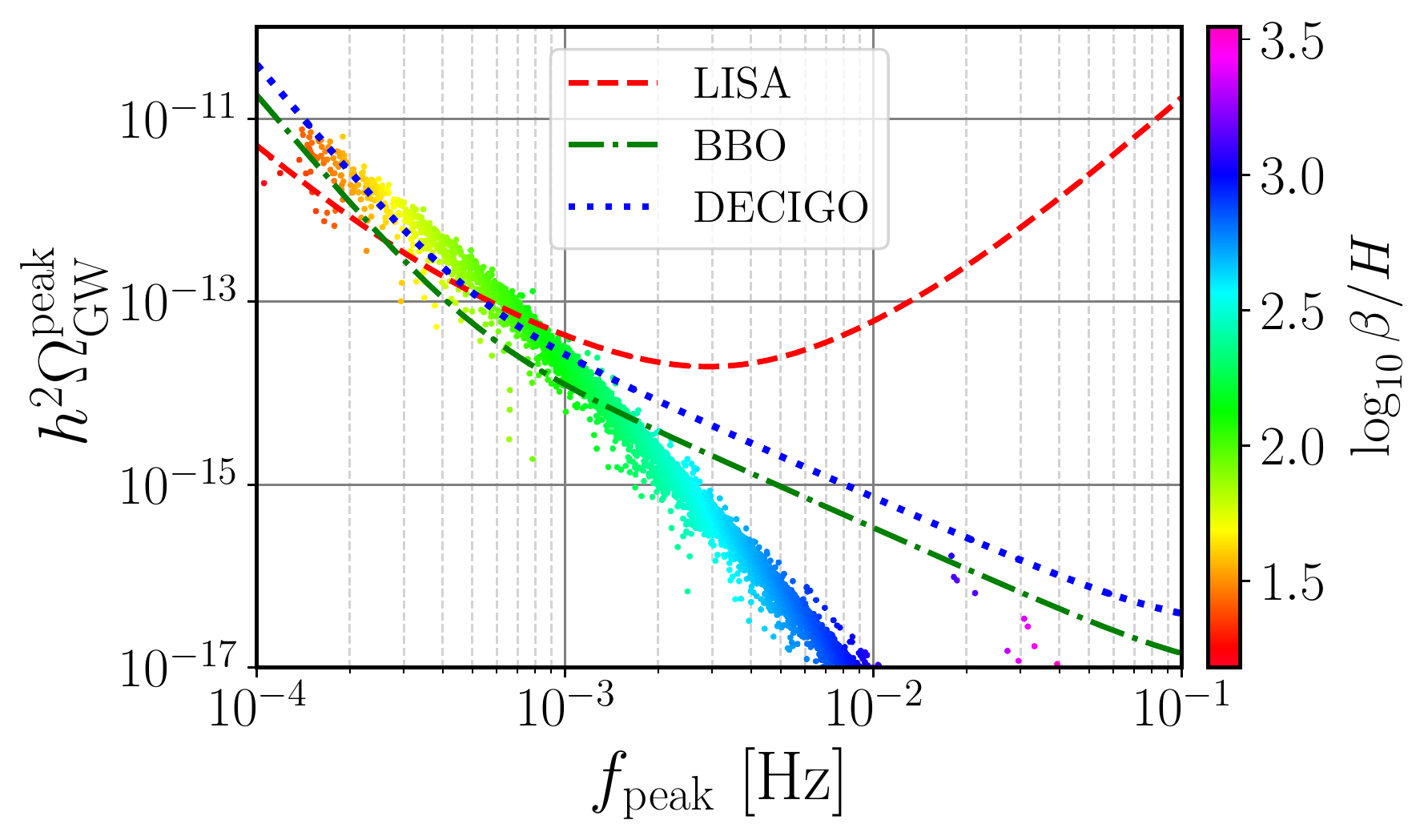} 
\caption{\footnotesize Scatter plots showing the strength (left) and inverse duration (right) of the phase transition in terms of the peak frequency and peak energy density amplitude of the SGWB.}
\label{fig:alpha_beta}
\end{figure}

We show in \cref{fig:alpha_beta} the SGWB peak amplitude and frequency for all generated points in terms of the phase transition parameters $\alpha$ and $\beta/H$ in the colour scale. Notice that the linear distribution of points is justified by \cref{eq:Opeak2} where, in a log-log scale, $h^2 \Omega_\mathrm{GW}^\mathrm{peak} \propto f_\mathrm{peak}^{-2}$ does indeed linearly depends on $-2 \log_{10} f_\mathrm{peak}$ with a negative slope.
The effect of including dimension-6 operators fundamentally changes the conclusions revisited in \cref{sec:4D}. In particular, for peak frequencies $f_\mathrm{peak} \lesssim 10^{-3}~\mathrm{Hz}$, the SGWB becomes observable at LISA. For such scenarios both the strength and the duration of the phase transition are considerably larger, i.e.~$\alpha \sim 0.1$ and $10 \lesssim \beta/H \lesssim 100$.

\begin{figure}[htb!]
\centering
\includegraphics[width=0.49\textwidth]{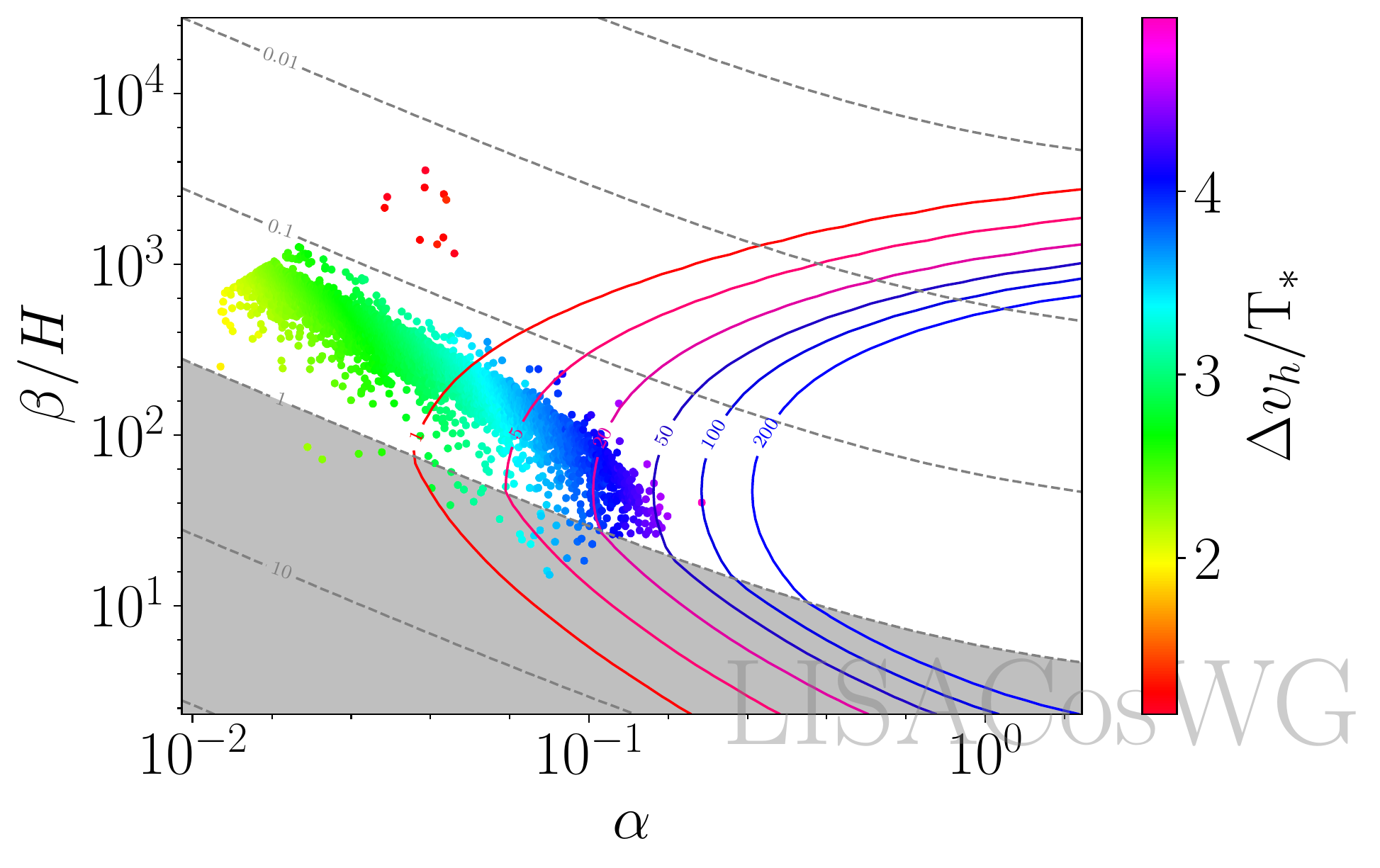}
\includegraphics[width=0.48\textwidth]{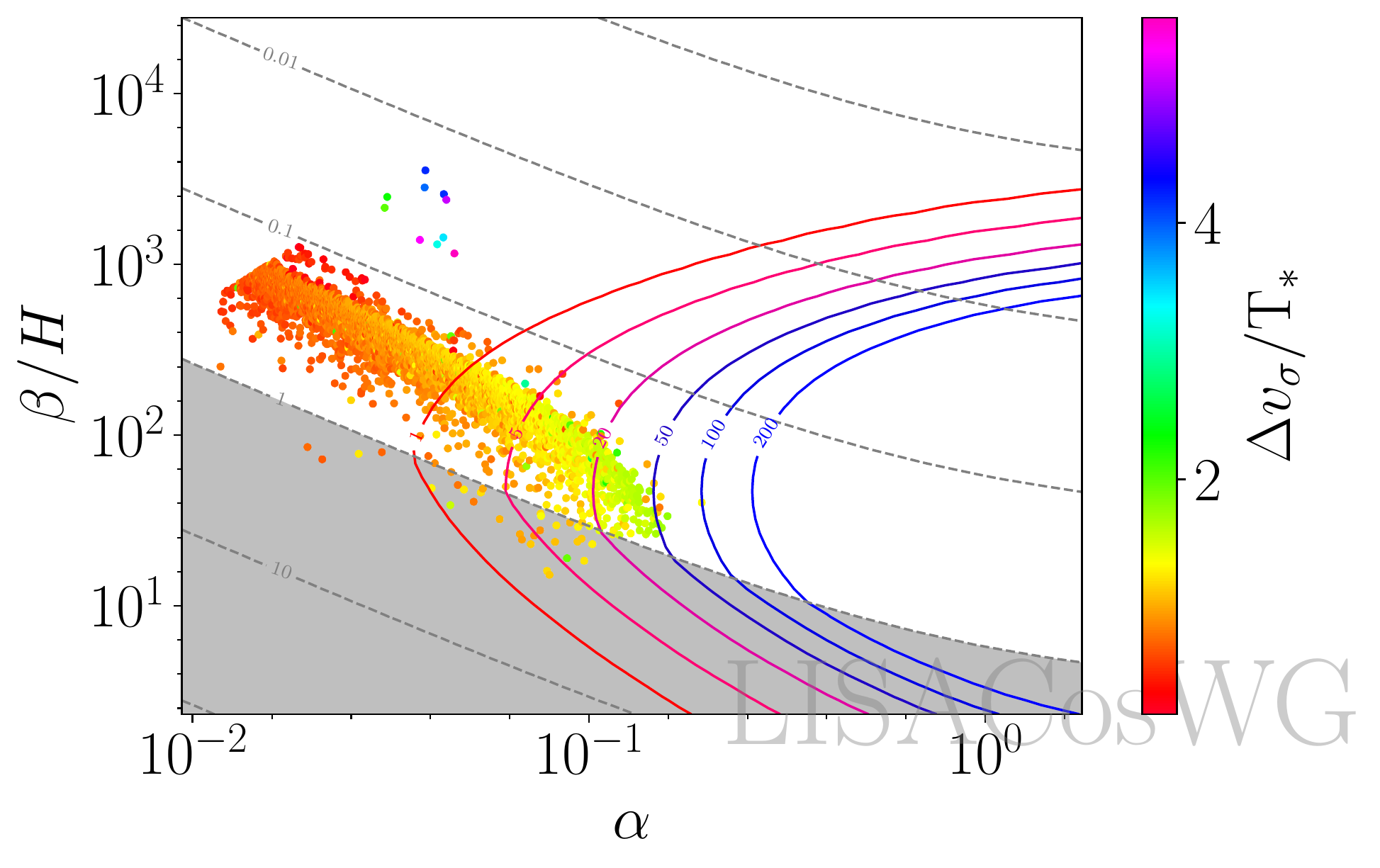} 
\caption{\footnotesize SNR plots showing the order parameters of the EW (left) and lepton number symmetry  (right) phase transitions. The coloured isolines represent the SNR at LISA while the grey dahsed ones denote the shock formation time. The grey shaded are represents the area where the sound waves treatment is mostly reliable. These plots were produced using the public software \texttt{PTPlot} \cite{Caprini:2019egz}.}
\label{fig:SNR1}
\end{figure}

In \cref{fig:SNR1} we show the Signal to Noise Ratio (SNR) for LISA with the colour scale indicating the electroweak (left) and lepton number symmetry (right) order parameters defined in \cref{eq:order}. 
The colored isolines represent the expected SNR values for a five year exposure time. The dashed contours display the shock formation time $\tau_\mathrm{sh}$ where the grey shaded area corresponds to an acoustic period lasting longer than the Hubble time where the sound waves treatment is mostly reliable~\cite{Ellis:2018mja,Hindmarsh:2017gnf}. Conversely, if $\tau_\mathrm{sh} << 1$, the turbulence effects may become important. However, none of the generated points feature a too small shock formation time with the majority having $\tau_\mathrm{sh} > 0.1$. The order parameters in the colour scales indicate that both the EW and the $\U{L}$ phase transitions must be simultaneously strong, with $\Delta v_h /T_\ast \approx 4$ and $\Delta v_\sigma / T_\ast \approx 2$, such that the SNR at LISA is larger than $10$.

The Majoron masses considered in this study span over 15 orders of magnitude, equally distributed as shown in \cref{tab2}. Whenever $m_J > 2 m_\nu$ the Majoron can decay in a pair of neutrinos with a rate given by \cref{eq:Jnunu} \cite{Reig:2019sok,Heeck:2017xbu,Boulebnane:2017fxw,Frigerio:2011in,Lattanzi:2014mia}. According to the combined analysis of Planck, WMAP, WiggleZ and BOSS \cite{Audren:2014bca}, the Majoron is very long lived and a dark matter candidate if $\Gamma(J \to \nu \nu) < 1.9 \times 10^{-19} s^{-1}~@~95\%~\mathrm{C.L.}$. This is typically achieved for large $\U{L}$ breaking scales, several orders of magnitude above $v_\sigma \sim \mathcal{O}(\mathrm{TeV})$ as considered in this work and motivated by what we believe being a natural scale in the EIS model (see \cref{eq:nu-light} and the discussion below). On the other hand, ultralight Majorons, $m_J < 2 m_\nu$, with negligible decay rates to photons are stable. Whether they can offer a good dark matter is not studied here and left for future work.

\subsection{Connection to collider observables}
\label{sec:cosmo-hep}

The potential discovery of a SGWB can become the first direct measurement of the Universe prior to the Big Bang Nucleosynthesis era and a breakthrough comparable to the Cosmic Microwave Background detection \cite{Penzias:1965wn}. Such an observation (or the lack of it), will pose constraints on NP models in the form of bounds (or upper limits) on the amount of allowed Gravitational radiation in the early Universe. It is therefore legitimate to expect correlations between collider and GW observables, in particular for models featuring Higgs portal interactions. In what follows we study how the scalar mixing angle $\sin \alpha_h$, the Higgs trilinear coupling modifier $\kappa_\lambda$ and the mass of a second visible scalar $m_{h_2}$, are related to the phase transition parameters and the peak amplitude $h^2 \Omega^\mathrm{peak}_\mathrm{GW}$ of the associated SGWB.

The presence of the 6D operator $(H^\dagger H)^3$, parameterized by $\delta_0$ in this work, can on its own induce FOPTs as discussed in \cite{Camargo-Molina:2021zgz,Postma:2020toi,Chala:2018ari}. However, recall that from \cref{fig:SNR1}, an observable SGWB requires both $\Delta v_h/T_\ast > 1$ and $\Delta v_\sigma / T_\ast > 1$, suggesting sizeable $\delta_2$ and/or $\delta_4$.
\begin{figure}[htb!]
\centering
\includegraphics[width=0.48\textwidth]{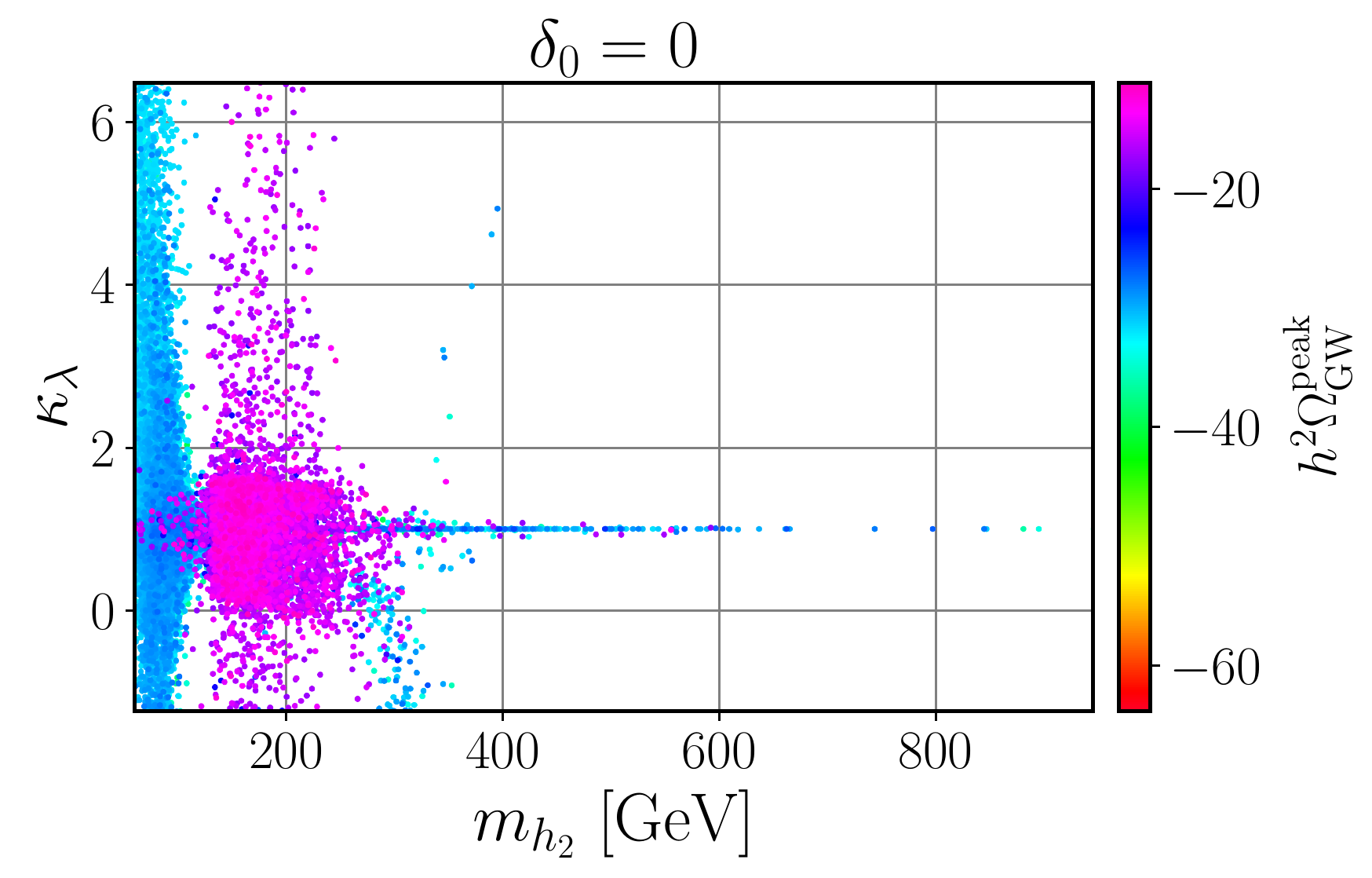}
\includegraphics[width=0.48\textwidth]{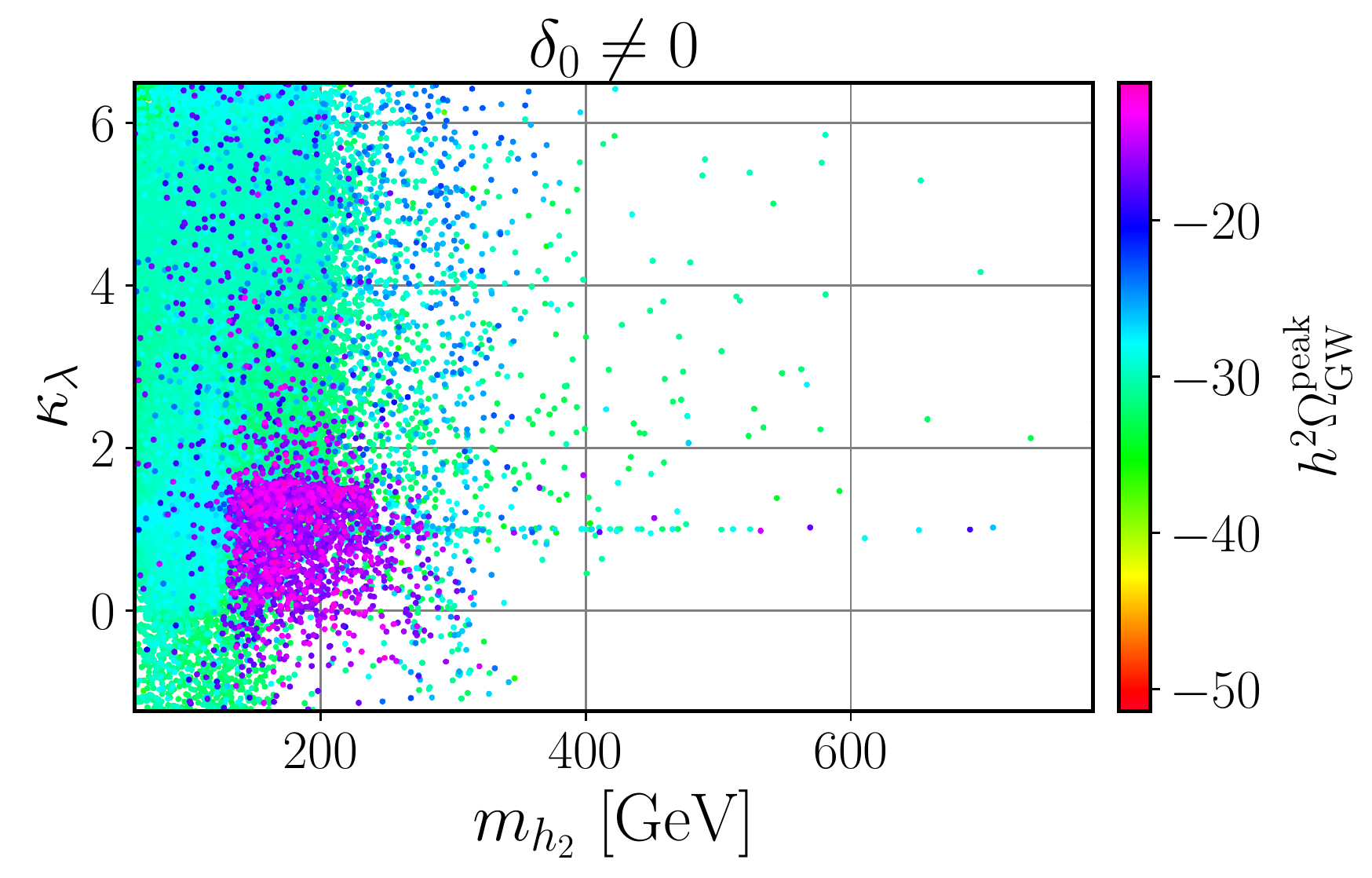}
\caption{\footnotesize Scatter plots showing the dependency of the Higgs trilinear coupling modifier in terms of the second CP-even Higgs boson mass and the energy density amplitude of the SGWB in the colour scale. In the left panel $\delta_0 = 0$ whereas in the right panel $\delta_0 \neq 0$.}
\label{fig:kl-mh2-Om}
\end{figure}
For a finer scrutiny we show in \cref{fig:kl-mh2-Om} two selections of data where in the left panel $\delta_0 = 0$ and in the right we require $\delta_0 \neq 0$. The magenta points represent the $(\kappa_\lambda,m_{h_2})$ region with SGWBs potentially observable at LISA, BBO and DECIGO. The values of $\kappa_\lambda$ comply with current CMS constraints \cite{CMS:2022dwd}, while the new CP-even Higgs boson couplings to the SM are suppressed by a small mixing angle $\abs{\sin \alpha_h} < 0.23$ \cite{Papaefstathiou:2022oyi,ATLAS:2021vrm}. As anticipated, $\delta_0 \neq 0$ significantly increases the area populated with FOPTs. However, a considerable subset of such points fill out the green region on the right panel where phase transitions are not strong enough to be within the sensitivity reach of future GW detectors. In both scenarios we have found that testable SGWB signals prefer $100 \lesssim m_{h_2}^2/ \mathrm{GeV} \lesssim 300$ and $0 \lesssim \kappa_\lambda \lesssim 2$.

\begin{figure}[htb!]
\centering
\includegraphics[width=0.48\textwidth]{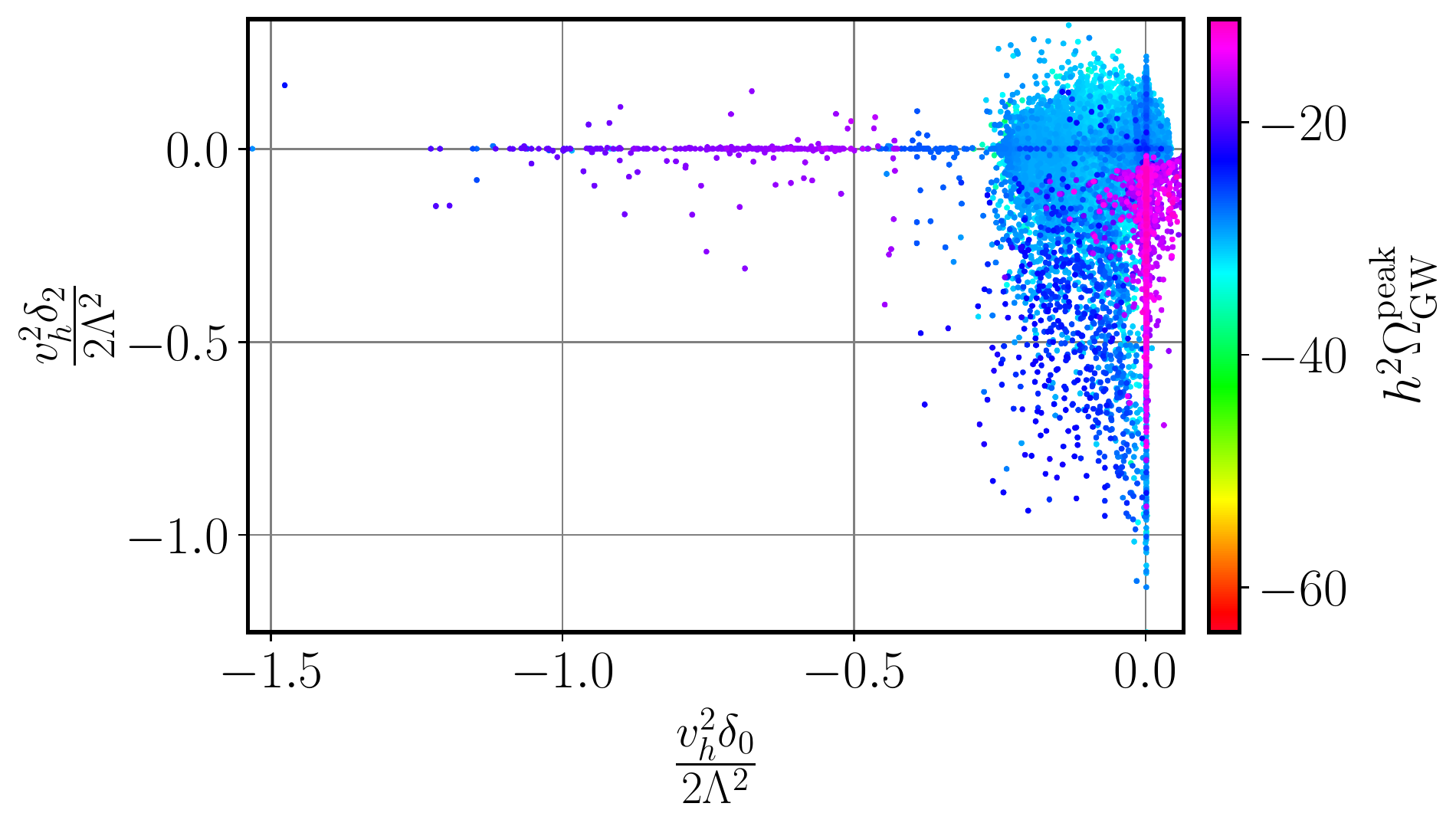}
\includegraphics[width=0.48\textwidth]{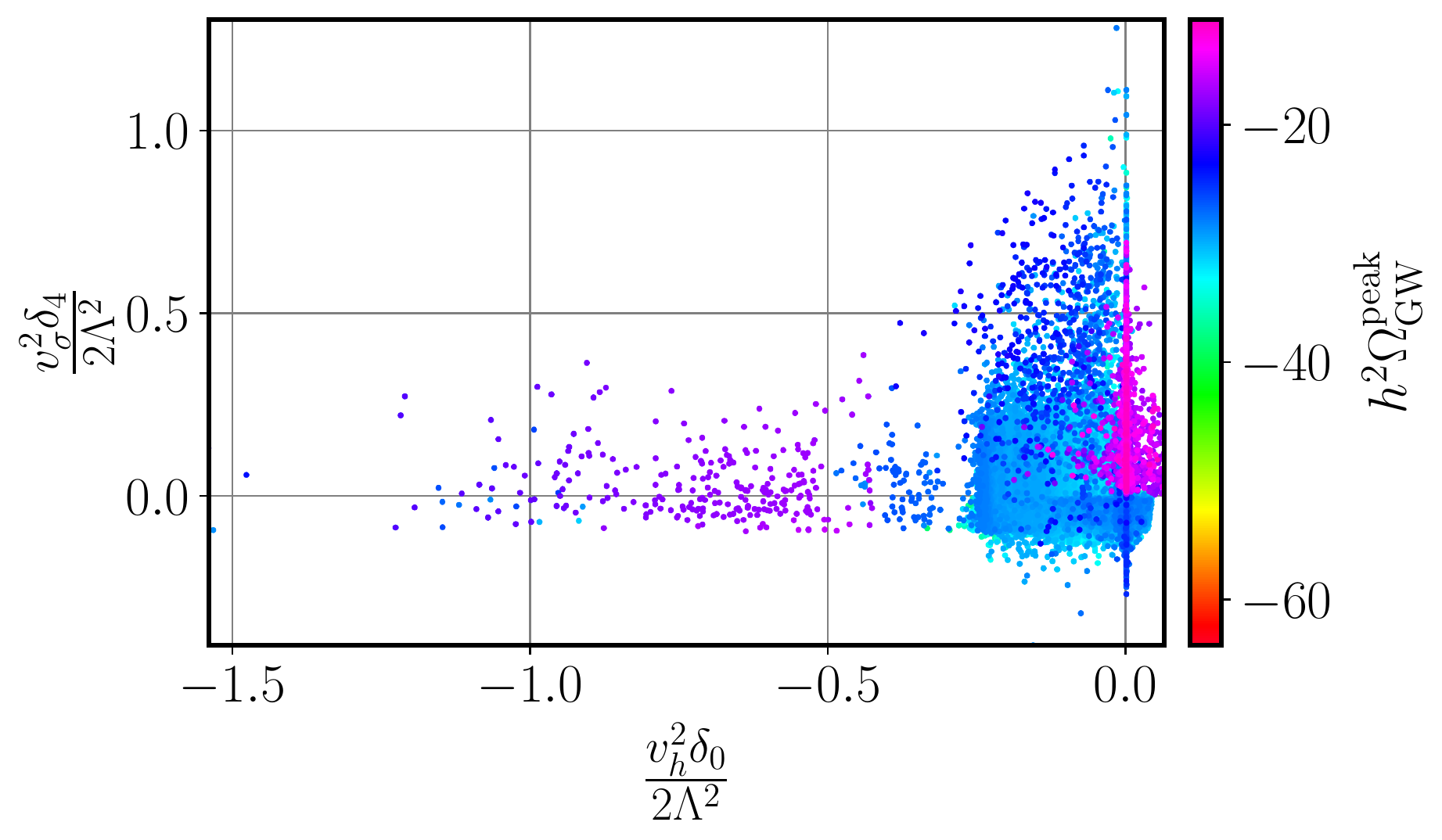}
\includegraphics[width=0.48\textwidth]{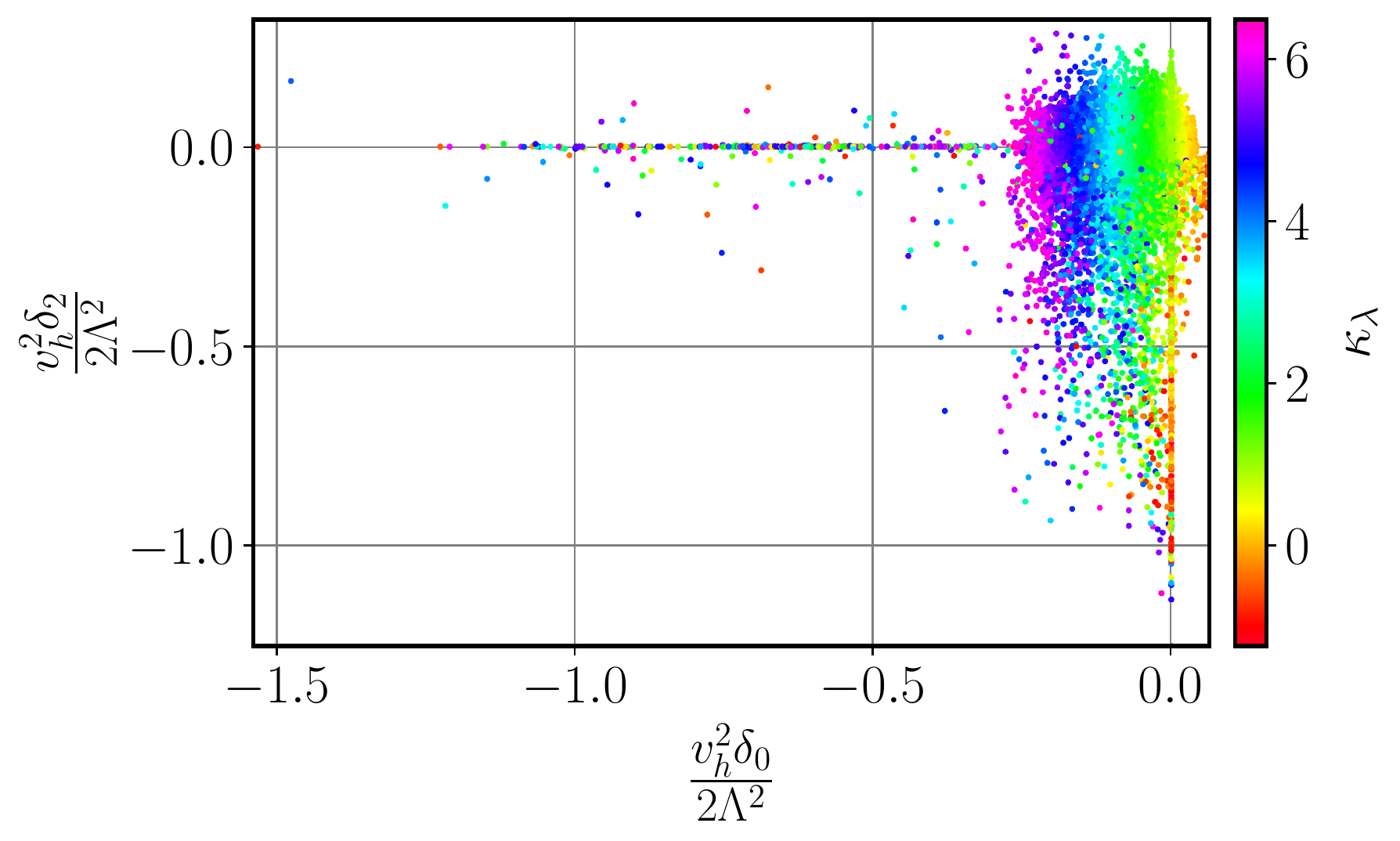}
\includegraphics[width=0.48\textwidth]{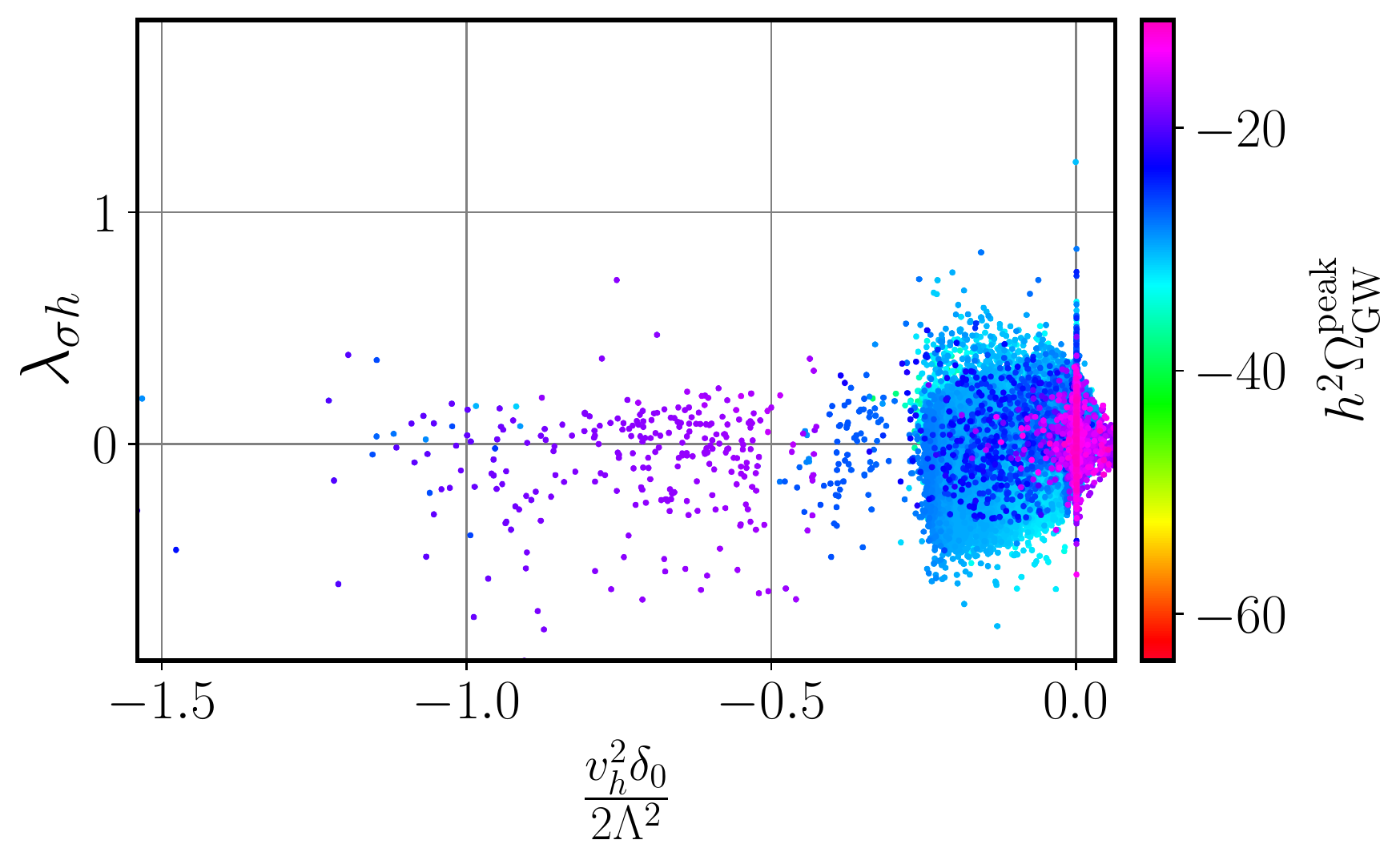}
\caption{\footnotesize Scatter plots showing the correlation between the strength of the Higgs self interaction $\tfrac{v_h^2 \delta_0}{2 \Lambda^2}$ and the effective portal couplings $\tfrac{v_h^2 \delta_2}{2 \Lambda^2}$ (both left panels), $\tfrac{v_\sigma^2 \delta_4}{2 \Lambda^2}$ (top-right) and $\lambda_{\sigma h}$ (bottom-right). The SGWB peak amplitude is shown in the colour scale of the top and bottom-right panels, while in the bottom-right we show the trilinear Higgs coupling modifier.}
\label{fig:d0d2d4}
\end{figure}

For completeness, we show in \cref{fig:d0d2d4} how the GW peak amplitude and the 6D Higgs self coupling $\tfrac{v_h^2 \delta_0}{2 \Lambda^2}$ are related to the effective portal interactions $\tfrac{v_h^2 \delta_2}{2 \Lambda^2}$ (top-left panel), $\tfrac{v_\sigma^2 \delta_4}{2 \Lambda^2}$ (top-right panel) and $\lambda_{\sigma h}$ (bottom-right). In the bottom-left panel we show the same parameter space projection as in the top-left with the Higgs trilinear coupling modifier in the colour scale.
The brighter magenta points, where the SGWB peak amplitude is maximized, populate a region with small $\tfrac{v_h^2\delta_0}{2 \Lambda^2}$. These also correspond to both magenta blobs in \cref{fig:kl-mh2-Om}, with the left panel highlighting those scenarios which overlap the $\tfrac{v_h^2\delta_0}{2 \Lambda^2} = 0$ axis. Larger absolute values of the 6D Higgs self interaction, in particular represented by the sparse purple region where $-1.0 \lesssim \tfrac{v_h^2\delta_0}{2 \Lambda^2} \lesssim -0.5 $, the peak amplitude of the SGWB was found to be smaller than $h^2 \Omega^\mathrm{peak}_\mathrm{GW} \lesssim \mathcal{O}(10^{-16})$, at least three orders of magnitude below the LISA reach. However, these are possibly accessible to a next generation of GW detectors such as BBO or DECIGO, sensitive to peak frequencies in the $\mathrm{mHz}$ to $\mathrm{Hz}$ range.

Larger absolute values of $\delta_0$ typically enhance the Higgs trilinear self coupling as one can see in the term proportional to $\cos^3 \alpha_h$ in \cref{111}. Such a leading order effect is visible in the colour gradient of the bottom-left panel of \cref{fig:d0d2d4}. The vertical separation between the sparsely and densely populated regions is due to the current CMS upper bound where $\kappa_\lambda < 6.5$ \cite{CMS:2022dwd}. Future measurements at colliders can impose stronger constraints as further discussed below.

The approximately symmetric distribution of points along the horizontal axis in the top and bottom-right panels of \cref{fig:d0d2d4}, particularly evident in the densely populated magenta-blue region, is a reflection of the combined effect of sizeable portal interactions $\tfrac{v_h^2 \delta_2}{2 \Lambda^2}$, $\tfrac{v_\sigma^2 \delta_4}{2 \Lambda^2}$ and $\lambda_{\sigma h}$, needed to induce FOPTs, with the required cancellation among them to keep the invisible Higgs decay branching ratio under control. We refer to the discussion related to \cref{eq:inv,eq:Lhjj,eq:BRinv} for further details. We also find that strong FOPTs with gravitational radiation observable in the form of a SGWB at LISA, requires non-vanishing $\delta_2 < 0$ and $\delta_4 > 0$.

\begin{figure}[htb!]
\centering
\includegraphics[width=0.48\textwidth]{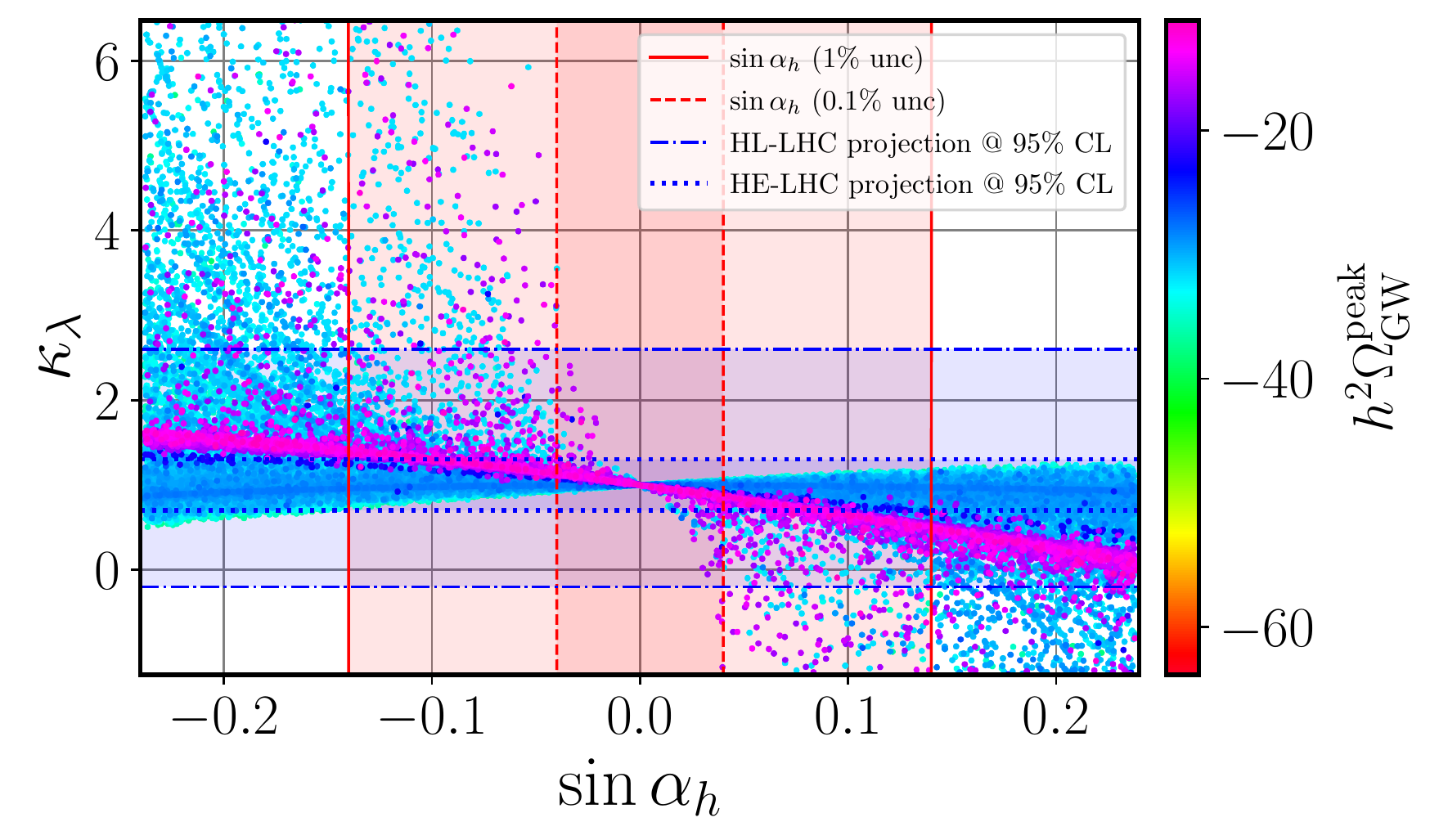}
\includegraphics[width=0.48\textwidth]{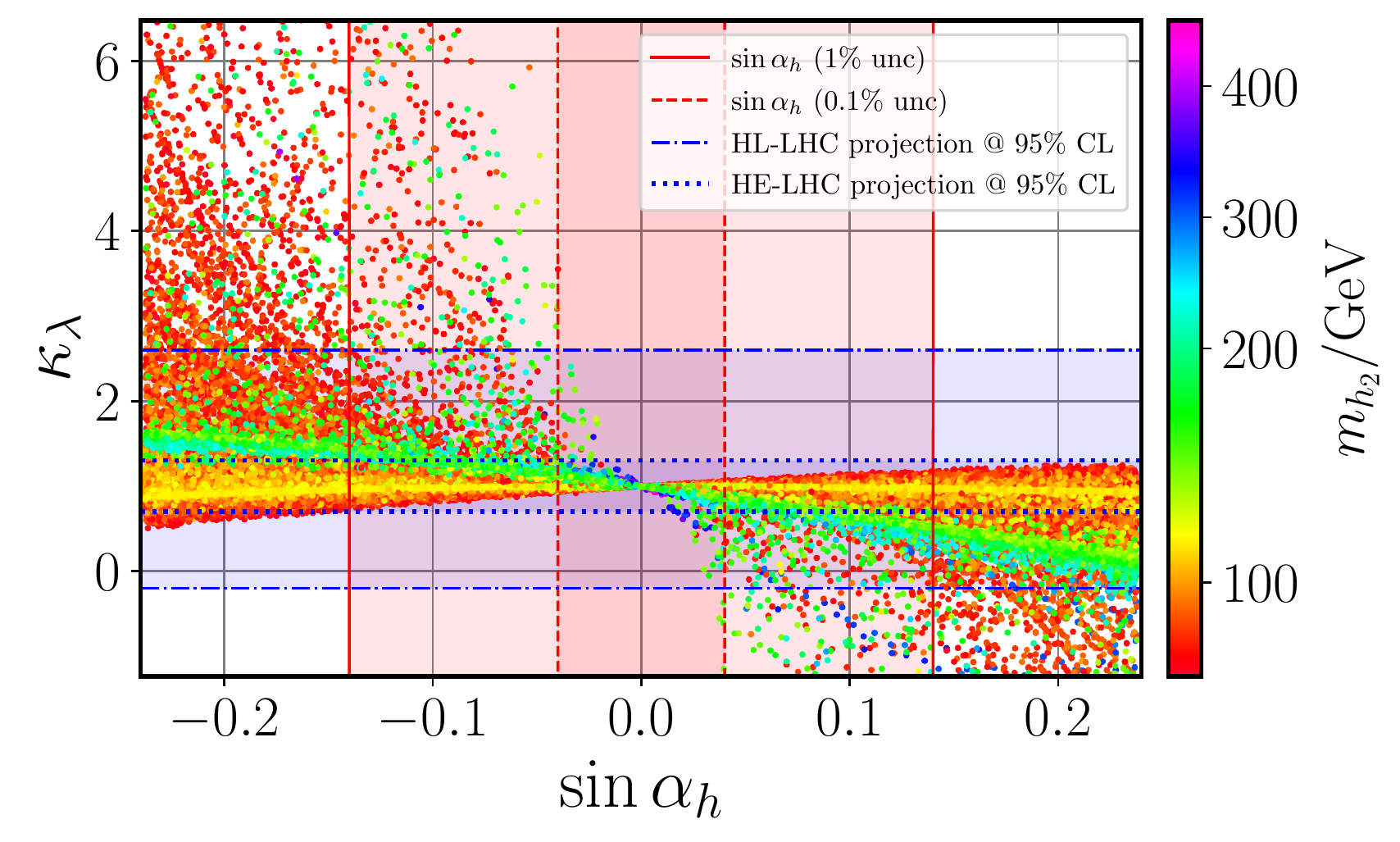}
\includegraphics[width=0.48\textwidth]{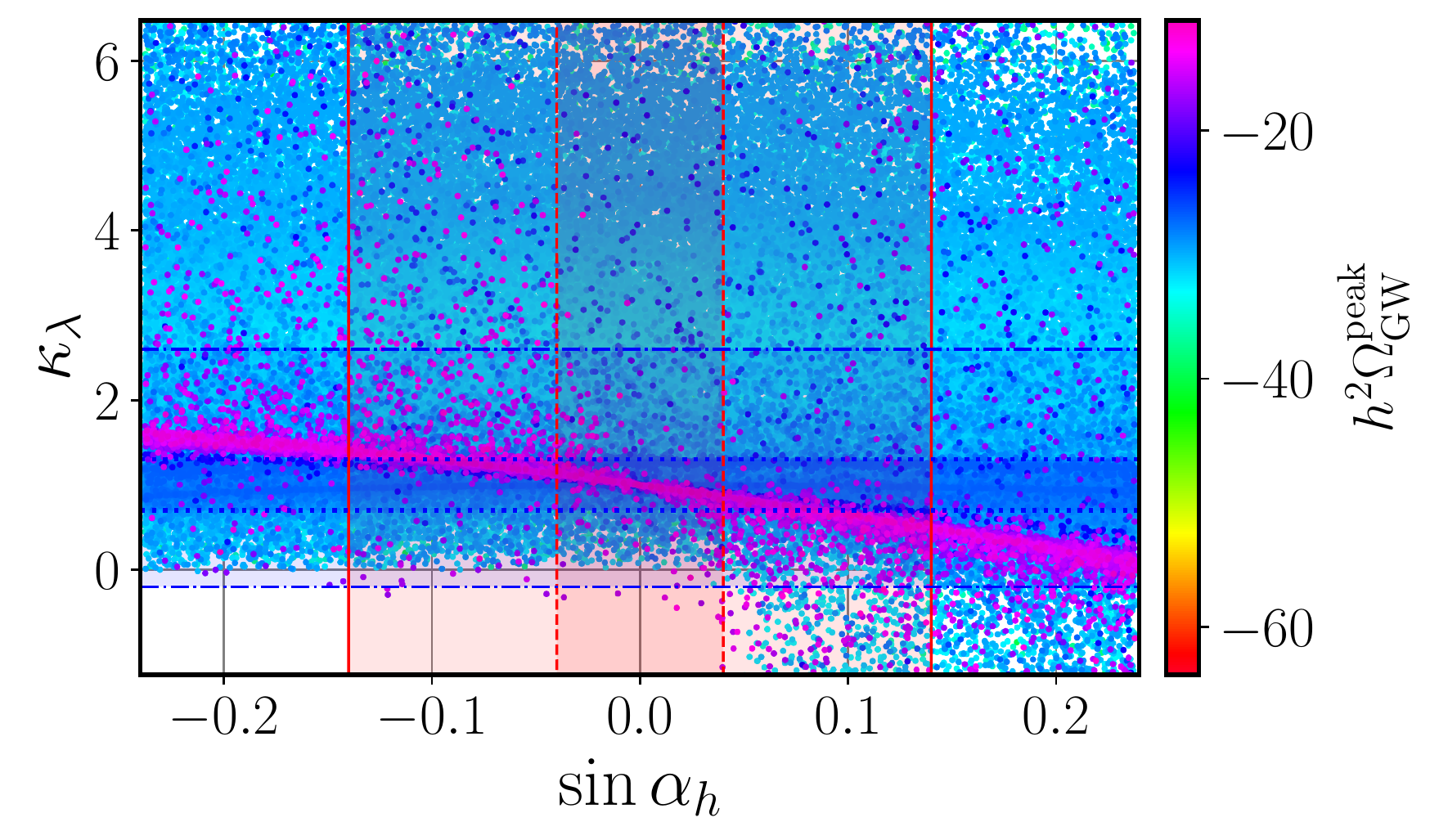}
\includegraphics[width=0.48\textwidth]{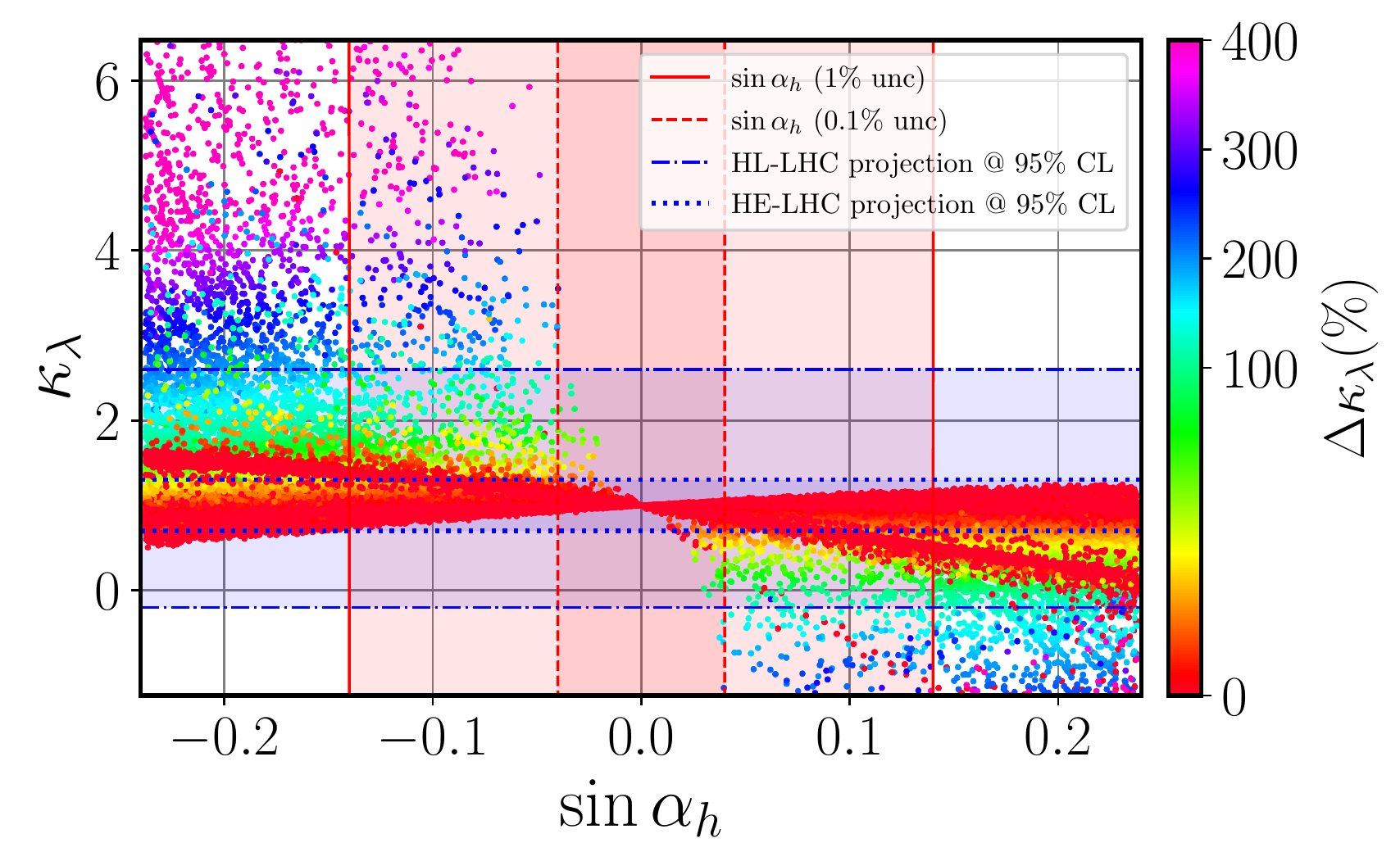}
\caption{\footnotesize Scatter plots showing the correlations between the Higgs trilinear coupling modifier $\kappa_\lambda$ and the scalar mixing angle $\sin \alpha_h$. On both left panels, where the top one includes only $\delta_0 = 0$ data while the bottom one features all generated viable points, the colour scale denotes the energy density peak amplitude of the SGWB. On the top-right panel the colour gradient describes the mass of the new CP-even Higgs boson, $h_2$, and on the bottom-right one it quantifies the size of the one-loop contribution to the Higgs trilinear self coupling as defined in \cref{eq:DeltaK}. The vertical red lines represent future constraints on $\sin \alpha_h$ assuming a precision of $1\%$ (solid lines) and $0.1\%$ (dashed lines) at future colliders \cite{Papaefstathiou:2022oyi}. The blue horizontal lines indicate the projected $95\%$ CL limits in $\kappa_\lambda$ measurements for the high-luminosity LHC (dot-dashed lines) and the future $\sqrt{s} = 27~\mathrm{TeV}$ high-energy upgrade (dotted lines) \cite{Goncalves:2018qas}. The regions under the darker blue and red shades correspond to the least constrained ones upon future measurements.}
\label{fig:kappa-sin}
\end{figure}
In \cref{fig:kappa-sin} we show the dependency of the Higgs trilinear self coupling modifier, $\kappa_\lambda$, in terms of the sine of the scalar mixing angle, $\sin \alpha_h$. On both left panels the colour scale represents the peak amplitude of the SGWB, $h^2 \Omega^\mathrm{peak}_\mathrm{GW}$, with the top-left panel including only those points generated with $\delta_0 = 0$ while for the bottom-left one no cut in $\delta_0$ was applied. Notice that the $\delta_0 \to 0$ limit offers cleaner results while capturing the key features relevant for our discussion. On the top-right panel the colour gradation describes the mass of the new CP-even Higgs boson while on the bottom-right it represents the size of the one-loop corrections on the trilinear Higgs coupling, defined as
\begin{equation}
    \Delta \kappa_\lambda (\%) = \abs{\frac{ \kappa_\lambda - \kappa_\lambda^\mathrm{tree}}{\kappa_\lambda^\mathrm{tree}}} \times 100\,.
    \label{eq:DeltaK}
\end{equation}

For the considered model, the vast majority of the scenarios testable at LISA populate the dense magenta band in the left panels with $0 < \kappa_\lambda < 2$. Comparing with the right plots we observe that such a region coincides with the green band on the top-right panel where $150 \lesssim m_{h_2}/ \mathrm{GeV} \lesssim 250$. The one-loop contributions to $\kappa_\lambda$ are of a sub-percent level as indicated by the red points in the bottom-right panel overlapping the magenta and green bands. For scenarios falling outside this region, sizeable one-loop corrections modifying the Higgs trilinear coupling up to a factor of four can enhance the strength of the phase transition. However, the latter can be significantly constrained, if not entirely ruled out, with future measurements of the scalar mixing angle (red vertical lines) and the Higgs trilinear self coupling (blue horizontal lines). Notice that in \cref{fig:kappa-sin}, the shaded regions between the red and blue lines correspond to the allowed parameter space projected for future measurements.

The results in \cref{fig:kappa-sin} are a good illustration of the potential interplay between collider and astrophysical measurements. For example, the hypothetical observation of a SGWB at LISA cannot, on its own, offer conclusive information about $\kappa_\lambda$, $\sin \alpha$ or $m_{h_2}$. However, with an increased precision in the determination of the mixing angle and trilinear Higgs coupling bounds at the high-luminosity (HL) or high-energy (HE) LHC upgrades, the viable parameter space becomes largely reduced as indicated by the shaded blue and red regions. Last but not least, in the decoupling limit, \textit{i.e.}~$\sin \alpha_h \to 0$ and $\kappa_\lambda \to 1$, only through GW experiments one can possibly test the considered Majoron model via parameter inference \cite{Caprini:2019pxz,Flauger:2020qyi}.

\subsection{Connection to the neutrino sector}
\label{sec:nu}

\begin{figure}[htb!]
\centering
\includegraphics[width=0.48\textwidth]{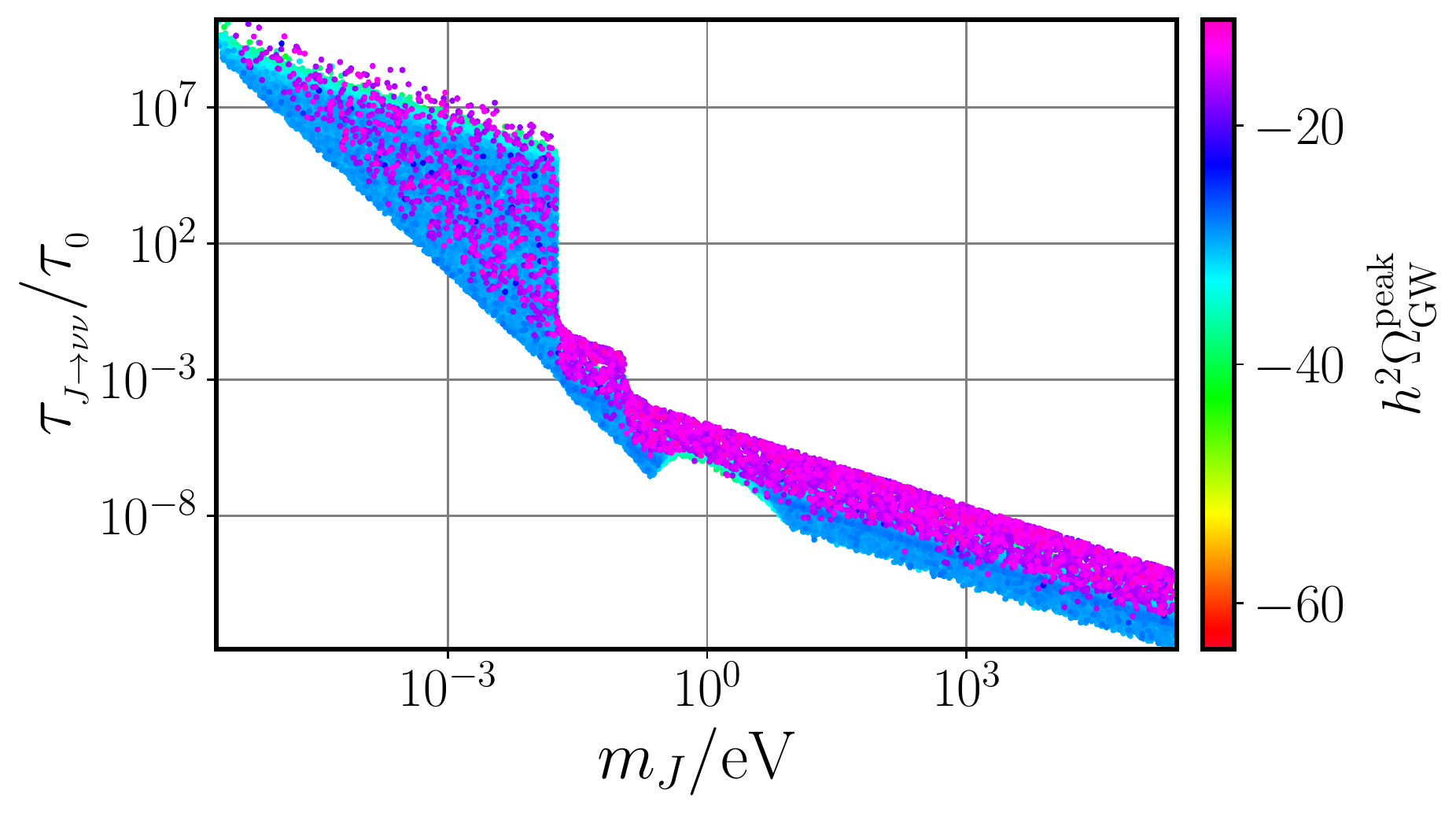}
\includegraphics[width=0.48\textwidth]{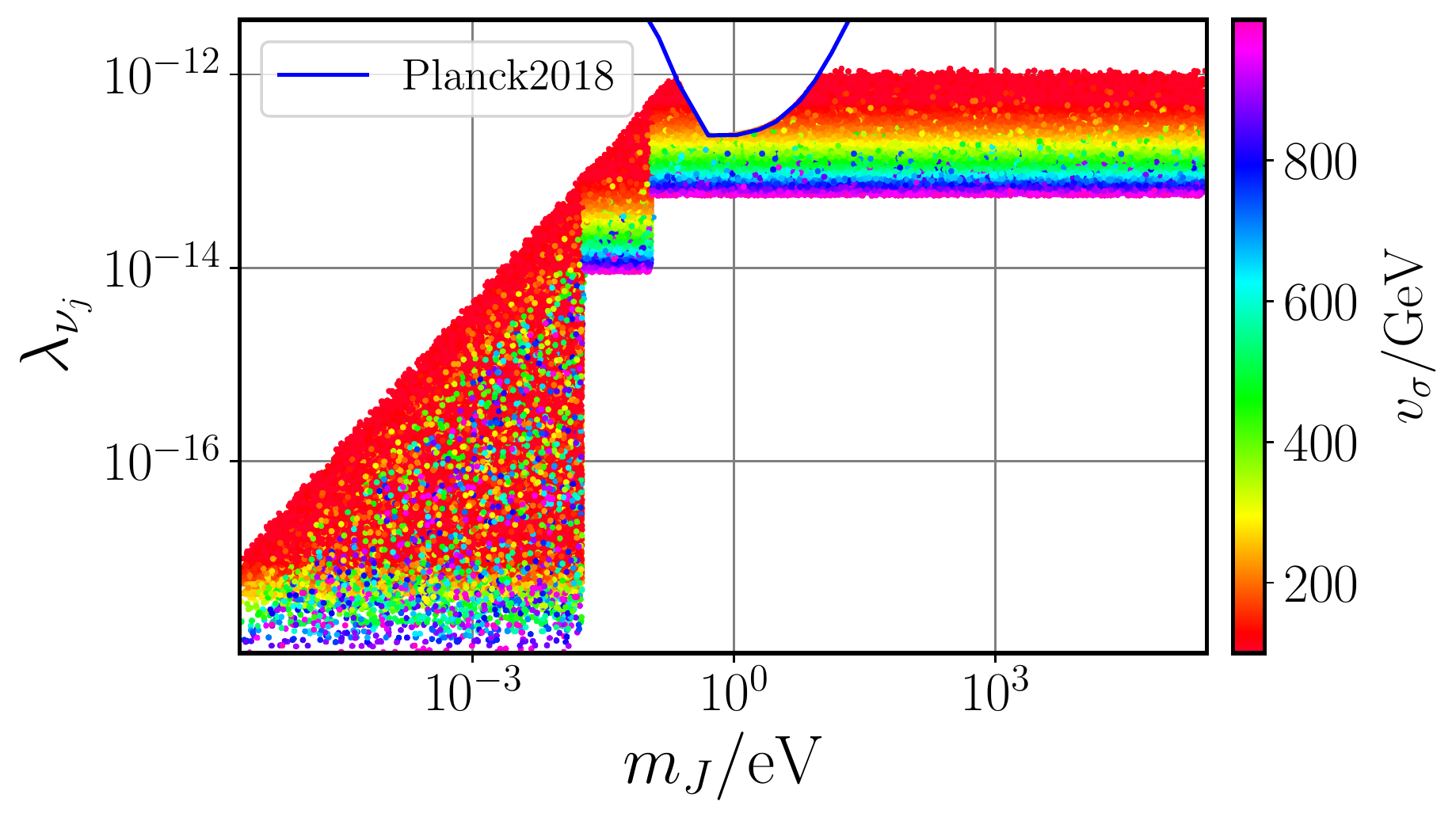}
\includegraphics[width=0.48\textwidth]{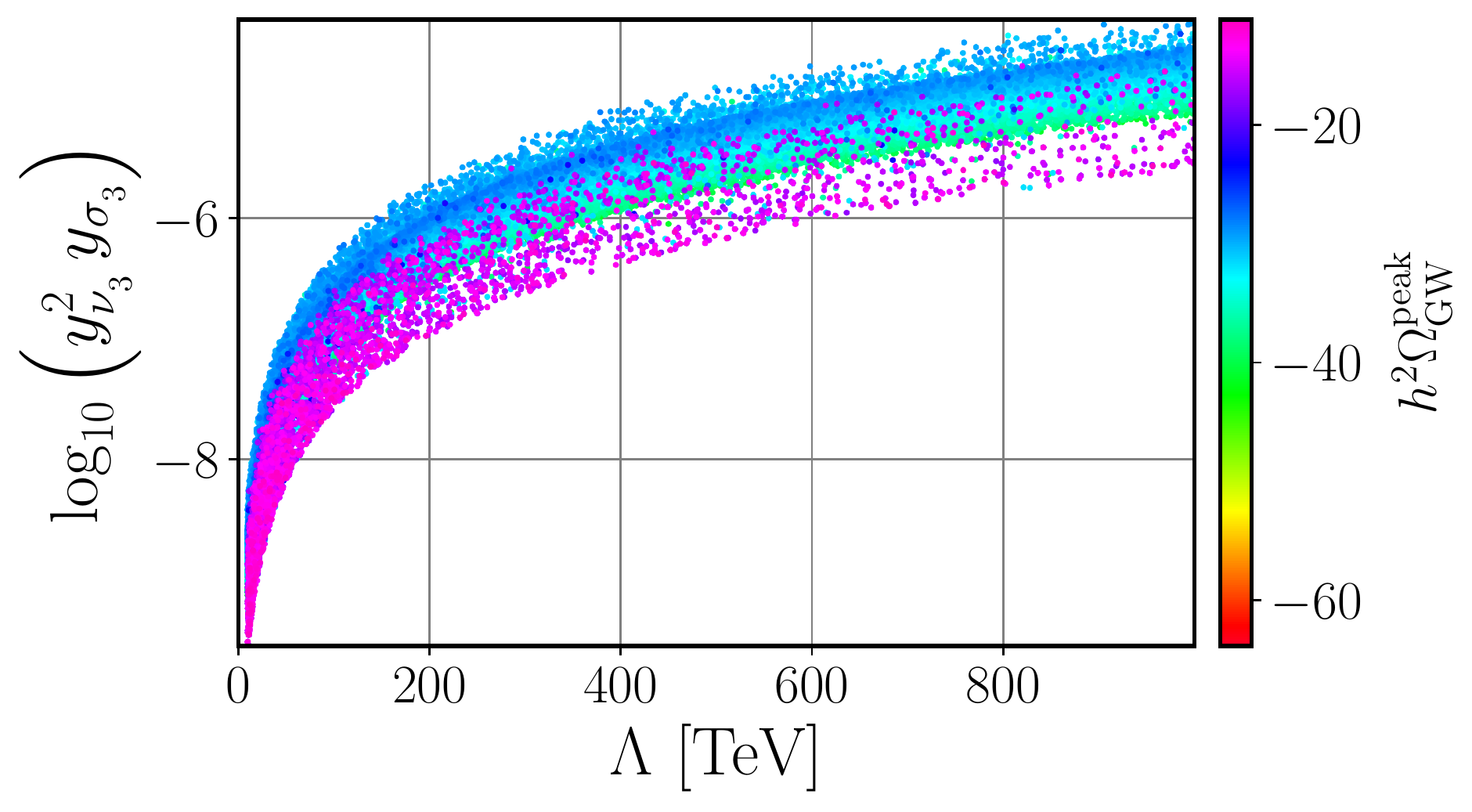}
\caption{\footnotesize Scatter plots showing the Majoron decay life-time normalized to the age of the Universe $\tau_0$ in terms of the Majoron mass and the SGWB peak amplitude (top-left), the strength of the neutrino coupling to Majorons versus the Majoron mass and the $\U{L}$ lepton number symmetry breaking scale (top-right), and the third generation nutrino Yukawa couplings in terms of the heavy neutrinos mass scale and the amplitude of the SGWB (bottom).}
\label{fig:nu}
\end{figure}
Compared with a number of astrophysical observations\cite{Bazzocchi:2008fh,Queiroz:2014yna,Lattanzi:2014mia}, we observe in our numerical results that there are several scenarios for which the Majoron is stable or long-lived. This is possible either when all $J \to \nu_j \nu_j$ channels are kinematically forbidden, or due to a large suppression caused by the lightest neutrino mass, \textit{i.e.}~$\lambda_{\nu_1} = m_1/v_\sigma$, with $m_1 \ll m_2 < m_3$. For scenarios with larger $\lambda_{\nu_j}$ constraints from CMB data must be taken care of. In particular, it was demonstrated in \cite{Escudero:2019gvw} that current \texttt{Planck2018} results \cite{Planck:2018vyg,Planck:2019nip} can provide an indirect probe to the $\U{L}$ lepton number symmetry breaking scale in the range $100~\mathrm{GeV}$ to $1~\mathrm{TeV}$, which precisely coincides with the values of $v_\sigma$ relevant for our analysis as shown in the colour scale of the top-right panel of \cref{fig:nu}. The excluded region at $95\%$ CL, illustrated with a blue contour, in which neutrinos and Majorons thermalize after BBN, features Majoron masses of approximately $0.1~\mathrm{eV}$ to $100~\mathrm{eV}$ and couplings to neutrinos in the range $10^{-13} \lesssim \lambda_{\nu_i} \lesssim 10^{-12}$. There are three distinct regions in the top-right panel whose separation results from kinematical thresholds. The rightmost one corresponds to $m_J > 2 m_{\nu_3}$ with the upper and lower flat boundaries explained by $\lambda^\mathrm{upper}_{\nu_3} = \tfrac{\max m_{\nu_3}}{\min v_\sigma} \approx \tfrac{0.1~\mathrm{eV}}{60~\mathrm{GeV}} \approx 1.7 \times 10^{-12}$ and  $\lambda^\mathrm{lower}_{\nu_3} = \tfrac{\min m_{\nu_3}}{\max v_\sigma} \approx \tfrac{0.06~\mathrm{eV}}{1~\mathrm{TeV}} \approx 6 \times 10^{-14}$, respectively. In the second domain $2 m_{\nu_2} < m_J < 2 m_{\nu_3}$ whereas in the third one $2 m_{\nu_1} < m_J < 2 m_{\nu_2}$. Scenarios with $m_J < 2 m_{\nu_1}$ are not represented in both top panels of \cref{fig:nu}.

In the top-left plot of \cref{fig:nu} we show the Majoron decay lifetime $\tau_{_{J\to \nu \nu}} = \Gamma^{-1}(J \to \nu \nu)$ normalized to the age of the Universe $\tau_0 = 13.787~\mathrm{Gyr}$ \cite{Nikolic:2020fom}. The Majoron decay width to a pair of neutrinos is given in \cref{eq:Jnunu}. Note that \texttt{Planck2018} only marginally constrains our parameter space and do not affect the magenta band where the amplitude of the SGWB peaks are within LISA sensitivity reach.

Last but not least, the bottom panel in \cref{fig:nu} shows the size of the product of third generation Yukawa couplings $y_{\nu_3}^2 y_{\sigma_3}$ against the heavy neutrino masses scale $\Lambda$. These parameters provide the leading contribution to the active neutrino mass $m_{\nu_3}$ according to \cref{eq:y-sig}. One can see that the two regions previously identified, the magenta band with strong FOPTs and observable SGWB, and the blue one where FOPTs are weak and beyond reach, partially overlap, with the former typically favouring smaller $y_{\nu_3}^2 y_{\sigma_3}$. A similar behaviour was found for the first and second generation Yukawa couplings.

\section{Conclusions}
\label{sec:con}

With the LISA mission scheduled to begin operations during the 2030 decade, and with the high-luminosity phase of the LHC expected to deliver at least an order of magnitude more data than so far collected, an opportunity to scrutinize a wealth of New Physics models in multiple channels is opening up. The key goal, and yet a great challenge, is to understand how to obtain information relevant for our favourite HEP models from SGWB measurements. In this article we have focused on a well motivated Majoron scenario equipped with an extended inverse seesaw mechanism and dimension-six effective operators in the scalar sector as a benchmark example for the type of physics to explore with LISA.

We have first verified that, in the absence of dimension-six operators, the only portal coupling in the theory is bounded to be small due to severe constraints from invisible Higgs decays. This results in a small potential barrier between the true and false vacua, resulting in weak FOPTs with too small SGWB peak amplitudes to be observable in a foreseeable future. A model with purely renormalizable operators is justified either if it is assumed to be UV complete or if new particles are several orders of magnitude heavier, thus decoupled. However, if New Physics is not significantly heavier, say 1 to 3 orders of magnitude larger, effective higher dimensional operators in the scalar sector play a game-changing role as we have demonstrated. In particular, the emergence of new portal-like interactions offers extra freedom to sufficiently enhance the potential barrier between the true and false vacua while keeping the invisible Higgs decay rate under control. This is achieved with not too small couplings of order $\mathcal{O}(0.1)$ to $\mathcal{O}(1)$, and a mild $1\%$ to $10\%$ cancellation among them.

We have also searched for correlations between collider and astrophysical observables and found that, for the EIS model, observable SGWB signatures at LISA indicate a preference for a triliniear Higgs self coupling modifier within the range $0 < \kappa_\lambda < 2$ and a new CP-even Higgs boson mass $m_{h_2} \approx (200 \pm 50)~\mathrm{GeV}$. Direct searches for new scalars at the LHC as well as improved cosmological bounds from CMB data, can further constrain the allowed parameter space and shed new light on the current picture, with relevance for the future LISA mission.


\bigskip

\acknowledgments
A.A.~work is supported by the Talent Scientific Research Program of College of Physics, Sichuan University, Grant No.1082204112427 \& the Fostering Program in Disciplines Possessing Novel Features for Natural Science of Sichuan University,  Grant No. 2020SCUNL209 \& 1000 Talent program of Sichuan province 2021. 
A.M.~wishes to acknowledge support by the Shanghai Municipality, through the grant No.~KBH1512299, by Fudan University, through the grant No.~JJH1512105, the Natural Science Foundation of China, through the grant No.~11875113, and by the Department of Physics at Fudan University, through the grant No.~IDH1512092/001.
This work was supported by the grants CERN/FIS-PAR/0021/2021, CERN/FIS-PAR/0019/2021, CERN/FIS-PAR/0024/2021, CERN/FIS-PAR/0025/2021 and PTDC/FIS-AST/3041/2020. A.P.M.~is supported by the Center for Research and Development in Mathematics and Applications (CIDMA) through the Portuguese Foundation for Science and Technology (FCT - Fundação para a Ciência e a Tecnologia), references UIDB/04106/2020 and UIDP/04106/2020, and by national funds (OE), through FCT, I.P., in the scope of the framework contract foreseen in the numbers 4, 5 and 6 of the article 23, of the Decree-Law 57/2016, of August 29, changed by Law 57/2017, of July 19. A.P.M.~also wants to thank Gongjun Choi, Jérémie Quevillon and Miguel Escudero Abenza for insightful discussions about the cosmology of Majorons and axion-like particles in general. A.P.M.~also acknowledges Rui Santos, Tania Robens and  Johannes Braathen for discussions about the trilinar Higgs coupling and invisible Higgs decays.
R.P.~is supported in part by the Swedish Research Council grant, contract number 2016-05996, as well as by the European Research Council (ERC) under the European Union's Horizon 2020 research and innovation programme (grant agreement No 668679).
J.V.~is supported by FCT under contracts UIDB/00618/2020, UIDP/00618/2020,
PTDC/FIS-PAR/31000/2017, PRT/BD/154191/2022, CERN/FIS-PAR/0025/2021.
\appendix

\section{One-loop expressions for the physical trilinear couplings}
\label{app:tri}

One-loop corrections to the coupling of the Higgs boson with a pair of Majorons solely result from heavy CP-even Higgs bosons in the zero external momentum approach. These are given by
\begin{equation}
    \begin{aligned}
         \lambda_{h_1 J J}^{h_2}  = & f_{J J h_2} \lambda_{J J h_1}^{(0)} \left(\lambda_{J J h_2}^{(0)}\right)^2
         +
         2  f_{J h_1 h_2} \lambda_{J J h_1}^{(0)} \lambda_{J J h_2}^{(0)} \lambda_{h_1 h_1 h_2}^{(0)}
         +
          f_{J h_2 h_2} \lambda_{h_1 h_2 h_2}^{(0)} \left(\lambda_{J J h_2}^{(0)}\right)^2
        \\
        +&
        2 (f_{J h_2} - 1) \lambda_{J J h_2}^{(0)} \lambda_{J J h_1 h_2}^{(0)}
        +
        (f_{h_1 h_2} - 1) \lambda_{h_1 h_1 h_2}^{(0)} \lambda_{J J h_1 h_2}^{(0)}
        +
        \frac12 (f_{h_2 h_2} - 1) \lambda_{h_1 h_2 h_2}^{(0)} \lambda_{J J h_2 h_2}^{(0)}
    \end{aligned}
    \label{eq:lhJJh2}
\end{equation}

For the trilinear Higgs coupling we start with the fermionic contributions from top-quark and heavy neutrino loops:
\begin{equation}
    \begin{aligned}
        \lambda_{h_1 h_1 h_1}^t =& -\left[\frac{12}{6} f_{ttt} \lambda^{(0)}_{tth_1} + \frac{18}{6} (f_{tt} - 1) \lambda^{(0)}_{tt h_1} \lambda^{(0)}_{tt h_1 h_1}\right]\,,
        \\
        \lambda_{h_1 h_1 h_1}^N \approx & -2 f_{NNN}\left[ \sum_{j = 1}^6 \left(\lambda^{(0)}_{N_j N_j h_1} \right)^3 
        + 3 \sum_{j = 1 }^3  \lambda^{(0)}_{N_{2j-1} N_{2j-1} h_1} \left( \lambda^{(0)}_{N_{2j-1} N_{2j} h_1}  \right)^2 \right. 
        \\
        &+ 
        \left. 3 \sum_{j = 1 }^3 \lambda^{(0)}_{N_{2j} N_{2j} h_1} \left( \lambda^{(0)}_{N_{2j-1} N_{2j} h_1}  \right)^2
        \right] 
        -3 \left( f_{NN} - 1\right) \left[ \sum_{j = 1}^6 \lambda^{(0)}_{N_j N_j h_1} \lambda^{(0)}_{N_j N_j h_1 h_1} 
        \right.
        \\
        &+
        \left.
        2 \sum_{j = 1 }^3 \lambda^{(0)}_{N_{2j-1} N_{2j} h_1} \lambda^{(0)}_{N_{2j-1} N_{2j} h_1 h_1}
        \right]
        \,.
    \end{aligned}
    \label{eq:lhhhN}
\end{equation}
Note that in \cref{eq:lhhhN} we have considered, to a good approximation, the limit of degenerate heavy neutrino masses motivated by $m_{N_{1,\ldots,6}} \approx \Lambda$. Last but not least, the scalar contributions to the Higgs trilinear coupling read as
\begin{equation}
    \begin{aligned}
          \lambda_{h_1 h_1 h_1}^{h_2} =& 3 f_{h_1 h_1 h_2} \lambda^{(0)}_{h_1 h_1 h_1} \left( \lambda^{(0)}_{h_1 h_1 h_2} \right)^2 
          + 3 f_{h_1 h_2 h_2} \lambda^{(0)}_{h_1 h_2 h_2} \left( \lambda^{(0)}_{h_1 h_1 h_2} \right)^2
          + f_{h_2 h_2 h_2} \left( \lambda^{(0)}_{h_2 h_2 h_2} \right)^2
          \\
          +& 3\left( f_{h_1 h_2} - 1\right) \lambda^{(0)}_{h_1 h_1 h_2} \lambda^{(0)}_{h_1 h_1 h_1 h_2}
          + \frac32 \left( f_{h_2 h_2} - 1\right) \lambda^{(0)}_{h_1 h_2 h_2} \lambda^{(0)}_{h_1 h_1 h_2 h_2}\,.
    \end{aligned}
    \label{eq:lhhhh2}
\end{equation}
The loop functions, as defined in \cite{Camargo-Molina:2016moz}, read as
\begin{equation}
    f_{a_1 \ldots a_N} = \sum_{x=1}^N \frac{m_{a_x}^2 \log\left(\frac{m_{a_x}^2}{\mu^2}\right)}{\Pi_{y \neq x} \left( m_{a_x}^2 - m_{a_y}^2 \right)}\,,
    \label{eq:f}
\end{equation}
whith $\mu$ the renormalization scale. While the $\lambda_{J J h_1}^{(0)}$ is given in \cref{eq:Lhjj}, all remaining tree-level couplings entering \cref{eq:lhJJh2,eq:lhhhN,eq:lhhhh2} read as:
%
\begin{equation}
    \begin{aligned}
        \lambda_{J J h_2}^{(0)} = \frac{\cos (\alpha_h ) v_{\sigma } \left(\delta _4 v_h^2+2 \Lambda ^2 \lambda _{\sigma }+3 \delta _6
   v_{\sigma }^2\right)-\sin (\alpha_h ) v_h \left(\delta _2 v_h^2+\Lambda ^2 \lambda _{\sigma h}+\delta _4 v_{\sigma }^2\right)}{\Lambda ^2}
    \end{aligned}
    \label{JJh2}
\end{equation}
\begin{equation}
    \begin{aligned}
        \lambda_{J J h_1 h_2}^{(0)} = \frac{\sin \left(2 \alpha _h\right) \left(\left(\delta _4-3 \delta _2\right) v_h^2+\Lambda ^2 \left(2
   \lambda _{\sigma }-\lambda _{\sigma h}\right)-\left(\delta _4-9 \delta _6\right) v_{\sigma
   }^2\right)+4 \delta _4 v_h v_{\sigma } \cos \left(2 \alpha _h\right)}{2 \Lambda ^2}
    \end{aligned}
    \label{JJh1h2}
\end{equation}
\begin{equation}
    \begin{aligned}
        \lambda_{J J h_2 h_2}^{(0)} &= \frac{1}{\Lambda^2} \left[\sin ^2\left(\alpha _h\right) \left(3 \delta _2 v_h^2+\Lambda ^2 \lambda _{\sigma h}+\delta _4 v_{\sigma }^2\right)+\cos ^2\left(\alpha _h\right) \left(\delta _4 v_h^2+2 \Lambda ^2
   \lambda _{\sigma }+9 \delta _6 v_{\sigma }^2\right)
   \right.
   \\
   &-
   \left.
   4 \delta _4 v_h v_{\sigma } \sin \left(\alpha
   _h\right) \cos \left(\alpha _h\right) \right]
    \end{aligned}
    \label{JJh2h2}
\end{equation}
\begin{equation}
    \begin{aligned}
    \scalemath{0.9}{
        \lambda_{h_1 h_1 h_1}^{(0)}} & \scalemath{0.9}{= \frac{3}{\Lambda^2} \left[v_{\sigma } \sin \left(\alpha _h\right) \cos ^2\left(\alpha _h\right) \left(3 \delta _2
   v_h^2+\Lambda ^2 \lambda _{\sigma h}+\delta _4 v_{\sigma }^2\right)
   +
   v_h \cos ^3\left(\alpha _h\right)
   \left(2 \Lambda ^2 \lambda _h+5 \delta _0 v_h^2+\delta _2 v_{\sigma }^2\right)
   \right.}
    \\
    &\scalemath{0.9}{+
    \left.
   v_h \sin ^2\left(\alpha
   _h\right) \cos \left(\alpha _h\right) \left(\delta _2 v_h^2+\Lambda ^2 \lambda _{\sigma h}+3
   \delta _4 v_{\sigma }^2\right)+v_{\sigma } \sin ^3\left(\alpha _h\right) \left(\delta _4 v_h^2+2
   \Lambda ^2 \lambda _{\sigma }+5 \delta _6 v_{\sigma }^2\right)\right]}
    \end{aligned}
    \label{111}
\end{equation}
\begin{equation}
    \begin{aligned}
    \scalemath{0.9}{
        \lambda_{h_2 h_2 h_2}^{(0)}} & \scalemath{0.9}{= \frac{3}{\Lambda^2} \left[-v_h \sin \left(\alpha _h\right) \cos ^2\left(\alpha _h\right) \left(\delta _2
   v_h^2+\Lambda ^2 \lambda _{\sigma h}+3 \delta _4 v_{\sigma }^2\right)
   +v_{\sigma } \cos ^3\left(\alpha
   _h\right) \left(\delta _4 v_h^2+2 \Lambda ^2 \lambda _{\sigma }+5 \delta _6 v_{\sigma
   }^2\right)
   \right.}
   \\
   &+
   \scalemath{0.9}{\left.
   v_{\sigma } \sin
   ^2\left(\alpha _h\right) \cos \left(\alpha _h\right) \left(3 \delta _2 v_h^2+\Lambda ^2 \lambda
   _{\sigma h}+\delta _4 v_{\sigma }^2\right)-v_h \sin ^3\left(\alpha _h\right) \left(2 \Lambda
   ^2 \lambda _h+5 \delta _0 v_h^2+\delta _2 v_{\sigma }^2\right)\right]}
    \end{aligned}
    \label{222}
\end{equation}
\begin{equation}
    \begin{aligned}
    \lambda_{h_1 h_1 h_2}^{(0)} & =\frac{1}{\Lambda^2} \left[-v_h \sin ^3\left(\alpha _h\right) \left(\delta _2 v_h^2+\Lambda ^2 \lambda _{\sigma h}+3
   \delta _4 v_{\sigma }^2\right)+v_{\sigma } \cos ^3\left(\alpha _h\right) \left(3 \delta _2
   v_h^2+\Lambda ^2 \lambda _{\sigma h}+\delta _4 v_{\sigma }^2\right)
   \right.
   \\
   &+
   \left.
   v_h \sin \left(\alpha
   _h\right) \cos ^2\left(\alpha _h\right) \left(2 \Lambda ^2 \left(\lambda _{\sigma h}-3
   \lambda _h\right)+\left(2 \delta _2-15 \delta _0\right) v_h^2-3 \left(\delta _2-2 \delta _4\right)
   v_{\sigma }^2\right)
   \right.
   \\
   &+
   \left.
   v_{\sigma } \sin ^2\left(\alpha _h\right) \cos \left(\alpha _h\right) \left(3
   \left(\delta _4-2 \delta _2\right) v_h^2+2 \Lambda ^2 \left(3 \lambda _{\sigma }-\lambda
   _{\sigma h}\right)+\left(15 \delta _6-2 \delta _4\right) v_{\sigma }^2\right)\right]
    \end{aligned}
    \label{112}
\end{equation}
\begin{equation}
    \begin{aligned}
    \lambda_{h_1 h_2 h_2}^{(0)} & = \frac{1}{\Lambda^2} \left[v_{\sigma } \sin ^3\left(\alpha _h\right) \left(3 \delta _2 v_h^2+\Lambda ^2 \lambda
   _{\sigma h}+\delta _4 v_{\sigma }^2\right)+v_h \cos ^3\left(\alpha _h\right) \left(\delta _2
   v_h^2+\Lambda ^2 \lambda _{\sigma h}+3 \delta _4 v_{\sigma }^2\right)
   \right.
   \\
   &+
   \left.
   v_{\sigma } \sin
   \left(\alpha _h\right) \cos ^2\left(\alpha _h\right) \left(3 \left(\delta _4-2 \delta _2\right)
   v_h^2+2 \Lambda ^2 \left(3 \lambda _{\sigma }-\lambda _{\sigma h}\right)+\left(15 \delta
   _6-2 \delta _4\right) v_{\sigma }^2\right)
   \right.
   \\
   &+
   \left.
   v_h \sin ^2\left(\alpha _h\right) \cos \left(\alpha
   _h\right) \left(2 \Lambda ^2 \left(3 \lambda _h-\lambda _{\sigma h}\right)+\left(15 \delta
   _0-2 \delta _2\right) v_h^2+3 \left(\delta _2-2 \delta _4\right) v_{\sigma }^2\right) \right]
    \end{aligned}
    \label{122}
\end{equation}
\begin{equation}
    \begin{aligned}
        \scalemath{0.9}{\lambda_{h_1 h_1 h_1 h_2}^{(0)} } &\scalemath{0.9}{= \frac{3}{8\Lambda^2} \left[-\sin \left(4 \alpha _h\right) \left(2 \Lambda ^2 \left(\lambda _h+\lambda _{\sigma }-\lambda _{\sigma h}\right)+\left(15 \delta _0-6 \delta
   _2+\delta _4\right) v_h^2
   \right. \right.}
   \\
   &+
   \scalemath{0.9}{\left. \left.
   \left(\delta _2-6 \delta _4+15 \delta _6\right) v_{\sigma }^2\right)+2 \sin \left(2 \alpha _h\right) \left(2 \Lambda ^2 \left(\lambda
   _{\sigma }-\lambda _h\right)+\left(\delta _4-15 \delta _0\right) v_h^2-\left(\delta _2-15 \delta _6\right) v_{\sigma }^2\right)
   \right.}
   \\
   &
   \scalemath{0.9}{+
   \left.
   8 \left(\delta _2-\delta
   _4\right) v_h v_{\sigma } \cos \left(4 \alpha _h\right)+8 \left(\delta _2+\delta _4\right) v_h v_{\sigma } \cos \left(2 \alpha _h\right)\right]}
    \end{aligned}
    \label{h1h1h1h2}
\end{equation}
\begin{equation}
    \begin{aligned}
        \scalemath{0.9}{\lambda_{h_1 h_1 h_2 h_2}^{(0)} } &= \frac{1}{\Lambda^2} \left[2 \Lambda ^2 \left(3 \left(\lambda _h+\lambda _{\sigma }\right)+\lambda _{\sigma h}\right)-3 \cos \left(4 \alpha _h\right) \left(2 \Lambda ^2
   \left(\lambda _h+\lambda _{\sigma }-\lambda _{\sigma h}\right)
   \right. \right.
   \\
   &+
   \left. \left.
   \left(15 \delta _0-6 \delta _2+\delta _4\right) v_h^2+\left(\delta _2-6 \delta _4+15
   \delta _6\right) v_{\sigma }^2\right)+24 \left(\delta _4-\delta _2\right) v_h v_{\sigma } \sin \left(4 \alpha _h\right)
   \right.
   \\
   &+
   \left.
   3 \left(15 \delta _0+2 \delta _2+\delta
   _4\right) v_h^2+3 \left(\delta _2+2 \delta _4+15 \delta _6\right) v_{\sigma }^2 \right]
    \end{aligned}
    \label{h1h1h2h2}
\end{equation}
\begin{equation}
    \begin{aligned}
        \lambda_{t t h_1}^{(0)} = v_h y_t^2 \cos \left(\alpha _h\right)
    \end{aligned}
    \label{tth1}
\end{equation}
\begin{equation}
    \begin{aligned}
        \lambda_{t t h_1 h_1}^{(0)} = y_t^2 \cos ^2\left(\alpha _h\right)
    \end{aligned}
    \label{tth1h1}
\end{equation}
\begin{equation}
    \begin{aligned}
    \scalemath{0.8}{
        \lambda_{N_{1} N_{1} h_1}^{(0)} = 8 v_{\sigma } y_{\sigma _1}^2 \sin \left(\alpha _h\right) \left(1-\frac{v_{\sigma } y_{\sigma _1}}{\sqrt{2 \Lambda ^2+v_{\sigma }^2 y_{\sigma _1}^2}}\right)-\frac{8
   \sin \left(\alpha _h\right) \left(\sqrt{2 \Lambda ^2 v_{\sigma }^2 y_{\sigma _1}^2+v_{\sigma }^4 y_{\sigma _1}^4}+v_{\sigma }^2 y_{\sigma _1}^2\right)}{v_{\sigma
   } \sqrt{\frac{v_{\sigma } y_{\sigma _1} \left(\sqrt{2 \Lambda ^2+v_{\sigma }^2 y_{\sigma _1}^2}+v_{\sigma } y_{\sigma _1}\right)}{\Lambda ^2}+2}
   \sqrt{\frac{\sqrt{2 \Lambda ^2 v_{\sigma }^2 y_{\sigma _1}^2+v_{\sigma }^4 y_{\sigma _1}^4}+v_{\sigma }^2 y_{\sigma _1}^2}{\Lambda ^2}+2}}
        }
    \end{aligned}
    \label{N1N1h1}
\end{equation}
\begin{equation}
    \begin{aligned}
    \scalemath{0.8}{
        \lambda_{N_{2} N_{2} h_1}^{(0)} = 2 v_{\sigma } y_{\sigma _1}^2 \sin \left(\alpha _h\right) \left(1+\frac{v_{\sigma } y_{\sigma _1}}{\sqrt{2 \Lambda ^2+v_{\sigma }^2 y_{\sigma _1}^2}}\right)+\frac{2 \sin \left(\alpha _h\right) \left(\sqrt{2 \Lambda ^2 v_{\sigma }^2 y_{\sigma _1}^2+v_{\sigma }^4 y_{\sigma _1}^4}-v_{\sigma }^2 y_{\sigma _1}^2\right)}{v_{\sigma } \sqrt{\frac{v_{\sigma } y_{\sigma _1} \left(v_{\sigma } y_{\sigma _1}-\sqrt{2 \Lambda ^2+v_{\sigma }^2 y_{\sigma _1}^2}\right)}{\Lambda ^2}+2} \sqrt{\frac{v_{\sigma }^2 y_{\sigma _1}^2-\sqrt{2 \Lambda ^2 v_{\sigma }^2 y_{\sigma _1}^2+v_{\sigma }^4 y_{\sigma _1}^4}}{\Lambda ^2}+2}}
        }
    \end{aligned}
    \label{N2N2h1}
\end{equation}
\begin{equation}
    \begin{aligned}
    \scalemath{0.8}{
        \lambda_{N_{3} N_{3} h_1}^{(0)} = 2 v_{\sigma } y_{\sigma _2}^2 \sin \left(\alpha _h\right) \left(1-\frac{v_{\sigma } y_{\sigma _2}}{\sqrt{2 \Lambda ^2+v_{\sigma }^2 y_{\sigma _2}^2}}\right)-\frac{2 \sin \left(\alpha _h\right) \left(\sqrt{2 \Lambda ^2 v_{\sigma }^2 y_{\sigma _2}^2+v_{\sigma }^4 y_{\sigma _2}^4}+v_{\sigma }^2 y_{\sigma _2}^2\right)}{v_{\sigma } \sqrt{\frac{v_{\sigma } y_{\sigma _2} \left(\sqrt{2 \Lambda ^2+v_{\sigma }^2 y_{\sigma _2}^2}+v_{\sigma } y_{\sigma _2}\right)}{\Lambda ^2}+2} \sqrt{\frac{\sqrt{2 \Lambda ^2 v_{\sigma }^2 y_{\sigma _2}^2+v_{\sigma }^4 y_{\sigma _2}^4}+v_{\sigma }^2 y_{\sigma _2}^2}{\Lambda ^2}+2}}
        }
    \end{aligned}
    \label{N3N3h1}
\end{equation}
\begin{equation}
    \begin{aligned}
    \scalemath{0.8}{
        \lambda_{N_{4} N_{4} h_1}^{(0)} = 2 v_{\sigma } y_{\sigma _2}^2 \sin \left(\alpha _h\right) \left(1+\frac{v_{\sigma } y_{\sigma _2}}{\sqrt{2 \Lambda ^2+v_{\sigma }^2 y_{\sigma _2}^2}}\right)+\frac{2 \sin \left(\alpha _h\right) \left(\sqrt{2 \Lambda ^2 v_{\sigma }^2 y_{\sigma _2}^2+v_{\sigma }^4 y_{\sigma _2}^4}-v_{\sigma }^2 y_{\sigma _2}^2\right)}{v_{\sigma } \sqrt{\frac{v_{\sigma } y_{\sigma _2} \left(v_{\sigma } y_{\sigma _2}-\sqrt{2 \Lambda ^2+v_{\sigma }^2 y_{\sigma _2}^2}\right)}{\Lambda ^2}+2} \sqrt{\frac{v_{\sigma }^2 y_{\sigma _2}^2-\sqrt{2 \Lambda ^2 v_{\sigma }^2 y_{\sigma _2}^2+v_{\sigma }^4 y_{\sigma _2}^4}}{\Lambda ^2}+2}}
        }
    \end{aligned}
    \label{N4N4h1}
\end{equation}
\begin{equation}
    \begin{aligned}
    \scalemath{0.80}{
        \lambda_{N_{5} N_{5} h_1}^{(0)} = 2 v_{\sigma } y_{\sigma _3}^2 \sin \left(\alpha _h\right) \left(1-\frac{v_{\sigma } y_{\sigma _3}}{\sqrt{2 \Lambda ^2+v_{\sigma }^2 y_{\sigma _3}^2}}\right)-\frac{2 \sin \left(\alpha _h\right) \left(\sqrt{2 \Lambda ^2 v_{\sigma }^2 y_{\sigma _3}^2+v_{\sigma }^4 y_{\sigma _3}^4}+v_{\sigma }^2 y_{\sigma _3}^2\right)}{v_{\sigma } \sqrt{\frac{v_{\sigma } y_{\sigma _3} \left(\sqrt{2 \Lambda ^2+v_{\sigma }^2 y_{\sigma _3}^2}+v_{\sigma } y_{\sigma _3}\right)}{\Lambda ^2}+2} \sqrt{\frac{\sqrt{2 \Lambda ^2 v_{\sigma }^2 y_{\sigma _3}^2+v_{\sigma }^4 y_{\sigma _3}^4}+v_{\sigma }^2 y_{\sigma _3}^2}{\Lambda ^2}+2}}
        }
    \end{aligned}
    \label{N5N5h1}
\end{equation}
\begin{equation}
    \begin{aligned}
    \scalemath{0.80}{
        \lambda_{N_{6} N_{6} h_1}^{(0)} = 2 v_{\sigma } y_{\sigma _3}^2 \sin \left(\alpha _h\right) \left(1+\frac{v_{\sigma } y_{\sigma _3}}{\sqrt{2 \Lambda ^2+v_{\sigma }^2 y_{\sigma _3}^2}}\right)+\frac{2 \sin \left(\alpha _h\right) \left(\sqrt{2 \Lambda ^2 v_{\sigma }^2 y_{\sigma _3}^2+v_{\sigma }^4 y_{\sigma _3}^4}-v_{\sigma }^2 y_{\sigma _3}^2\right)}{v_{\sigma } \sqrt{\frac{v_{\sigma } y_{\sigma _3} \left(v_{\sigma } y_{\sigma _3}-\sqrt{2 \Lambda ^2+v_{\sigma }^2 y_{\sigma _3}^2}\right)}{\Lambda ^2}+2} \sqrt{\frac{v_{\sigma }^2 y_{\sigma _3}^2-\sqrt{2 \Lambda ^2 v_{\sigma }^2 y_{\sigma _3}^2+v_{\sigma }^4 y_{\sigma _3}^4}}{\Lambda ^2}+2}}
        }
    \end{aligned}
    \label{N6N6h1}
\end{equation}
\begin{equation}
    \begin{aligned}
        \lambda_{N_{1} N_{2} h_1}^{(0)} &= \frac{16 v_{\sigma } y_{\sigma _1}^2 \sin \left(\alpha _h\right)}{\sqrt{\frac{v_{\sigma } y_{\sigma _1} \left(v_{\sigma } y_{\sigma _1}-\sqrt{2 \Lambda ^2+v_{\sigma }^2 y_{\sigma _1}^2}\right)}{\Lambda ^2}+2} \sqrt{\frac{v_{\sigma } y_{\sigma _1} \left(\sqrt{2 \Lambda ^2+v_{\sigma }^2 y_{\sigma _1}^2}+v_{\sigma } y_{\sigma _1}\right)}{\Lambda ^2}+2}}
        \\
        &
        +\frac{4 \sin \left(\alpha _h\right) \left(\sqrt{2 \Lambda ^2 v_{\sigma }^2 y_{\sigma _1}^2+v_{\sigma }^4 y_{\sigma _1}^4}-v_{\sigma }^2 y_{\sigma _1}^2\right)}{v_{\sigma } \sqrt{\frac{v_{\sigma } y_{\sigma _1} \left(\sqrt{2 \Lambda ^2+v_{\sigma }^2 y_{\sigma _1}^2}+v_{\sigma } y_{\sigma _1}\right)}{\Lambda ^2}+2} \sqrt{\frac{v_{\sigma }^2 y_{\sigma _1}^2-\sqrt{2
   \Lambda ^2 v_{\sigma }^2 y_{\sigma _1}^2+v_{\sigma }^4 y_{\sigma _1}^4}}{\Lambda ^2}+2}}
   \\
   &
   -\frac{4 \sin \left(\alpha _h\right) \left(\sqrt{2 \Lambda ^2 v_{\sigma }^2 y_{\sigma _1}^2+v_{\sigma }^4 y_{\sigma _1}^4}+v_{\sigma }^2 y_{\sigma _1}^2\right)}{v_{\sigma } \sqrt{\frac{v_{\sigma } y_{\sigma _1} \left(v_{\sigma } y_{\sigma _1}-\sqrt{2 \Lambda ^2+v_{\sigma }^2 y_{\sigma _1}^2}\right)}{\Lambda ^2}+2} \sqrt{\frac{\sqrt{2 \Lambda ^2 v_{\sigma }^2 y_{\sigma _1}^2+v_{\sigma }^4 y_{\sigma _1}^4}+v_{\sigma }^2 y_{\sigma _1}^2}{\Lambda ^2}+2}}
    \end{aligned}
    \label{N1N2h1}
\end{equation}
\begin{equation}
    \begin{aligned}
        \lambda_{N_{3} N_{4} h_1}^{(0)} &= \frac{16 v_{\sigma } y_{\sigma _2}^2 \sin \left(\alpha _h\right)}{\sqrt{\frac{v_{\sigma } y_{\sigma _2} \left(v_{\sigma } y_{\sigma _2}-\sqrt{2 \Lambda ^2+v_{\sigma }^2 y_{\sigma _2}^2}\right)}{\Lambda ^2}+2} \sqrt{\frac{v_{\sigma } y_{\sigma _2} \left(\sqrt{2 \Lambda ^2+v_{\sigma }^2 y_{\sigma _2}^2}+v_{\sigma } y_{\sigma _2}\right)}{\Lambda ^2}+2}}
        \\
        &
        +\frac{4\sin \left(\alpha _h\right) \left(\sqrt{2 \Lambda ^2 v_{\sigma }^2 y_{\sigma _2}^2+v_{\sigma }^4 y_{\sigma _2}^4}-v_{\sigma }^2 y_{\sigma _2}^2\right)}{v_{\sigma } \sqrt{\frac{v_{\sigma } y_{\sigma _2} \left(\sqrt{2 \Lambda ^2+v_{\sigma }^2 y_{\sigma _2}^2}+v_{\sigma } y_{\sigma _2}\right)}{\Lambda ^2}+2} \sqrt{\frac{v_{\sigma }^2 y_{\sigma _2}^2-\sqrt{2 \Lambda ^2
   v_{\sigma }^2 y_{\sigma _2}^2+v_{\sigma }^4 y_{\sigma _2}^4}}{\Lambda ^2}+2}}
   \\
   &
   -\frac{4\sin \left(\alpha _h\right) \left(\sqrt{2 \Lambda ^2 v_{\sigma }^2 y_{\sigma _2}^2+v_{\sigma }^4 y_{\sigma _2}^4}+v_{\sigma }^2 y_{\sigma _2}^2\right)}{v_{\sigma } \sqrt{\frac{v_{\sigma } y_{\sigma _2} \left(v_{\sigma } y_{\sigma _2}-\sqrt{2 \Lambda ^2+v_{\sigma }^2 y_{\sigma _2}^2}\right)}{\Lambda ^2}+2} \sqrt{\frac{\sqrt{2 \Lambda ^2 v_{\sigma }^2 y_{\sigma _2}^2+v_{\sigma }^4 y_{\sigma _2}^4}+v_{\sigma }^2 y_{\sigma _2}^2}{\Lambda ^2}+2}}
    \end{aligned}
    \label{N3N4h1}
\end{equation}
\begin{equation}
    \begin{aligned}
        \lambda_{N_{5} N_{6} h_1}^{(0)} &= \frac{16 v_{\sigma } y_{\sigma _3}^2 \sin \left(\alpha _h\right)}{\sqrt{\frac{v_{\sigma } y_{\sigma _3} \left(v_{\sigma } y_{\sigma _3}-\sqrt{2 \Lambda ^2+v_{\sigma }^2 y_{\sigma _3}^2}\right)}{\Lambda ^2}+2} \sqrt{\frac{v_{\sigma } y_{\sigma _3} \left(\sqrt{2 \Lambda ^2+v_{\sigma }^2 y_{\sigma _3}^2}+v_{\sigma } y_{\sigma _3}\right)}{\Lambda ^2}+2}}
        \\
        &
        +\frac{4\sin \left(\alpha _h\right) \left(\sqrt{2 \Lambda ^2 v_{\sigma }^2 y_{\sigma _3}^2+v_{\sigma }^4 y_{\sigma _3}^4}-v_{\sigma }^2 y_{\sigma _3}^2\right)}{v_{\sigma } \sqrt{\frac{v_{\sigma } y_{\sigma _3} \left(\sqrt{2 \Lambda ^2+v_{\sigma }^2 y_{\sigma _3}^2}+v_{\sigma } y_{\sigma _3}\right)}{\Lambda ^2}+2} \sqrt{\frac{v_{\sigma }^2 y_{\sigma _3}^2-\sqrt{2 \Lambda ^2
   v_{\sigma }^2 y_{\sigma _3}^2+v_{\sigma }^4 y_{\sigma _3}^4}}{\Lambda ^2}+2}}
   \\
   &
   -\frac{4\sin \left(\alpha _h\right) \left(\sqrt{2 \Lambda ^2 v_{\sigma }^2 y_{\sigma _3}^2+v_{\sigma }^4 y_{\sigma _3}^4}+v_{\sigma }^2 y_{\sigma _3}^2\right)}{v_{\sigma } \sqrt{\frac{v_{\sigma } y_{\sigma _3} \left(v_{\sigma } y_{\sigma _3}-\sqrt{2 \Lambda ^2+v_{\sigma }^2 y_{\sigma _3}^2}\right)}{\Lambda ^2}+2} \sqrt{\frac{\sqrt{2 \Lambda ^2 v_{\sigma }^2 y_{\sigma _3}^2+v_{\sigma }^4 y_{\sigma _3}^4}+v_{\sigma }^2 y_{\sigma _3}^2}{\Lambda ^2}+2}}
    \end{aligned}
    \label{N5N6h1}
\end{equation}
\begin{equation}
    \begin{aligned}
        \lambda_{N_{1} N_{1} h_1 h_1}^{(0)} = 8 y_{\sigma _1}^2 \sin ^2\left(\alpha _h\right) \left(1-\frac{v_{\sigma } y_{\sigma _1}}{\sqrt{2 \Lambda ^2+v_{\sigma }^2 y_{\sigma _1}^2}}\right)
    \end{aligned}
    \label{N1N1h1h1}
\end{equation}
\begin{equation}
    \begin{aligned}
        \lambda_{N_{2} N_{2} h_1 h_1}^{(0)} = 8 y_{\sigma _1}^2 \sin ^2\left(\alpha _h\right) \left(1+\frac{v_{\sigma } y_{\sigma _1}}{\sqrt{2 \Lambda ^2+v_{\sigma }^2 y_{\sigma _1}^2}}\right)
    \end{aligned}
    \label{N2N2h1h1}
\end{equation}
\begin{equation}
    \begin{aligned}
        \lambda_{N_{3} N_{3} h_1 h_1}^{(0)} = 8 y_{\sigma _2}^2 \sin ^2\left(\alpha _h\right) \left(1-\frac{v_{\sigma } y_{\sigma _2}}{\sqrt{2 \Lambda ^2+v_{\sigma }^2 y_{\sigma _2}^2}}\right)
    \end{aligned}
    \label{N3N3h1h1}
\end{equation}
\begin{equation}
    \begin{aligned}
        \lambda_{N_{4} N_{4} h_1 h_1}^{(0)} = 8 y_{\sigma _2}^2 \sin ^2\left(\alpha _h\right) \left(1+\frac{v_{\sigma } y_{\sigma _2}}{\sqrt{2 \Lambda ^2+v_{\sigma }^2 y_{\sigma _2}^2}}\right)
    \end{aligned}
    \label{N4N4h1h1}
\end{equation}
\begin{equation}
    \begin{aligned}
        \lambda_{N_{5} N_{5} h_1 h_1}^{(0)} = 8 y_{\sigma _3}^2 \sin ^2\left(\alpha _h\right) \left(1-\frac{v_{\sigma } y_{\sigma _3}}{\sqrt{2 \Lambda ^2+v_{\sigma }^2 y_{\sigma _3}^2}}\right)
    \end{aligned}
    \label{N5N5h1h1}
\end{equation}
\begin{equation}
    \begin{aligned}
        \lambda_{N_{6} N_{6} h_1 h_1}^{(0)} = 8 y_{\sigma _3}^2 \sin ^2\left(\alpha _h\right) \left(1+\frac{v_{\sigma } y_{\sigma _3}}{\sqrt{2 \Lambda ^2+v_{\sigma }^2 y_{\sigma _3}^2}}\right)
    \end{aligned}
    \label{N6N6h1h1}
\end{equation}
\begin{equation}
    \begin{aligned}
        \lambda_{N_{1} N_{2} h_1 h_1}^{(0)} = \frac{16 y_{\sigma _1}^2 \sin ^2\left(\alpha _h\right)}{\sqrt{\frac{v_{\sigma } y_{\sigma _1} \left(v_{\sigma } y_{\sigma _1}-\sqrt{2 \Lambda ^2+v_{\sigma }^2 y_{\sigma _1}^2}\right)}{\Lambda ^2}+2} \sqrt{\frac{v_{\sigma } y_{\sigma _1} \left(\sqrt{2 \Lambda ^2+v_{\sigma }^2 y_{\sigma
   _1}^2}+v_{\sigma } y_{\sigma _1}\right)}{\Lambda ^2}+2}}
    \end{aligned}
    \label{N1N2h1h1}
\end{equation}
\begin{equation}
    \begin{aligned}
        \lambda_{N_{3} N_{4} h_1 h_1}^{(0)} = \frac{16 y_{\sigma _2}^2 \sin ^2\left(\alpha _h\right)}{\sqrt{\frac{v_{\sigma } y_{\sigma _2} \left(v_{\sigma } y_{\sigma _2}-\sqrt{2 \Lambda ^2+v_{\sigma }^2 y_{\sigma _2}^2}\right)}{\Lambda ^2}+2} \sqrt{\frac{v_{\sigma } y_{\sigma _2} \left(\sqrt{2 \Lambda ^2+v_{\sigma }^2 y_{\sigma
   _2}^2}+v_{\sigma } y_{\sigma _2}\right)}{\Lambda ^2}+2}}
    \end{aligned}
    \label{N3N4h1h1}
\end{equation}
\begin{equation}
    \begin{aligned}
        \lambda_{N_{5} N_{6} h_1 h_1}^{(0)} = \frac{16 y_{\sigma _3}^2 \sin ^2\left(\alpha _h\right)}{\sqrt{\frac{v_{\sigma } y_{\sigma _3} \left(v_{\sigma } y_{\sigma _3}-\sqrt{2 \Lambda ^2+v_{\sigma }^2 y_{\sigma _3}^2}\right)}{\Lambda ^2}+2} \sqrt{\frac{v_{\sigma } y_{\sigma _3} \left(\sqrt{2 \Lambda ^2+v_{\sigma }^2 y_{\sigma
   _3}^2}+v_{\sigma } y_{\sigma _3}\right)}{\Lambda ^2}+2}}
    \end{aligned}
    \label{N5N6h1h1}
\end{equation}

\bibliographystyle{JHEP}
\bibliography{biblio}

\end{document}